\documentclass[12pt,a4paper]{article}
\pdfoutput=1

\usepackage{latexsym}
\usepackage{array}
\usepackage{epsfig,color,colordvi}
\usepackage{jheppub}

\usepackage{graphicx}
\usepackage{autobreak}
\allowdisplaybreaks

\usepackage{floatrow}
\DeclareFloatFont{tiny}{\tiny}
\def\colour4colour#1{\Blue{#1}}

\newcolumntype{M}{>{$\displaystyle} c <{$}}
\newcolumntype{P}[1]{>{\centering\arraybackslash}p{#1}}
\newcommand{\btz}{\beta_0}

\renewcommand{\theequation}{\thesection.\arabic{equation}}

\newcommand{\hspn}{{\hspace{-4mm}}}

\newcommand{\beq}{\begin{equation}}
\newcommand{\eeq}{\end{equation}}
\newcommand{\bea}{\begin{eqnarray}}
\newcommand{\eea}{\end{eqnarray}}
\newcommand{\nn}{\nonumber}
\newcommand{\MSb}{$\overline{\mbox{MS}}$}
\newcommand{\ra}{\rightarrow}
\newcommand{\als}{\alpha_{\rm s}}
\newcommand{\ars}{a_{\rm s}}

\newcommand{\Lqr}{L_{\,\rm qr\,}}
\newcommand{\Lqrs}{L_{\,\rm qr\,}^{\,2}}
\newcommand{\Lqrt}{L_{\,\rm qr\,}^{\,3}}
\newcommand{\Lqrf}{L_{\,\rm qr\,}^{\,4}}
\newcommand{\Lfr}{L_{\,\rm fr\,}}
\newcommand{\Lfrs}{L_{\,\rm fr\,}^{\,2}}
\newcommand{\Lfrt}{L_{\,\rm fr\,}^{\,3}}
\newcommand{\Lfrf}{L_{\,\rm fr\,}^{\,4}}
\newcommand{\DD}{{\cal D}}
\newcommand{\nt}{\widetilde{N}}

\def\gg0#1DIS{{\tilde{g}_{0#1}}}
 \def\g0#1DISN{{g_{0#1}}}
 \def\z#1{{\zeta_{#1}}}
 \newcommand{\floo}{fl_{11}}
\newcommand{\dabcton}{\frac{d_{abc}d^{abc}}{n_c}}
\def\GE{{\gamma_{E}^{}}}
\newcommand{\Ca}{C_A}
\newcommand{\Cf}{C_F}
\newcommand{\nf}{n_f}

\def\z#1{{\zeta_{#1}}}
\def\zss{\zeta_2^{\,2}}

\def\zss{\zeta_2^{\,2}}
\def\zst{{\zeta_{2}^{\,3}}}

\def\zts{{\zeta_{3}^{\,2}}}

\def\nc{{n_c}}
\def\ncs{{n_{c}^{2}}}
\def\nct{{n_{c}^{3}}}
\def\ncf{{n_{c}^{4}}}

\def\ca{{C^{}_{\!A}}}
\def\cas{{C^{\,2}_{\!A}}}
\def\cat{{C^{\,3}_{\!A}}}

\def\cf{{C^{}_F}}
\def\cfs{{C^{\, 2}_F}}
\def\cft{{C^{\, 3}_F}}
\def\cff{{C^{\, 4}_F}}

\def\nf{{n^{}_{\! f}}}
\def\nfz{{n^{\,0}_{\! f}}}
\def\nfo{{n^{\,1}_{\! f}}}
\def\nfs{{n^{\,2}_{\! f}}}
\def\nft{{n^{\,3}_{\! f}}}

\def\cqonlnN{{g_{01}^{\rm DIS}}}
\def\cqsnlnN{{g_{02}^{\rm DIS}}}
\def\cqtnlnN{{g_{03}^{\rm DIS}}}
\def\fnoq{{f_{01}^{\rm q}}}
\def\fnsq{{f_{02}^{\rm q}}}
\def\fntq{{f_{03}^{\rm q}}}

\def\dabcnc{{{d^{abc}d_{abc}}\over{n_c}}}

\def\floo{fl_{11}}

\def\dfFAnc{{\frac{d_F^{\,abcd}d_A^{\,abcd}}{n_F }}}
\def\dfFFnc{{\frac{d_F^{\,abcd}d_F^{\,abcd}}{n_F }}}

\def\dfFA{d^{\,(4)}_{F\!A}}
\def\dfFF{d^{\,(4)}_{F\!F}}

\def\DfRAnc{{\frac{d^{\,(4)}_{F\!A}}{\nc^{}}}}
\def\DfRRnc{{\frac{d^{\,(4)}_{F\!F}}{\nc^{}}}}

\title{Soft corrections to inclusive deep-inelastic scattering at four loops and beyond}
\author[a]{Goutam Das,}
\author[b]{Sven-Olaf Moch,}
\author[c]{Andreas Vogt}
\affiliation[a]{Theoretische Physik 1, Naturwissenschaftlich-Technische Fakult{\"a}t, Universit{\"a}t Siegen,
Walter-Flex-Strasse 3, 57068 Siegen, Germany}
\affiliation[b]{II. Institute for Theoretical Physics, Hamburg University, D-22761 Hamburg, Germany}
\affiliation[c]{Department of Mathematical Sciences, University of Liverpool, Liverpool L69 3BX, United Kingdom}
\emailAdd{goutam.das@uni-siegen.de}
\emailAdd{sven-olaf.moch@desy.de}
\emailAdd{andreas.vogt@liverpool.ac.uk}
\abstract{
We study the threshold corrections for inclusive deep-inelastic scattering (DIS) and their all-order resummation.
Using recent results for the QCD form factor, related anomalous dimensions 
and Mellin moments of DIS structure functions at four loops we derive 
the complete soft and collinear contributions to the DIS Wilson coefficients at four loops. 
For a general $SU(n_c)$ gauge group the results are exact in the large-$n_c$
approximation and for QCD with $n_c=3$ we present precise approximations.
We extend the threshold resummation exponent $G^N$ in Mellin-$N$ space 
to the fifth logarithmic (N$^4$LL) order collecting the terms $\alpha_{\rm s}^{\,3} (\alpha_{\rm s} \ln N)^n$ 
to all orders in the strong coupling constant $\alpha_{\rm s}$. 
We study the numerical effect of the N$^4$LL corrections using both the fully
exponentiated form and the expansion of the coefficient function in towers of logarithms. 
As a byproduct, we derive a numerical result for the complete pole structure of the QCD form factor
in the parameter of dimensional regularization $\varepsilon$ at four loops.
}

\begin{document}

\preprint{SI-HEP-2019-21, DESY 19-088, LTH 1205}
\maketitle

\section{Introduction}
\label{sec:intro}

The cross section for inclusive deep-inelastic scattering (DIS) 
of charged leptons or neutrinos off a nucleon target 
can be expressed in terms of structure functions, 
which factorize into the product of operator matrix elements for the nucleon
under consideration and the DIS Wilson coefficients.
The latter parametrize the hard partonic scattering cross section 
driven by short-distance physics and are calculable in perturbative Quantum Chromodynamics (QCD) 
order by order in the strong coupling constant $\als$. 
However, only a few terms in this expansion can be calculated completely.
Currently, the QCD corrections at next-to-next-to-next-to-leading order (N$^3$LO) 
for the neutral- and charged-current DIS structure functions $F_1$,
$F_2$ and $F_3$ are known~\cite{Vermaseren:2005qc,Moch:2008fj}.
In the exceptional region of phase space, on the other hand, 
when the Bjorken scaling variable $x$ is close to unity, $x \to 1$, 
the DIS Wilson coefficients develop large Sudakov logarithms at all orders in the form 
$\als^n \ln^{l}(1-x)/(1-x)_+$, with $l=2n-1, \dots, 0$. 
These are a consequence of the constrained phase space available 
for the real emission of soft and collinear gluons.

\vspace{1mm}
The DIS Wilson coefficients are subject to an additional factorization for
$x \to 1$ (or $N \to \infty$ in Mellin-$N$ space) which allows for resummation
of those Sudakov double logarithms~\cite{Sterman:1986aj,Catani:1989ne,Catani:1990rp,Catani:1996yz,Contopanagos:1996nh}.
Based on the fixed-order QCD results up to N$^3$LO for the DIS structure functions~\cite{Vermaseren:2005qc,Moch:2008fj}
this resummation has been performed to next-to-next-to-next-to-leading logarithmic (N$^3$LL) accuracy
in Mellin-$N$ space~\cite{Moch:2005ba}, where the resummation takes the form of an exponentiation which organizes the respective 
logarithms $\als^n \ln^{l}N$, with $l=2n, \dots, 1$ 
(see also~\cite{Manohar:2003vb,Chay:2005rz,Idilbi:2006dg,Becher:2006mr} for related work in the Soft-Collinear Effective Theory).

\vspace{1mm}
From a phenomenological point of view, resummation in DIS is essential to
obtain reliable QCD predictions in the (small) region of phase space dominated by such large threshold logarithms. 
This is of particular relevance since DIS structure functions 
also receive power corrections beyond the QCD factorization at leading twist
and such higher-twist contributions are supposed to be significant at large $x$.
Therefore, good control of the perturbative series at leading twist is important to
distinguish those effects in the analyses of DIS data, see for instance~\cite{Accardi:2016ndt}.

\vspace{1mm}
From a field-theoretic point of view, on the other hand, the threshold limit
in DIS is of interest, since the factorization of the DIS Wilson coefficients 
for \mbox{$x \to 1$} links a number of fundamental quantities in QCD with each other 
due to the universality of the soft and collinear limit.
In particular, this comprises the QCD form factor which features an 
all-order exponentiation in dimensional regularization 
with $d=4-2\varepsilon$ dimensions~\cite{Collins:1980ih,Sen:1981sd,Korchemsky:1988hd,Magnea:1990zb,Magnea:2000ss,Moch:2005id}
and the QCD splitting functions, currently fully known to next-to-next-to-leading
order (NNLO)~\cite{Moch:2004pa,Vogt:2004mw}.
The necessary cancellation of soft and collinear singularities in $\varepsilon$
in inclusive observables due to the Kinoshita-Lee-Nauenberg theorem~\cite{Kinoshita:1962ur,Lee:1964is}
offers a constructive approach to the DIS Wilson coefficients near threshold~\cite{Moch:2005ky}, 
see also~\cite{Laenen:2005uz,Ravindran:2005vv,Ravindran:2006cg}.

\vspace{1mm}
Recent progress in the computation of QCD corrections has been obtained 
in particular for the quark form factor~\cite{Lee:2016ixa} and the non-singlet
quark splitting functions~\cite{Moch:2017uml}, which have been calculated at four-loop order 
in the limit of large-$\nc$ for a general $SU(\nc)$ gauge group.
In addition, a low number of fixed Mellin moments for the 
DIS structure functions are available at four loops~\cite{Ruijl:2016pkm}, 
as well as information on the $\nf$-dependence of the QCD form factor~\cite{Ruijl:2016pkm,Lee:2017mip,Grozin:2018vdn,Henn:2019rmi,Bruser:2019auj}
and on new colour factors (quartic Casimirs)~\cite{Moch:2018wjh,Henn:2019rmi,Lee:2019zop,vonManteuffel:2019wbj,Henn:2019swt,Huber:2019fxe}.
Taken together~\cite{Das:2019uvh}, these results allow us to extend the threshold resummation exponent $G^N$ 
\vspace{1mm}
for DIS Wilson coefficients in Mellin-$N$ space to the fifth logarithmic (N$^4$LL) order.
At this accuracy we collect and resum in $G^N$ the terms $\als^{\,3} (\als \ln N)^n$ to all orders in $\als$. 
To that end, we extract the resummation exponent $B^{\rm DIS}$ of the quark jet function 
collecting final-state collinear emissions at the fourth order in $\als$,
and we address the term $\als^4/(1-x)_+$ in the fixed-order expansion of the DIS Wilson coefficients at four loops.
With the help of the available four-loop QCD results, 
we can provide exact expressions in the large-$\nc$ limit and adequate numerical
approximations for full QCD with $\nc=3$.
As a byproduct, we also derive a numerical result including full colour dependence 
for the single pole $1/\varepsilon$ of the dimensional regulated QCD form factor at four loops.


\vspace{1mm}
This article is organized as follows. 
In Sec.~\ref{sec:theory} we recall the general structure of the threshold resummation for DIS 
and provide the required formulae to perform resummation to N$^4$LL accuracy. 
In Sec.~\ref{sec:resumcoeff} we describe the extraction of the resummation
coefficients and the calculation of the DIS Wilson coefficient in the soft and
collinear limit. We present a numerical study for the exponentiated resummed 
DIS Wilson coefficients and their tower expansion in Sec.~\ref{sec:numerical} 
and summarize in Sec.~\ref{sec:summary}. 
Exact expressions for the DIS Wilson coefficients are given in Appendix~\ref{app:coeff}.

\section{Theoretical framework}
\label{sec:theory}

The general structure of the resummed DIS Wilson coefficients 
takes the following form in the Mellin $N$-space~\cite{Sterman:1986aj,Catani:1989ne}, 
\beq
\label{eq:cNres}
  C_{a,q}^{\,N}(Q^2) \: =\: 
  g_{0}^{a}(Q^2) \cdot \exp\, [G^N(Q^2)] \: + \: 
  {\cal O}(N^{-1}\ln^n N) \:\: ,
\eeq
for the neutral- or charged-current DIS structure functions $F_a$, where $a=1,2,3$. 
In Mellin space they factorize in a sum over all partons $i$ as $F_a^{\,N}(Q^2) = C_{a, i}^{\,N}(Q^2) f_i^{\,N}$
with $f_i^{\,N}$ denoting the Mellin transformed parton distributions $f_i(x)$
for the parton $i$, and the Born contributions to the are normalized as $C_{a,q}^{\,N}|_{\rm LO} = 1$.
Here the resummation exponent $G^N$ contains all 
terms of the form $\ln^{\,k} N$ to all orders in $\als$, and
the pre-factor $g_0$ encompasses the $N$- independent terms.
$G^N$ and $g_0$ depend on the physical hard scale $Q^2$ 
($\,= -q^2$ in DIS, with $q$ the four-momentum of the exchanged gauge boson), 
and on the renormalization and factorization scales $\mu$ and $\mu_f$, suppressed for brevity.
For inclusive DIS $G^N$ reads in the notation of~\cite{Catani:1998tm,Moch:2005ba}
\bea
\label{eq:GNdec}
  G^N \;\; & = & 
    \ln \Delta_{\,\rm q} \: + \: \ln J_{\rm q} \: + \: 
    \ln \Delta^{\,\rm int}_{\,\rm DIS} \:\: ,
\eea
where the radiation factors ($\Delta_{\,\rm q}, J_{\rm q}, \Delta^{\,\rm int}_{\,\rm DIS}$) 
are given by well-known integrals over functions of the running coupling. 
The collinear soft-gluon radiation off an initial-state quark is collected by
\beq
\label{eq:dint}
  \ln \Delta_{\,\rm q} (Q^2,\, \mu_{\!f}^{\:\!2}) \: = \: \int_0^1 \! dz \,
  \frac{z^{N-1}-1}{1-z} \,\int_{\mu_{\!f}^{\:\!2}}^{(1-z)^2 Q^2}
  \frac{dq^2}{q^2}\, A^{\rm q}(\als(q^2)) 
\:\: ,
\eeq
with the light-like cusp anomalous dimension $A^{\rm q}(\als)$ addressed below.
The collinear emissions from an `unobserved' final-state quark are summarized in the jet function,
\beq
\label{eq:Jint}
  \ln J_{\rm q} (Q^2) \: = \: \int_0^1 \! dz \,\frac{z^{N-1}-1}{1-z}
  \, \left[ \int_{(1-z)^2 Q^2}^{(1-z) Q^2} \frac{dq^2}{q^2}\,
  A^{\rm q}(\als(q^2)) + B^{\rm DIS} (\als([1-z] Q^2)) \right] 
\eeq
including the additional function $B^{\rm DIS}(\als)$. 
All expansions in terms of the strong coupling $\als$ are normalized as
\beq
\label{eq:asexpdef}
A^{\rm q}(\als) 
\,=\, \sum_{n=1}^{\infty} \left( \frac{\als}{4\pi}\right)^{n}\, A_n
\,\equiv\, \sum_{n=1}^{\infty} \ars^{n}\, A_n
\eeq
etc. 
Any process-dependent contributions from large-angle soft gluons emissions are
contained in the function $\Delta^{\rm int}_{\rm DIS}$ which, however, evaluates to
unity, i.e., $\Delta^{\rm int}_{\rm DIS} = 1$, since the corresponding evolution 
kernels vanish at all order in $\als$ for inclusive DIS \cite{Forte:2002ni,Gardi:2002xm}.
Thus, the last term in eq.~(\ref{eq:GNdec}) is absent for inclusive DIS.
The running of the strong coupling is governed by the QCD beta function 
\beq
\label{eq:beta}
\beta(\ars) = - \sum_{n=0}^{\infty} \ars^{n+2}\, \beta_n \, ,
\eeq
with $\beta_0=(11/3) \ca - (2/3) \nf$ and so on.

\vspace{1mm}
The evaluation of the integrals in eqs.~(\ref{eq:dint}) and (\ref{eq:Jint}) 
proceeds in a standard manner, see for instance~\cite{Moch:2005ba,Catani:2003zt}, 
leading to the following expression of the resummation exponent $G^N$, 
\beq
\label{eq:GN}
G^N = \ln \widetilde{N} g_1^{\rm DIS}(\lambda) + g_2^{\rm DIS}(\lambda) + \ars g_3^{\rm DIS}(\lambda) + \ars^2 g_4^{\rm DIS}(\lambda) + \ars^3 g_5^{\rm DIS}(\lambda)
\, ,
\eeq
where $\lambda = \btz \ars \ln \widetilde{N} $ with $\widetilde{N} = N \exp(\gamma_{E}^{})$, 
where $\gamma_{E}^{} \simeq 0.57721566$ is the Euler-Mascheroni constant.
The number of terms on the right hand side of $G^N$ provides the resummation accuracy, namely
LL, NLL \dots, respectively. 
For notational brevity we have organized the logarithms in Mellin space as $\ln \widetilde{N}$.
Eq.~(\ref{eq:GN}) is accurate to N$^4$LL order and the coefficient
$g_5^{\,\rm DIS}(\lambda)$ collects all large logarithms up to $\als^{\,3} (\als \ln \widetilde{N})^n$ in $G^N$.
The complete results for $g_i^{\rm DIS}$ with $i \leq 5$, including for consistency
the well-known lower order
results~\cite{Sterman:1986aj,Catani:1989ne,Vogt:2000ci,Catani:2003zt,Moch:2005ba}, read
\bea
  g_1^{\,\rm DIS}(\lambda) & = & 
  (1+\lambda^{-1} \* L_{\lambda_1}-L_{\lambda_1}) \* A_1
\label{eq:g1n}
\:\: ,
\nonumber\\[1ex]
  g_2^{\,\rm DIS}(\lambda) & = & 
  -(L_{\lambda_1}+\lambda) \* A_2
  + L_{\lambda_1} \* B^{\rm DIS}_1
  +(L_{\lambda_1}+L_{\lambda_1}^2/2+\lambda) \* A_1 \* \beta_1
  +\Lqr \* L_{\lambda_1} \* A_1
  +\Lfr \* \lambda \* A_1
\label{eq:g2n}
\:\: ,
\nonumber\\[1ex]
  g_3^{\,\rm DIS}(\lambda) & = & 
  -(1-\lambda_1^{-1}+\lambda) \* A_3/2
  +(1-\lambda_1^{-1}) \* ( B^{\rm DIS}_2 - A_1 \* \zeta_2/2 )
\nonumber\\
&&\mbox{}
  -(1-\lambda_1^{-1}-2 \* \lambda_1^{-1} \* L_{\lambda_1}-\lambda_1^{-1} \* L_{\lambda_1}^2+2 \* L_{\lambda_1}+\lambda) \* A_1 \* \beta_1^2/2
\nonumber\\
&&\mbox{}
  -(1-\lambda_1^{-1}-\lambda_1^{-1} \* L_{\lambda_1}) \* B^{\rm DIS}_1 \* \beta_1
  +(3 - 3 \* \lambda_1^{-1}-2 \* \lambda_1^{-1} \* L_{\lambda_1} + \lambda) \* A_2 \* \beta_1/2
\nonumber\\
&&\mbox{}
  -(1-\lambda_1^{-1}-2 \* L_{\lambda_1}-\lambda) \* A_1 \* \beta_2/2
\nonumber\\
&&\mbox{}
  + \Lqr \* \bigl\{
    (1-\lambda_1^{-1}) \* (A_2 - B^{\rm DIS}_1)
    - (1-\lambda_1^{-1}-\lambda_1^{-1} \* L_{\lambda_1}) \* A_1 \* \beta_1
    \bigr\}
\nonumber\\
&&\mbox{}
  - \Lqrs \* (1-\lambda_1^{-1}) \* A_1/2
  + \Lfr \* \lambda \* A_2
  - \Lfrs \* \lambda \* A_1/2
\label{eq:g3n}
\:\: ,
\nonumber\\[1ex]
  g_4^{\rm DIS}(\lambda) & = & 
  -(1-\lambda_1^{-2}+2 \* \lambda) \* A_4/6
  +(1-\lambda_1^{-2}) \* (B^{\rm DIS}_3/2 + B^{\rm DIS}_1 \* \zeta_2/2 - A_2 \* \zeta_2/2 - A_1 \* \zeta_3/3)
\nonumber\\
&&\mbox{}
  - (1-\lambda_1^{-2}-2 \* \lambda_1^{-2} \* L_{\lambda_1}) \* B^{\rm DIS}_2 \* \beta_1/2
  + (5 - 5 \*\lambda_1^{-2}-6 \* \lambda_1^{-2} \* L_{\lambda_1}+4 \* \lambda) \* A_3 \* \beta_1/12
\nonumber\\
&&\mbox{}
  - (11+\lambda_1^{-2}-6 \* \lambda_1^{-2} \* L_{\lambda_1}-6 \* \lambda_1^{-2} \* L_{\lambda_1}^2-12 \* \lambda_1^{-1}+4 \* \lambda) \* A_2 \* \beta_1^2/12
\nonumber\\
&&\mbox{}
  - \lambda_1^{-2} \* L_{\lambda_1} \* A_1 \* \beta_1 \* \zeta_2/2
  - (1-\lambda_1^{-2}-6 \* L_{\lambda_1}-4 \* \lambda) \* A_1 \* \beta_3/12
\nonumber\\
&&\mbox{}
  + (2+\lambda_1^{-2}-3 \* \lambda_1^{-1}+\lambda) \* A_2 \* \beta_2/3
  +(2+\lambda_1^{-2}+3/2 \* \lambda_1^{-2} \* L_{\lambda_1}
\nonumber\\
&&\mbox{}
     -1/2 \* \lambda_1^{-2} \* L_{\lambda_1}^3-3 \* \lambda_1^{-1}
     -3 \* \lambda_1^{-1} \* L_{\lambda_1}+3/2 \* L_{\lambda_1}+\lambda) \* A_1 \* \beta_1^3/3
\nonumber\\
&&\mbox{}
  -(7+5 \* \lambda_1^{-2}+6 \* \lambda_1^{-2} \* L_{\lambda_1}-12 \* \lambda_1^{-1}-12 \* \lambda_1^{-1} \* L_{\lambda_1}+12 \* L_{\lambda_1}+8 \* \lambda) \* A_1 \* \beta_1 \* \beta_2/12
\nonumber\\
&&\mbox{}
  +(1+\lambda_1^{-2}-\lambda_1^{-2} \* L_{\lambda_1}^2-2 \* \lambda_1^{-1}) \* B^{\rm DIS}_1 \* \beta_1^2/2
  -(1+\lambda_1^{-2}-2 \* \lambda_1^{-1}) \* B^{\rm DIS}_1 \* \beta_2/2
\nonumber\\
&&\mbox{}
  + \Lqr \* \bigl\{
    (1-\lambda_1^{-2}) \* (A_3/2 - B^{\rm DIS}_2 + A_1 \* \zeta_2/2)
    -(1-\lambda_1^{-2}-2 \* \lambda_1^{-2} \* L_{\lambda_1}) \* A_2 \* \beta_1/2
\nonumber\\
&&\mbox{}
    +(1+\lambda_1^{-2}-\lambda_1^{-2} \* L_{\lambda_1}^2-2 \* \lambda_1^{-1}) \* A_1 \* \beta_1^2/2
    -(1+\lambda_1^{-2}-2 \* \lambda_1^{-1}) \* A_1 \* \beta_2/2
\nonumber\\
&&\mbox{}
    -\lambda_1^{-2} \* L_{\lambda_1} \* B^{\rm DIS}_1 \* \beta_1
         \bigr\}
  - \Lqrs \* \bigl\{
    (1-\lambda_1^{-2}) \* (A_2/2 - B^{\rm DIS}_1/2)
    +(\lambda_1^{-2} \* L_{\lambda_1}) \* A_1 \* \beta_1/2
    \bigr\}
\nonumber\\
&&\mbox{}
  + \Lqrt \* (1-\lambda_1^{-2}) \* A_1/6
  + \Lfr \* \lambda \* A_3
  - \Lfrs \* \lambda \* (A_2 + A_1 \* \beta_1/2)
  + \Lfrt \* \lambda \* A_1/3
\label{eq:g4n}
\, .
\eea
For brevity, we use here the short-hand notations 
$\lambda_1=1-\lambda$,
$L_{\lambda_1}=\ln(1-\lambda)$, 
$L_{\rm qr} = \ln (Q^2/\mu^{\,2})$
and $\,L_{\rm fr} = \ln (\mu_f^{\,2}/\mu^{\,2})\,$. 
Also we suppress factors of $\beta_0$ which can be restored easily 
by the substitutions 
$A_k \ra A_k / \beta_0^{\,k}$,  $B^{\rm DIS}_k \ra B^{\rm DIS}_k /\beta_0^{\,k}$ 
and $\beta_k \ra \beta_k /\beta_0^{\,k+1}$. 
In addition, $g_3^{}$ needs to be multiplied by $\beta_0$ and $g_4^{}$ by $\beta_0^{\,2}$. 

\vspace{1mm}
The new function $g_5^{}$ (computed with methods of~\cite{Moch:2005ba} and
to be multiplied by $\beta_0^{\,3}$) is given by 
\bea
\label{eq:g5n}
{\lefteqn{
  g_5^{\rm DIS}(\lambda) \,\,=\,\, }}
\nonumber\\
& & 
          (1-\lambda_1^{-3}) \* (B^{\rm DIS}_4/3 + B^{\rm DIS}_2 \* \zeta_2 
         +2 \* B^{\rm DIS}_1 \* \zeta_3/3
         -A_3 \* \zeta_2/2
         -2 \* A_2 \* \zeta_3/3
         -9 \* A_1 \* \zeta_4/8)
\nonumber\\
&&\mbox{}
         -(1-\lambda_1^{-3}+3 \* \lambda) \* A_5/12
         +(1-\lambda_1^{-3}+2 \* \lambda_1^{-3} \* L_{\lambda_1}) \* (B^{\rm DIS}_1 \* \beta_1 \* \zeta_2/2
         -A_1 \* \beta_1 \* \zeta_3/3)
\nonumber\\
&&\mbox{}
         -(1-\lambda_1^{-3}-3 \* \lambda_1^{-3} \* L_{\lambda_1}) \* B^{\rm DIS}_3 \* \beta_1/3
         +(7-7 \* \lambda_1^{-3}-12 \* \lambda_1^{-3} \* L_{\lambda_1}+9 \* \lambda) \* A_4 \* \beta_1/36
\nonumber\\
&&\mbox{}
         -(1-\lambda_1^{-3}-12 \* L_{\lambda_1}-9 \* \lambda) \* A_1 \* \beta_4/36
         +(5+\lambda_1^{-3}-6 \* \lambda_1^{-1}+3 \* \lambda) \* A_2 \* \beta_3/12
\nonumber\\
&&\mbox{}
         -(5-5 \* \lambda_1^{-3}+18 \* \lambda_1^{-2}-18 \* \lambda_1^{-1}+12 \* L_{\lambda_1}+9 \* \lambda) \* A_1 \* \beta_2^2/36
\nonumber\\
&&\mbox{}
         -(7-\lambda_1^{-3}+2 \* \lambda_1^{-3} \* L_{\lambda_1}+6 \* \lambda_1^{-3} \* L_{\lambda_1}^2+2 \* \lambda_1^{-3} \* L_{\lambda_1}^3-\lambda_1^{-3} \* L_{\lambda_1}^4+6 \* \lambda_1^{-2}-6 \* \lambda_1^{-2} \* L_{\lambda_1}^2-12 \* \lambda_1^{-1}
\nonumber\\
&&\mbox{}
         -6 \* \lambda_1^{-1} \* L_{\lambda_1}+4 \* L_{\lambda_1}+3 \* \lambda) \* A_1 \* \beta_1^4/12
         +(41-5 \* \lambda_1^{-3}+12 \* \lambda_1^{-3} \* L_{\lambda_1}+18 \* \lambda_1^{-3} \* L_{\lambda_1}^2+36 \* \lambda_1^{-2}
\nonumber\\
&&\mbox{}
         -18 \* \lambda_1^{-2} \* L_{\lambda_1}^2-72 \* \lambda_1^{-1}-36 \* \lambda_1^{-1} \* L_{\lambda_1}+36 \* L_{\lambda_1}+27 \* \lambda) \* A_1 \* \beta_1^2 \* \beta_2/36
\nonumber\\
&&\mbox{}
         -(2+\lambda_1^{-3}-3 \* \lambda_1^{-1}) \* B^{\rm DIS}_1 \* \beta_3/6
         +(1+\lambda_1^{-3}-2 \* \lambda_1^{-2}+\lambda) \* A_3 \* \beta_2/4
\nonumber\\
&&\mbox{}
         -(7+2 \* \lambda_1^{-3}+3 \* \lambda_1^{-3} \* L_{\lambda_1}-9 \* \lambda_1^{-1}-9 \* \lambda_1^{-1} \* L_{\lambda_1}+12 \* L_{\lambda_1}+9 \* \lambda) \* A_1 \* \beta_1 \* \beta_3/18
\nonumber\\
&&\mbox{}
         -(20+7 \* \lambda_1^{-3}+12 \* \lambda_1^{-3} \* L_{\lambda_1}-9 \* \lambda_1^{-2}-18 \* \lambda_1^{-2} \* L_{\lambda_1}-18 \* \lambda_1^{-1}+9 \* \lambda) \* A_2 \* \beta_1 \* \beta_2/18
\nonumber\\
&&\mbox{}
         -(13+5 \* \lambda_1^{-3}-12 \* \lambda_1^{-3} \* L_{\lambda_1}-18 \* \lambda_1^{-3} \* L_{\lambda_1}^2-18 \* \lambda_1^{-2}+9 \* \lambda) \* A_3 \* \beta_1^2/36
\nonumber\\
&&\mbox{}
         +(25+11 \* \lambda_1^{-3}+24 \* \lambda_1^{-3} \* L_{\lambda_1}-12 \* \lambda_1^{-3} \* L_{\lambda_1}^3-18 \* \lambda_1^{-2}-36 \* \lambda_1^{-2} \* L_{\lambda_1}-18 \* \lambda_1^{-1}+9 \* \lambda) \* A_2 \* \beta_1^3/36
\nonumber\\
&&\mbox{}
         +(2+\lambda_1^{-3}+3 \* \lambda_1^{-3} \* L_{\lambda_1}-3 \* \lambda_1^{-2} \* L_{\lambda_1}-3 \* \lambda_1^{-1}) \* B^{\rm DIS}_1 \* \beta_1 \* \beta_2/3
         - \lambda_1^{-3} \* L_{\lambda_1} \* A_2 \* \beta_1 \* \zeta_2
\nonumber\\
&&\mbox{}
         -(2+\lambda_1^{-3}+6 \* \lambda_1^{-3} \* L_{\lambda_1}+3 \* \lambda_1^{-3} \* L_{\lambda_1}^2-2\*\lambda_1^{-3} \* L_{\lambda_1}^3-6 \* \lambda_1^{-2} \* L_{\lambda_1}-3 \* \lambda_1^{-1}) \* B^{\rm DIS}_1 \* \beta_1^3/6
\nonumber\\
&&\mbox{}
         +(1+2 \* \lambda_1^{-3}-3 \* \lambda_1^{-3} \* L_{\lambda_1}^2-3 \* \lambda_1^{-2}) \* B^{\rm DIS}_2 \* \beta_1^2/3
         -(1+2 \* \lambda_1^{-3}-3 \* \lambda_1^{-2}) \* B^{\rm DIS}_2 \* \beta_2/3
\nonumber\\
&&\mbox{}
         -(\lambda_1^{-3}+\lambda_1^{-3} \* L_{\lambda_1}-\lambda_1^{-3} \* L_{\lambda_1}^2-\lambda_1^{-2}) \* A_1 \* \beta_1^2 \* \zeta_2/2
         +(\lambda_1^{-3}-\lambda_1^{-2}) \* A_1 \* \beta_2 \* \zeta_2/2
\nonumber\\
&&\mbox{}
      + \Lqr \* \bigl\{
          (1-\lambda_1^{-3}) \* (A_4/3 - B^{\rm DIS}_3 - B^{\rm DIS}_1 \* \zeta_2 
         + A_2 \* \zeta_2 + 2 \* A_1 \* \zeta_3/3)
\nonumber\\
&&\mbox{}
         - (2+\lambda_1^{-3}-3 \* \lambda_1^{-1}) \* A_1 \* \beta_3/6
         + (2+\lambda_1^{-3}+3 \* \lambda_1^{-3} \* L_{\lambda_1}-3 \* \lambda_1^{-2} \* L_{\lambda_1}-3 \* \lambda_1^{-1}) \* A_1 \* \beta_1 \* \beta_2/3
\nonumber\\
&&\mbox{}
         + (1-\lambda_1^{-3}+2 \* \lambda_1^{-3} \* L_{\lambda_1}) \* A_1 \* \beta_1 \* \zeta_2/2
         - (1-\lambda_1^{-3}-3 \* \lambda_1^{-3} \* L_{\lambda_1}) \* A_3 \* \beta_1/3
\nonumber\\
&&\mbox{}
         -(2+\lambda_1^{-3}+6 \* \lambda_1^{-3} \* L_{\lambda_1}+3 \* \lambda_1^{-3} \* L_{\lambda_1}^2-2\*\lambda_1^{-3} \* L_{\lambda_1}^3-6 \* \lambda_1^{-2} \* L_{\lambda_1}-3 \* \lambda_1^{-1}) \* A_1 \* \beta_1^3/6
\nonumber\\
&&\mbox{}
         +(1+2 \* \lambda_1^{-3}-3 \* \lambda_1^{-3} \* L_{\lambda_1}^2-3 \* \lambda_1^{-2}) \* A_2 \* \beta_1^2/3 
         +(\lambda_1^{-3}-\lambda_1^{-2}) \* B^{\rm DIS}_1 \* \beta_2
         -2 \* (\lambda_1^{-3} \* L_{\lambda_1}) \* B^{\rm DIS}_2 \* \beta_1
\nonumber\\
&&\mbox{}
         -(1+2 \* \lambda_1^{-3}-3 \* \lambda_1^{-2}) \* A_2 \* \beta_2/3 
         -(\lambda_1^{-3}+\lambda_1^{-3} \* L_{\lambda_1}-\lambda_1^{-3} \* L_{\lambda_1}^2-\lambda_1^{-2}) \* B^{\rm DIS}_1 \* \beta_1^2
         \bigr\}
\nonumber\\
&&\mbox{}
  + \Lqrs \* \bigl\{
     (1-\lambda_1^{-3}) \* (B^{\rm DIS}_2
    -A_3/2
    -A_1 \* \zeta_2/2)
    +(1-\lambda_1^{-3}+2 \* \lambda_1^{-3} \* L_{\lambda_1}) \* B^{\rm DIS}_1 \* \beta_1/2
\nonumber\\
&&\mbox{}
    -(\lambda_1^{-3}+\lambda_1^{-3} \* L_{\lambda_1}-\lambda_1^{-3} \* L_{\lambda_1}^2-\lambda_1^{-2}) \* A_1 \* \beta_1^2/2
    +(\lambda_1^{-3}-\lambda_1^{-2}) \* A_1 \* \beta_2/2
    -(\lambda_1^{-3} \* L_{\lambda_1}) \* A_2 \* \beta_1
    \bigr\}
\nonumber\\
&&\mbox{}
  + \Lqrt \* \bigl\{
    (1-\lambda_1^{-3}) \* (A_2 - B^{\rm DIS}_1)/3
    +(1-\lambda_1^{-3}+2 \* \lambda_1^{-3} \* L_{\lambda_1}) \* A_1 \* \beta_1/6
    \bigr\}
\nonumber\\
&&\mbox{}
  - \Lqrf \* (1-\lambda_1^{-3}) \* A_1/12
  + \Lfr \* \lambda \* A_4
  - \Lfrs \* \lambda \* (3 \* A_3/2 + A_2 \* \beta_1 + A_1 \* \beta_2/2)
\nonumber\\
&&\mbox{}
  + \Lfrt \* \lambda \* (A_2 + 5 \* A_1 \* \beta_1/6)
  - \Lfrf \* \lambda \* A_1/4
\, .
\qquad
\eea
Note that for resummation to N$^4$LL accuracy the function $g_5^{\rm DIS}(\lambda)$ 
needs the five-loop coefficient $\beta_4$ of the QCD beta function in eq.~(\ref{eq:beta}) which is
available due to \cite{Baikov:2016tgj,Herzog:2017ohr,Luthe:2017ttg,Chetyrkin:2017bjc},
as well as the cusp anomalous dimension up to five loops, i.e., 
the coefficient $A_5$ which has recently been estimated \cite{Herzog:2018kwj} (see below).
In addition, one also needs the evolution kernel $B^{\rm DIS}(\als)$ 
of the jet function in eq.~(\ref{eq:Jint}) to four-loop order, 
i.e., the term $B^{\rm DIS}_4$, which will be addressed below.
We will collect and discuss all necessary resummation coefficients in the next section.

\section{Resummation coefficients}
\label{sec:resumcoeff}

We present results for a general $SU(\nc)$ gauge group with $\nc$ colours and $\nf$ massless fermions; the 
QCD expressions can always be recovered by setting $\nc = 3$.
In this way, the resummation coefficients are expressible in terms of the usual 
$SU(\nc)$ quadratic Casimir factors $\ca=\nc$ and $\cf=(\ncs-1)/(2\nc)$,
i.e., $\ca~=~3$, $\cf~=~4/3\,$ in QCD.
Starting from three-loop order in perturbation theory, higher group invariants enter, 
such as the square of the symmetric part of the trace of three $SU(\nc)$
generators $T_{\!F}^{\:\!a}$ in the fundamental representation,
\beq
\label{eq:dabc}
\dabcnc = \frac{1}{16\,\ncs}\: (\ncs - 1) (\ncs - 4)
\; ,
\eeq
that is $d^{abc}d_{abc}/\nc = 5/18$ in QCD, cf.~\cite{Moch:2015usa} for conventions on the normalization. 
At four loops we also have for the first time contributions with quartic colour factors 
$d^{\,(4)}_{xy} \;\equiv\; d_x^{\,abcd} d_y^{\,abcd}$, where $x,y$ labels the representations with generators $T_r^a$ 
and
\beq
\label{d4def}
  d_{r}^{\,abcd} \; =\; \frac{1}{6}\: {\rm Tr} \, ( \, 
   T_{r}^{a\,} T_{r}^{b\,} T_{r}^{c\,} T_{r}^{d\,}
   + \,\mbox{ five $bcd$ permutations}\, ) 
   \; ,
\eeq
which leads to
\beq
\label{d4SUn}
  \DfRAnc \!=\! 
  \frac{1}{48}\:  ( \ncs - 1 )  ( \ncs + 6 ) 
\;, \qquad
  \DfRRnc  \!=\!
  \frac{1}{96\,\nct}\: ( \ncs - 1 ) ( \ncf - 6\,\ncs + 18 )
\;. \\
\eeq
In QCD these factors evaluate to $d^{\,(4)}_{F\!A}/\nc = 5/2$ and $d^{\,(4)}_{F\!F}/\nc = 5/36$.

\subsection{Splitting functions at large $x$ }
\label{sec:splittingfn}

The coefficients of the cusp anomalous dimension $A_n$ appear in the 
large-$x$ expansion of the diagonal parts of the splitting functions. 
For a quark field, we consider the large-$x$ behaviour of the non-singlet splitting functions 
$P_{\rm ns}^{\,(n-1)}(x)$ in the \MSb\ scheme at $n$-loops, 
recall eq.~(\ref{eq:asexpdef}) for the normalization of the expansion in powers of $\als$.
Disregarding terms that vanish for $x \to 0$, these can be written 
\cite{ Korchemsky:1988si,Albino:2000cp}
\beq
\label{xto1Lnc}
  P_{\,\rm ns}^{\,(n-1)}(x) \:\: = \:\;
        \frac{A_n}{(1-x)_+}
  \,+\, B_n^{\:\!\rm q} \, \delta(1-x)
\, ,
\eeq
where $1/(1-x)_+$ on the right hand side represents the usual plus-distribution;
$B_n^{\:\!\rm q}$ is sometimes referred to as virtual anomalous dimension.

\vspace{1mm}
The cusp anomalous dimension of a quark field $A^{\rm q}$ 
is known up to third order for quite a long time~\cite{Moch:2004pa} and the
perturbative expansion reads according to eq.\ (\ref{eq:asexpdef}),
\bea
\label{eq:Aqexp}
  A_1 &=&
  4\* \cf
  \nn \, , \\
  A_2 &=& 
  8\* \cf \* \biggl\{ \biggl( \frac{67}{18} - \zeta_2 \biggr) \* \ca - \frac{5}{9}\*\nf \biggr\} 
\nn \, , \\
  A_3 &=&
  16\* \cf\* \biggl\{ \cas \* \left( \frac{245}{24} - \frac{67}{9} \* \zeta_2
    + \frac{11}{6} \* \zeta_3 + \frac{11}{2} \* \zeta_4\right) 
  + \cf \* \nf \* \biggl( -  \frac{55}{24}  + 2\*\zeta_3 \biggr) 
\nn
\\ & &  \mbox{}
  + \ca \* \nf\* \biggl( - \frac{209}{108}
  + \frac{10}{9} \* \zeta_2 - \frac{7}{3} \* \zeta_3 \biggr)
  + \nfs \* \biggl( - \frac{1}{27} \biggr) \biggr\}
\:\: .
\eea

The four-loop contribution $A_4$ has been of subject to intensive recent studies.
Combining all available results it reads for quark fields, 
\bea
\label{eq:A4q}
A_4 &=& 
         \cf \* \cat \* \left(
            \frac{84278}{81}
          - \frac{88400}{81}\*\zeta_2
          + \frac{20944}{27}\*\zeta_3
          + 1804\*\zeta_4
          - \frac{352}{3}\*\zeta_2\*\zeta_3
          - \frac{3608}{9}\*\zeta_5
\nonumber
\right.
\\
&& \nonumber
\left. \mbox{}
          - 16\*\zeta_3^2
          - \frac{2504}{3}\*\zeta_6
          \right)
       + \DfRAnc \* \left(
         - 128\*\zeta_2 
         + \frac{128}{3}\*\zeta_3 
         + \frac{3520}{3}\*\zeta_5 
         - 384\*\zeta_3^2 
         - 992\*\zeta_6
       \right)
\\
&& \nonumber
\mbox{}
       + \cft\*\nf \* \left(
            \frac{572}{9}
          + \frac{592}{3}\*\zeta_3
          - 320\*\zeta_5
          \right)
\\
&& \nonumber \mbox{}
       + \cfs\*\ca\*\nf \* \left(
          - \frac{34066}{81}
          + \frac{440}{3}\*\zeta_2
          + \frac{3712}{9}\*\zeta_3
          - 176\*\zeta_4
          - 128\*\zeta_2\*\zeta_3
          + 160\*\zeta_5
          \right)
\\
&& \nonumber \mbox{}
       + \cf\*\cas\*\nf \* \left(
          - \frac{24137}{81}
          + \frac{20320}{81}\*\zeta_2
          - \frac{23104}{27}\*\zeta_3
          - \frac{176}{3}\*\zeta_4
          + \frac{448}{3}\*\zeta_2\*\zeta_3
          + \frac{2096}{9}\*\zeta_5
          \right)
\\
&& \nonumber \mbox{}
       + \nf \* \DfRRnc \* \left(
            256\*\zeta_2 
          - \frac{256}{3}\*\zeta_3 
          - \frac{1280}{3}\*\zeta_5 
        \right)
       + \cfs\*\nfs \* \left(
            \frac{2392}{81}
          - \frac{640}{9}\*\zeta_3
          + 32\*\zeta_4
          \right)
\\
&&  \mbox{}
       + \cf\*\ca\*\nfs \* \left(
            \frac{923}{81}
          - \frac{608}{81}\*\zeta_2
          + \frac{2240}{27}\*\zeta_3
          - \frac{112}{3}\*\zeta_4
          \right)
       - \cf\*\nft \* \left(
            \frac{32}{81}
          - \frac{64}{27}\*\zeta_3
          \right)
\, . \;
\eea
This expression combines available exact results in the large-$\nc$ limit of QCD~\cite{Lee:2016ixa,Moch:2017uml},
as well as for the terms proportional to $\nf$~\cite{Grozin:2018vdn,Henn:2019rmi}, 
to $\nfs$~\cite{Davies:2016jie,Lee:2017mip} and 
to $\nft$~\cite{Gracey:1994nn,Beneke:1995pq}, the latter have been known for a long time.
The quartic colour factors have been completed in exact form recently~\cite{Henn:2019rmi,Lee:2019zop,Henn:2019swt}, 
and found to be in agreement with numerical estimates~\cite{Moch:2017uml,Moch:2018wjh}.

\vspace{1mm}
Finally, the five-loop quark cusp anomalous dimension in QCD 
has been estimated in \cite{Herzog:2018kwj} 
as 
\beq
\label{eq:A5q}
  A_5 \:=\: ( 1.7 \,\pm\, 0.5 \:,\; 
              1.1 \,\pm\, 0.5 \:,\;
              0.7 \,\pm\, 0.5 ) \cdot 10^{\,5}
\quad \mbox{for} \quad 
  \nf \:=\: 3\,,\:4\,,\:5
\; .
\eeq

Inserting the numerical values in eqs.~(\ref{eq:Aqexp})--(\ref{eq:A5q}) one obtains
the perturbative expansion for $A^{\rm q}(\als)$ 
according to eq.~(\ref{eq:asexpdef}) through five loops 
in powers of $\als$ for the physically relevant values of $\nf$ as~\cite{Herzog:2018kwj}  
\bea
\label{eq:AqExpQ}
  A^{\rm q}(\nf\!=\!3) &=&
  0.42441\:\als \:
  ( 1  +  0.7266\,\als +  0.7341\,\als^2 + 0.665\,\als^3
       +  ( 1.3 \pm 0.4) \als^4 \, + \, \ldots )
\; , \nn \\[1mm]
  A^{\rm q}(\nf\!=\!4) &=& 
  0.42441\:\als \:
  ( 1  +  0.6382\,\als  +  0.5100\,\als^2 + 0.317\,\als^3
       +  ( 0.8 \pm 0.4 ) \als^4  \, + \, \ldots )
\; , \nn \\[1mm]
  A^{\rm q}(\nf\!=\!5) &=&
  0.42441\:\als \:
  ( 1  +  0.5497\,\als  +  0.2840\, \als^2 + 0.013\,\als^3
       +  (0.5 \pm 0.4 ) \als^4  \, + \, \ldots )
\, . \nn \\
\quad
\eea

The $\delta(1-x)$ parts of the $n$-loop non-singlet splitting function $P_{\rm ns}^{\,(n-1)}(x)$, 
i.e. the coefficients $B_n^{\,\rm q}$ in eq.~(\ref{xto1Lnc}) 
are also known exactly to three loops~\cite{Moch:2004pa}, 
\bea
\label{eq:Bqexp}
  B_1^{\rm q} &=&
  3\* \cf
  \nn \, , \\
  B_2^{\rm q} &=& 
    4 \* \cf \* \biggl\{ \cf \* \left( 
      \frac{3}{8} 
      - 3 \* \zeta_2 
      + 6 \* \zeta_3 
    \right)
  + \ca \* \left( 
    \frac{17}{24} 
    + \frac{11}{3} \* \zeta_2 
    - 3 \* \zeta_3 
  \right)
  - \nf \* \left( 
    \frac{1}{12} 
    + \frac{2}{3} \* \zeta_2 
  \right) 
  \biggr\}
\nn \, , \\
  B_3^{\rm q} &=&
  16  \*  \cf  \* \biggl\{ \cfs  \*  \left( 
    \frac{29}{32} 
    + \frac{9}{8} \* \zeta_2 
    + \frac{17}{4} \* \zeta_3 
    + 9 \* \zeta_4 
    - 2 \* \zeta_2 \* \zeta_3 
    - 15 \* \zeta_5 
  \right)
\nonumber
\\ & & \mbox{} 
  + \cf \* \ca \*  \left( 
    \frac{151}{64} 
    - \frac{205}{24} \* \zeta_2 
    + \frac{211}{12} \* \zeta_3 
    - \frac{247}{24} \* \zeta_4
    + \zeta_2 \* \zeta_3 
    + \frac{15}{2} \* \zeta_5 
  \right)
\\ & &  \mbox{}
  - \cf \* \nf  \*  \left( 
    \frac{23}{16} 
    - \frac{5}{12} \* \zeta_2 
    + \frac{17}{6} \* \zeta_3 
    - \frac{29}{12} \* \zeta_4
  \right)
  + \ca \* \nf  \*  \left( 
    \frac{5}{4} 
    - \frac{167}{54} \* \zeta_2 
    + \frac{25}{18} \* \zeta_3 
    + \frac{1}{8} \* \zeta_4
  \right)
\nonumber
\\ & &  \mbox{}
  - \cas \*  \left( 
    \frac{1657}{576} 
    - \frac{281}{27} \* \zeta_2 
    + \frac{97}{9} \* \zeta_3 
    + \frac{5}{16} \* \zeta_4
    - \frac{5}{2} \* \zeta_5 
  \right)
  - \nfs  \*  \left( 
    \frac{17}{144} 
    - \frac{5}{27} \* \zeta_2 
    + \frac{1}{9} \* \zeta_3 
  \right)
  \biggr\}
\nonumber
\, .
\eea
At four loops the exact result in the large-$\nc$ limit of QCD
with an overall factor of $\cf$ reads~\cite{Moch:2017uml} 
\bea
\label{eq:B4qLnc}
\nonumber 
B_4^{\rm q}\biggl|_{{\rm L}\nc} &=& 
         \cf\*\nct \* \biggl(
       - \frac{1379569}{5184}
       + \frac{24211}{27}\*\zeta_2 
       - \frac{9803}{162}\*\zeta_3 
       - \frac{9382}{9}\*\zeta_4
       + \frac{838}{9}\*\zeta_2\*\zeta_3
       + 1002\*\zeta_5 
\\
&& \nonumber \mbox{}
       + \frac{16}{3}\*\zeta_3^2 
       + 135\*\zeta_6 
       - 80\*\zeta_2\*\zeta_5 
       + 32\*\zeta_3\*\zeta_4 
       - 560\*\zeta_7 
          \biggr)
\\
&& \nonumber \mbox{}
       + \cf\*\ncs\*\nf \* \biggl(
         \frac{353}{3} 
       - \frac{85175}{162}\*\zeta_2 
       - \frac{137}{9}\*\zeta_3 
       + \frac{16186}{27}\*\zeta_4 
       - \frac{584}{9}\*\zeta_2\*\zeta_3
       - \frac{248}{3}\*\zeta_5 
\\
&& \nonumber \mbox{}
       - \frac{16}{3}\*\zeta_3^2 
       - 144\*\zeta_6 
          \biggr)
\\
&& \nonumber \mbox{}
       + \cf\*\nc\*\nfs \* \left(
       - \frac{127}{18} 
       + \frac{5036}{81}\*\zeta_2 
       - \frac{932}{27}\*\zeta_3 
       - \frac{1292}{27}\*\zeta_4 
       + \frac{160}{9}\*\zeta_2\*\zeta_3
       + \frac{32}{3}\*\zeta_5 
          \right)
\\
&&  \mbox{}
       + \cf\*\nft \* \left(
       - \frac{131}{81} 
       + \frac{32}{81}\*\zeta_2
       + \frac{304}{81}\*\zeta_3 
       - \frac{32}{27}\*\zeta_4 
          \right)
\, .
\eea

\begin{table}[t!]
  \vspace*{-1mm}
  \centering
  \renewcommand{\arraystretch}{1.2}
  \begin{tabular}{MMMMM}
    \hline 
    \cff                 &
    \cft\, \ca           &
    \cfs\, \cas            &
    \cf\, \cat             &
    d^{\,(4)}_{F\!A}/n_{\!F}^{} 
\\
    196.5 \pm 1.& 
    -687.5 \pm 1.5 & 
    1219.5 \pm 2. & 
    295.7 \pm 0.5 & 
    -998.0 \pm 0.2   
\\[1ex]
    \hline
    \nf\, \cft           & 
    \nf\,\cfs\ca         & 
    \nf\,\cf\cas         & 
    \nf\,d^{\,(4)}_{F\!F}/n_{\!F}^{} & 
\\
    80.780 \pm 0.005 & 
    -455.247 \pm 0.005 & 
    -274.466 \pm 0.01 & 
    -143.6 \pm 0.2 & 
\\[1ex]
    \hline
    \cf\, \nct 
    & &
    \nf\, \cf\, \ncs 
    & &
\\
    716.5577
    & &
   -484.8864
    & &
\\[1ex]
    \hline
    \nfs\,\cfs           & 
    \nfs\,\cf\ca         & 
    &
    \nft\,\cf            & 
\\
    -5.775288 & 
    51.03056 & & 
    2.261237 & 
\\[1ex]
    \hline
  \end{tabular}
  \vspace*{1mm}
  \caption{\small{ 
  Numerical values for the colour coefficients of the $\delta(1-x)$ part $B_4^{\rm q}$ in eq.~(\ref{xto1Lnc})
  at fourth order. 
  All exact values for the coefficients have been 
  rounded to seven digits and the errors are correlated due to the known exact
  results at $\nf^0$ and $\nf^1$ in the large-$\nc$ limit given in
  the third row.
  }}
    \label{tab:Bq}
\end{table}

\vspace*{1mm}
The full colour dependence can be parametrized with coefficients 
$b_{4,\, {\cff}}^{\rm q}$, $b_{4,\, {\cft\*\ca}}^{\rm q}$, $b_{4,\, {\cfs\*\cas}}^{\rm q}$,
$b_{4,\, {\nf\*\cft}}^{\rm q}$, $b_{4,\, {\nf\*\cfs\*\ca}}^{\rm q}$, $b_{4,\, \dfFA}^{\rm q}$, 
and $b_{4,\, \dfFF}^{\rm q}$, which are given in numerical form in Tab.~\ref{tab:Bq}. 
The numerical values in Tab.~\ref{tab:Bq} have been improved considerably
compared to~\cite{Moch:2017uml}, thanks to the computation of more Mellin
moments of the corresponding anomalous dimension and the exact result for $A_4$ in eq.~(\ref{eq:A4q}).
All parts proportional to $\nfs$ and $\nft$ are known in analytic form~\cite{Gracey:1994nn,Davies:2016jie}. 
The result for $B_4^{\rm q}$ can then be written as
\bea
\label{eq:B4q}
B_4^{\rm q} &=& 
         \cff \* ~b_{4,\, {\cff}}^{\rm q}~
       + \cft \* \ca \* ~b_{4,\, {\cft\*\ca}}^{\rm q}~
       + \cfs \* \cas \* ~b_{4,\, {\cfs\*\cas}}^{\rm q}~
       + \cf \* \cat \* \left(
          - \frac{1379569}{5184}
          + \frac{24211}{27} \* \zeta_2
\right.
\nonumber
\\
&& \nonumber \mbox{}
          - \frac{9803}{162} \* \zeta_3
          - \frac{9382}{9} \* \zeta_4
          + \frac{838}{9} \* \zeta_2 \* \zeta_3
          + 1002 \* \zeta_5
          + \frac{16}{3} \* \zeta_3^2
          + 135 \* \zeta_6
          - 80 \* \zeta_2 \* \zeta_5
          + 32 \* \zeta_3 \* \zeta_4
\\
&& \nonumber \mbox{}
\left.
          - 560 \* \zeta_7
	  -\frac{1}{2}\* ~b_{4,\, {\cfs\*\cas}}^{\rm q}~
	  -\frac{1}{4}\* ~b_{4,\, {\cft\*\ca}}^{\rm q}~
	  -\frac{1}{8}\* ~b_{4,\, {\cff}}^{\rm q}~
	  -\frac{1}{24}\* ~b_{4,\, \dfFA}^{\rm q}~
          \right)
       + \DfRAnc \* ~b_{4,\, \dfFA}^{\rm q}~
\\
&& \nonumber \mbox{}
       + \cft\*\nf \* ~b_{4,\, {\nf\*\cft}}^{\rm q}~
       + \cfs\*\ca\*\nf \* ~b_{4,\, {\nf\*\cfs\*\ca}}^{\rm q}~
       + \cf\*\cas\*\nf \* \left(
            \frac{353}{3}
          - \frac{85175}{162} \* \zeta_2
          - \frac{137}{9} \* \zeta_3
\right.
\\
&& \nonumber \mbox{}
          + \frac{16186}{27} \* \zeta_4
          - \frac{584}{9} \* \zeta_2 \* \zeta_3
          - \frac{248}{3} \* \zeta_5
          - \frac{16}{3} \* \zeta_3^2
          - 144 \* \zeta_6
	  -\frac{1}{2}\* ~b_{4,\, {\nf\*\cfs\*\ca}}^{\rm q}~
	  -\frac{1}{4}\* ~b_{4,\, {\nf\*\cft}}^{\rm q}~
\\
&& \nonumber
\left. \mbox{}
	  -\frac{1}{48}\* ~b_{4,\, \dfFF}^{\rm q}~
          \right)
       + \nf \* \DfRRnc \* ~b_{4,\, \dfFF}^{\rm q}~
       + \cfs\*\nfs \* \left(
          - \frac{188}{27}
          + \frac{1244}{27} \* \zeta_2
          + \frac{56}{27} \* \zeta_3
\right.
\\
&& \nonumber
\left. \mbox{}
          - \frac{2104}{27} \* \zeta_4
          - \frac{160}{9} \* \zeta_2 \* \zeta_3
          + \frac{368}{9} \* \zeta_5
          \right)
       + \cf\*\ca\*\nfs \* \left(
          - \frac{193}{54}
          + \frac{3170}{81} \* \zeta_2
          - \frac{320}{9} \* \zeta_3
          - \frac{80}{9} \* \zeta_4
\right.
\\
&& 
\left. \mbox{}
          + \frac{80}{3} \* \zeta_2 \* \zeta_3
          - \frac{88}{9} \* \zeta_5
          \right)
       + \cf\*\nft \* \left(
          - \frac{131}{81}
          + \frac{32}{81} \* \zeta_2
          + \frac{304}{81} \* \zeta_3
          - \frac{32}{27} \* \zeta_4
          \right)
\, .
\eea
Taken together, eqs.~(\ref{eq:Bqexp}) and (\ref{eq:B4q}) lead to 
the following perturbative expansion for $B^{\rm q}(\als)$ 
through four loops,
\bea
\label{eq:BqExpQ}
  B^{\rm q}(\nf\!=\!3) &=&
  0.31831\:\als \:
  ( 1  +  0.9971\,\als +  1.2412\,\als^2 + 1.080\,\als^3
       + \, \ldots )
\; , \nn \\
  B^{\rm q}(\nf\!=\!4) &=& 
  0.31831\:\als \:
  ( 1  +  0.8719\,\als  +  0.9783\,\als^2 + 0.566\,\als^3
       \, + \, \ldots )
\; , \nn \\
  B^{\rm q}(\nf\!=\!5) &=&
  0.31831\:\als \:
  ( 1  +  0.7467\,\als  +  0.7191\, \als^2 + 0.109\,\als^3
       \, + \, \ldots )
\, , \quad
\eea
where the error on the four-loop result from the numerical uncertainty 
in respective coefficients in Tab.~\ref{tab:Bq} 
is of ${\cal O}(10^{-4})$, i.e., beyond the accuracy quoted here.

\vspace*{1mm}
In order to obtain the resummation coefficient $B^{\rm DIS}_4$  
in eq.~(\ref{eq:g5n}) relevant at N$^4$LL accuracy 
for the jet function in eq.~(\ref{eq:Jint}) we have to combine 
the results summarized above with 
those for the QCD form factor and the DIS cross sections in the soft and 
collinear limit at four-loop order; 
this will be done next.

\subsection{Quark form factor}
\label{sec:formfactor}

The quark form factor summarizes the QCD corrections to the vertex 
of a photon with virtuality $Q^2$ and a massless external 
quark$\,$/$\,$anti-quark, the relevant amplitude being
\beq
\label{eq:FFphoton}
\Gamma_\mu  \: = \: {\rm i} e_{\rm q}\, \bigl( {\bar \psi}\, \gamma_{\mu\,} \psi \bigr)\, {\cal F}\! (\als,Q^2)\, ,
\eeq
where $e_{\rm q}$ denotes the quark's electric charge and the scalar function ${\cal F}$ the quark form factor. 
As we are interested in neutral- and charged-current DIS, we note that analogous expressions hold 
for general vector or axial-vector currents with the appropriate replacements 
of the couplings $e_{\rm q} \to v_{\rm q}$ and $e_{\rm q} \to a_{\rm q}$,
i.e., for DIS including $Z$-boson or $W^\pm$-boson exchange.
The latter case is distinguished by certain diagram classes involving higher
group theory invariants which will be addressed below.

\vspace*{1mm}
${\cal F}$ is gauge invariant, but divergent. 
It has been computed in dimensional regularization with $d=4-2\varepsilon$ 
at four-loop order in the large-$\nc$ limit of QCD~\cite{Lee:2016ixa}.
In addition, all terms proportional to $\nfs$~\cite{Lee:2017mip}
and the quartic colour factor $d^{\,(4)}_{F\!F}/\nc$ at four loops are also known~\cite{Lee:2019zop,vonManteuffel:2019wbj}.
For lower order results to sufficient depth in $\varepsilon$, see, e.g., \cite{Moch:2005id,Baikov:2009bg,Gehrmann:2010ue,Gehrmann:2010tu}.
The exponentiation of ${\cal F}$
is achieved by solving the well-known evolution equations~\cite{Collins:1980ih,Sen:1981sd,Korchemsky:1988hd,Magnea:1990zb,Magnea:2000ss}
in $d=4-2\varepsilon$ dimensions,
\beq
\label{eq:ffdeq}
Q^2 {\partial \over \partial Q^2} \ln {\cal F}\!\left(\als , {Q^2 \over \mu^2}, 
\varepsilon\right) \:  = \:  
  {1 \over 2} \bigg[\: K(\als ,\varepsilon) 
+  \: G\left({Q^2 \over \mu^2},\als ,\varepsilon \right) \bigg]\, ,
\eeq
where $\mu$ again represents the renormalization scale. 
Here $K$ is a counter term containing all the poles in $\varepsilon$ whereas the function $G$ is finite in the limit $\varepsilon \to 0$. 
The functions $G$ and $K$ follow the renormalization group equations~\cite{Collins:1980ih},
\bea
\label{eq:Gdeq}
\left(\mu^2 {\partial \over \partial \mu^2} + \beta(\als ,\varepsilon)\, 
{\partial \over \partial \als } \right) 
G\left({Q^2 \over \mu^2},\als ,\varepsilon \right) & = & ~~~ A^{\rm q}(\als ) \, ,
\nonumber \\[1mm]
\left(\mu^2 {\partial \over \partial \mu^2} + \beta(\als ,\varepsilon)\, 
{\partial \over \partial \als } \right) K(\als ,\varepsilon) & = & - A^{\rm q}(\als ) \, ,
\eea
with the standard cusp anomalous dimension $A^{\rm q}$ discussed above. 

\vspace*{1mm}
The solution of eqs.~(\ref{eq:ffdeq}) and (\ref{eq:Gdeq}) order by order in
perturbation theory is straightforward, see, e.g., \cite{Magnea:2000ss,Moch:2005id,Ravindran:2005vv,Ravindran:2006cg}, 
and provides a perturbative expansion of the bare (unrenormalized) quark form
factor in terms of the bare strong coupling $a_s^{\rm{b}}$ 
(recall the normalization $\ars = \als/(4\pi)$ in eq.~(\ref{eq:asexpdef})),
as 
\beq
\label{eq:finiteorderff}
  {\cal F}^{\rm{b}}(a_s^{\rm{b}},Q^2) \: = \: 
  1 + \sum\limits_{n=1}^\infty\, \bigl(a_s^{\rm{b}} \bigr)^n \, 
  \biggl({Q^2 \over \mu^2}\biggl)^{\! -n\varepsilon}\, {\cal F}_n^{}\, \, .
\eeq

\vspace*{1mm}
In terms of the $n$-th order coefficients $A_n$ in eqs.~(\ref{eq:Aqexp}) and (\ref{eq:A4q}) 
and the coefficients $G_n(\varepsilon)$ of the function $G$ in eq.~(\ref{eq:Gdeq}), 
which are still functions of the parameter $\varepsilon$, 
the expansion of eq.~(\ref{eq:finiteorderff}) up to four loops leads to \cite{Moch:2005id,Ravindran:2005vv},
\bea
  {\cal F}_1^{} & = & 
          - {1 \over 2} \* {1 \over \varepsilon^2} \* A_1
          - {1 \over 2} \* {1 \over \varepsilon} \* G_1
\label{eq:ff1loop}
\, ,  \\[1mm]
  {\cal F}_2^{} & = & 
            {1 \over 8} \* {1 \over \varepsilon^4} \* \bigl(A_1\bigr)^2
          + {1 \over 8} \* {1 \over \varepsilon^3} \* A_1 \* ( 
            2 \* G_1
          - \beta_0
          )
          + {1 \over 8} \* {1 \over \varepsilon^2} \* \biggl(
            \bigl(G_1\bigr)^2 
          - A_2 
          - 2 \* \beta_0 \* G_1
          \biggr)
          - {1 \over 4} \* {1 \over \varepsilon} \* G_2
\label{eq:ff2loop}
\, ,  \\[1mm]
  {\cal F}_3^{} & = & 
          - {1 \over 48} \* {1 \over \varepsilon^6} \* \bigl(A_1\bigr)^3
          - {1 \over 16} \* {1 \over \varepsilon^5} \* \bigl(A_1\bigr)^2 \* ( 
            G_1
          - \beta_0
          )
          - {1 \over 144} \* {1 \over \varepsilon^4} \* A_1 \* \biggl(
            9 \* \bigl(G_1\bigr)^2 
          - 9 \* A_2 
          - 27 \* \beta_0 \* G_1 
\nonumber\\
& &\mbox{}
          + 8 \* \beta_0^2
          \biggr)
          - {1 \over 144} \* {1 \over \varepsilon^3} \* \biggl( 
            3 \* \bigl(G_1\bigr)^3 
          - 9 \* A_2 \* G_1
          - 18 \* A_1 \* G_2
          + 4 \* \beta_1 \* A_1 
          - 18 \* \beta_0 \* \bigl(G_1\bigr)^2 
          + 16 \* \beta_0 \* A_2 
\nonumber\\
& &\mbox{}
          + 24 \* \beta_0^2 \* G_1
          \biggr)
          + {1 \over 72} \* {1 \over \varepsilon^2} \* \biggl(
            9 \* G_1 \* G_2
          - 4 \* A_3 
          - 6 \* \beta_1 \* G_1
          - 24 \* \beta_0 \* G_2
          \biggr)
          - {1 \over 6} \* {1 \over \varepsilon} \* G_3
\label{eq:ff3loop}
\, ,  \\[1mm]
  {\cal F}_4^{} & = & 
            {1 \over 384} \* {1 \over \varepsilon^8} \* \bigl(A_1\bigr)^4
          + {1 \over 192} \* {1 \over \varepsilon^7} \* \bigl(A_1\bigr)^3 \* (
            2 \* G_1
          - 3 \* \beta_0
          )
          + {1 \over 1152} \* {1 \over \varepsilon^6} \* \bigl(A_1\bigr)^2 \* \biggl(
            18 \* \bigl(G_1\bigr)^2 
          - 18 \* A_2 
\nonumber\\
& &\mbox{}
          - 72 \* \beta_0 \* G_1
          + 41 \* \beta_0^2
          \biggr)
          + {1 \over 576} \* {1 \over \varepsilon^5} \* A_1 \* \biggl(
            6 \* \bigl(G_1\bigr)^3 
          - 18 \* A_2 \* G_1
          - 18 \* A_1 \* G_2
          + 8 \* \beta_1 \* A_1 
\nonumber\\
& &\mbox{}
          - 45 \* \beta_0 \* \bigl(G_1\bigr)^2 
          + 41 \* \beta_0 \* A_2 
          + 82 \* \beta_0^2 \* G_1
          - 18 \* \beta_0^3
          \biggr)
          + {1 \over 1152} \* {1 \over \varepsilon^4} \* \biggl(
            3 \* \bigl(G_1\bigr)^4 
          - 18 \* A_2 \* \bigl(G_1\bigr)^2 
\nonumber\\
& &\mbox{}
          + 9 \* \bigl(A_2\bigr)^2 
          - 72 \* A_1 \* G_1 \* G_2
          + 32 \* A_1 \* A_3
          + 64 \* \beta_1 \* A_1 \* G_1
          - 36 \* \beta_0 \* \bigl(G_1\bigr)^3 
          + 100 \* \beta_0 \* A_2 \* G_1
\nonumber\\
& &\mbox{}
          + 228 \* \beta_0 \* A_1 \* G_2
          - 48 \* \beta_0 \* \beta_1 \* A_1 
          + 132 \* \beta_0^2 \* \bigl(G_1\bigr)^2 
          - 108 \* \beta_0^2 \* A_2 
          - 144 \* \beta_0^3 \* G_1
          \biggr)
\nonumber\\
& &\mbox{}
          + {1 \over 288} \* {1 \over \varepsilon^3} \* \biggl( 
          - 9 \* \bigl(G_1\bigr)^2 \* G_2
          + 8 \* A_3 \* G_1
          + 9 \* A_2 \* G_2
          + 24 \* A_1 \* G_3
          - 3 \* \beta_2 \* A_1 
          + 12 \* \beta_1 \* \bigl(G_1\bigr)^2 
\nonumber\\
& &\mbox{}
          - 9 \* \beta_1 \* A_2 
          + 66 \* \beta_0 \* G_1 \* G_2
          - 27 \* \beta_0 \* A_3 
          - 48 \* \beta_0 \* \beta_1 \* G_1
          - 108 \* \beta_0^2 \* G_2
          \biggr)
          + {1 \over 96} \* {1 \over \varepsilon^2} \* \biggl(
            3 \* \bigl(G_2\bigr)^2 
\nonumber\\
& &\mbox{}
          + 8 \* G_1 \* G_3
          - 3 \* A_4 
          - 4 \* \beta_2 \* G_1
          - 12 \* \beta_1 \* G_2
          - 36 \* \beta_0 \* G_3
          \biggr)
          - {1 \over 8} \*  {1 \over \varepsilon} \* G_4
\label{eq:ff4loop}
\, .
\eea

As observed at lower fixed orders \cite{Ravindran:2004mb,Moch:2005tm} and generalized in~\cite{Dixon:2008gr} 
the function $G$ in eq.~(\ref{eq:Gdeq}) generates the subleading poles in $\varepsilon$ at each order. 
$G$ is the sum of three terms:
twice the coefficient $B^{\rm q}$ in eq.~(\ref{xto1Lnc}) of the $\delta(1-x)$ part in the relevant parton splitting
function, the single-logarithmic anomalous dimension of the eikonal form factor, 
and a term associated with the QCD beta function.
Thus, the perturbative coefficients $G_n(\varepsilon)$ satisfy the following relations
\bea
\label{eq:Gff}
G_1 &=& 2\*B_1^{\rm q} + f_1^{\rm q}  + \varepsilon\*f_{01}^{\rm q}
\, ,\nonumber \\[1mm]
G_2 &=& 2\*B_2^{\rm q} + (f_2^{\rm q} + \beta_0\*f_{01}^{\rm q}) + \varepsilon\*f_{02}^{\rm q}
\, ,\nonumber \\[1mm]
G_3 &=& 2\*B_3^{\rm q} + (f_3^{\rm q} + \beta_1\*f_{01}^{\rm q} + \beta_0\*f_{02}^{\rm q}) + \varepsilon\*f_{03}^{\rm q}
\, ,\nonumber \\[1mm]
G_4 &=& 2\*B_4^{\rm q} + (f_4^{\rm q} + \beta_2\*f_{01}^{\rm q} + \beta_1\*f_{02}^{\rm q} + \beta_0\*f_{03}^{\rm q}) + \varepsilon\*f_{04}^{\rm q}
\, ,
\eea
where the functions $f_{0n}^{\rm q}(\varepsilon)$ at $n$ loops are polynomials in $\varepsilon$.
For consistency, their lower order expressions are needed to sufficient depth in $\varepsilon$ 
in eqs.~(\ref{eq:ff1loop})--(\ref{eq:ff4loop}).

\vspace*{1mm}
The finite coefficients $f_n^{\rm q}$ in the eq.~(\ref{eq:Gff}) are related to
the anomalous dimension of the eikonal form factor~\cite{Dixon:2008gr}, 
see also the recent work~\cite{Falcioni:2019nxk}. 
In analogy to the cusp anomalous dimensions $A^{\rm q}$, they exhibit 
a maximally non-Abelian colour structure. 
This implies that the $\nf$-independent colour factor for quarks in $f_n^{\rm q}$ is 
proportional to $\cf\ca^{(n-1)}$ up to three loops, 
where the explicit results read~\cite{Moch:2005tm}
\bea
\label{eq:softfqexp}
  f_1^{\rm q} &=&
  0
  \, ,
  \nn \\
  f_2^{\rm q} &=&
  \cf \* \biggl\{
  \ca\* \left( \frac{808}{27} - \frac{22}{3} \* \zeta_2 - 28 \* \zeta_3 \right)
  + \nf\* \left( - \frac{112}{27} + \frac{4}{3} \* \zeta_2 \right)
  \biggr\}
  \, ,
  \nn \\
  f_3^{\rm q} &=&
  \cf \* \biggl\{
  \cas \* \left( 
    \frac{136781}{729} 
    - \frac{12650}{81} \* \zeta_2 
    - \frac{1316}{3} \* \zeta_3 
    + 176 \* \zeta_4
    + \frac{176}{3} \* \zeta_2 \* \zeta_3 
    + 192 \* \zeta_5 
  \right)
\nonumber\\
& & \mbox{}
  + \cf\* \nf \* \left(  
    - \frac{1711}{27} 
    + 4 \* \zeta_2 
    + \frac{304}{9} \* \zeta_3 
    + 16 \* \zeta_4
  \right)
  + \ca\* \nf \* \left( 
    - \frac{11842}{729} 
    + \frac{2828}{81} \* \zeta_2 
\right.
\nonumber\\
&& 
\left. \mbox{}
    + \frac{728}{27} \* \zeta_3 
    - 48 \* \zeta_4
  \right)
  + \nfs \* \left( 
    - \frac{2080}{729} 
    - \frac{40}{27} \* \zeta_2 
    + \frac{112}{27} \* \zeta_3
  \right)
  \biggr\}
\, .
\eea

With the currently available results for $B_4^{\rm q}$ in eq.~(\ref{eq:B4q})
and $G_4$ in eq.~(\ref{eq:Gff}) in the planar limit we can determine $f_4^{\rm q}$ for quarks completely 
in the large-$\nc$ limit as 
\bea
\label{eq:f4qLnc}
\nonumber 
f_4^{\rm q}\biggl|_{{\rm L}\nc} &=& 
         \cf\*\nct \* \biggl(
            \frac{9364079}{6561}
          - \frac{1186735}{729} \* \zeta_2
          - \frac{837988}{243} \* \zeta_3
          + \frac{115801}{27} \* \zeta_4
          + \frac{11896}{9} \* \zeta_2 \* \zeta_3
\\
&& \nonumber \mbox{}
          + 3952 \* \zeta_5
          - \frac{4796}{9} \* \zeta_3^2
          - \frac{129547}{54} \* \zeta_6
          - 416 \* \zeta_2 \* \zeta_5
          - 720 \* \zeta_3 \* \zeta_4
          - 1700 \* \zeta_7
          \biggr)
\\
&& \nonumber \mbox{}
       + \cf\*\ncs\*\nf \* \biggl(
          - \frac{247315}{432}
          + \frac{412232}{729} \* \zeta_2
          + \frac{102205}{243} \* \zeta_3
          - \frac{7589}{6} \* \zeta_4
          - \frac{824}{9} \* \zeta_2 \* \zeta_3
\\
&& \nonumber \mbox{}
          - \frac{740}{9} \* \zeta_5
          + \frac{2816}{9} \* \zeta_3^2
          + \frac{15611}{27} \* \zeta_6
          \biggr)
\\
&& \nonumber \mbox{}
       + \cf\*\nc\*\nfs \* \left(
            \frac{329069}{17496}
          - \frac{22447}{729} \* \zeta_2
          + \frac{25300}{243} \* \zeta_3
          + \frac{140}{3} \* \zeta_4
          - \frac{176}{9} \* \zeta_2 \* \zeta_3
          - \frac{856}{9} \* \zeta_5
          \right)
\\
&&  \mbox{}
       + \cf\*\nft \* \left(
          - \frac{16160}{6561}
          - \frac{16}{81} \* \zeta_2
          - \frac{400}{243} \* \zeta_3
          + \frac{128}{27} \* \zeta_4
          \right)
\, .
\qquad\quad
\eea
With these results the perturbative expansion of $f^{\rm q}$ through four loops reads
\bea
\label{eq:fqExpQ}
  f^{\rm q}(\nf\!=\!3) &=&
  - 0.44960\:\als^2 \:  ( 1  +  0.9300\,\als + (1.086)\big|_{{\rm L}\nc}\, \als^2 \: 
       + \, \ldots )
\; , \nn \\
  f^{\rm q}(\nf\!=\!4) &=&
  - 0.46610\:\als^2 \:  ( 1  +  0.8043\,\als + (0.753)\big|_{{\rm L}\nc}\, \als^2 \: 
       + \, \ldots )
\; , \nn \\
  f^{\rm q}(\nf\!=\!5) &=&
  - 0.48261 \:\als^2 \: ( 1  +  0.6881\,\als + (0.438)\big|_{{\rm L}\nc}\, \als^2 \: 
       + \, \ldots )
\, , 
\quad
\eea
where we have used the full QCD coefficients $f_n^{\rm q}$ in eq.~(\ref{eq:softfqexp}) 
and, as indicated, the large-$\nc$ result of eq.~(\ref{eq:f4qLnc}). 
Based on experience with other anomalous dimensions and lower orders, 
the latter is expected to provide a very good approximation of full QCD for
the coefficients of individual powers of $\nf$, 
as well as the complete result for the
four loop term, putting $\nf = 3$, $4$ or $5$.

\vspace*{1mm}
With the results of Sec.~\ref{sec:splittingfn} and using all available results
for the quark form factor we can, in addition, also
determine the complete colour decomposition of the four-loop coefficient $f_4^{\rm q}$.
Given its relation to the anomalous dimension of the eikonal form factor, 
we assume here that $f_4^{\rm q}$ exhibits the same generalized maximal non-Abelian property 
in the presence of quartic colour factors 
as the cusp anomalous dimension $A_4$, see~\cite{Moch:2018wjh}.
This assumption, supported also by recent studies of amplitude
factorization in~\cite{Falcioni:2019nxk},
implies the absence of the colour coefficients $\cff$, $\cft\ca$ and $\cfs\cas$ in
$f_4^{\rm q}$, which leads to the result
\bea
\label{eq:f4q}
\nonumber 
f_4^{\rm q} &=& 
         \cf \* \cat \* \biggl(
            \frac{9364079}{6561}
          - \frac{1186735}{729} \* \zeta_2
          - \frac{837988}{243} \* \zeta_3
          + \frac{115801}{27} \* \zeta_4
          + \frac{11896}{9} \* \zeta_2 \* \zeta_3
\\
&& \nonumber \mbox{}
          + 3952 \* \zeta_5
          - \frac{4796}{9} \* \zeta_3^2
          - \frac{129547}{54} \* \zeta_6
          - 416 \* \zeta_2 \* \zeta_5
          - 720 \* \zeta_3 \* \zeta_4
          - 1700 \* \zeta_7
	  -\frac{1}{24}\* ~f_{4,\, \dfFA}^{\rm q}
          \biggr)
\\
&& \nonumber \mbox{}
       + \DfRAnc \* ~f_{4,\, \dfFA}^{\rm q}~ 
       + \cft\*\nf \* ~f_{4,\, {\nf\*\cft}}^{\rm q}~
       + \cfs\*\ca\*\nf \* ~f_{4,\, {\nf\*\cfs\*\ca}}^{\rm q}~
       + \cf\*\cas\*\nf \* \biggl(
          - \frac{243859}{432} 
\\
&& \nonumber \mbox{}
          + \frac{389228}{729}\*\zeta_2 
          + \frac{105193}{243}\*\zeta_3 
          - \frac{22667}{18}\*\zeta_4
          - \frac{848}{9}\*\zeta_2\*\zeta_3 
          - \frac{860}{27}\*\zeta_5 
          + \frac{2740}{9}\*\zeta_3^2 
          + \frac{5179}{9}\*\zeta_6
\\
&& \nonumber \mbox{}
          + \frac{1}{24}\* ~b_{4,\, \dfFF}^{\rm q}~
	  -\frac{1}{2}\* ~f_{4,\, {\nf\*\cfs\*\ca}}^{\rm q}~
	  -\frac{1}{4}\* ~f_{4,\, {\nf\*\cft}}^{\rm q}~
          \biggr)
       + \nf \* \DfRRnc \* \biggl(
           - 384 
           + \frac{4544}{3}\*\zeta_2 
\\
&& \nonumber \mbox{}
           - \frac{5312}{9}\*\zeta_3 
           - \frac{800}{3}\*\zeta_4
           + 128\*\zeta_2\*\zeta_3 
           - \frac{21760}{9}\*\zeta_5 
           + \frac{1216}{3}\*\zeta_3^2 
           + \frac{1184}{9}\*\zeta_6
          - 2\* ~b_{4,\, \dfFF}^{\rm q}~
          \biggr)
\\
&& \nonumber \mbox{}
       + \cfs\*\nfs  \* \biggl(
            \frac{16733}{486} 
          - \frac{172}{9}\*\zeta_2 
          - \frac{4568}{81}\*\zeta_3 
          + \frac{64}{9}\*\zeta_4
          + \frac{32}{3}\*\zeta_2\*\zeta_3 
          + \frac{304}{9}\*\zeta_5 
          \biggr)
\\
&& \nonumber \mbox{}
       + \cf\*\ca\*\nfs \* \left(
            \frac{27875}{17496} 
          - \frac{15481}{729}\*\zeta_2 
          + \frac{32152}{243}\*\zeta_3 
          + \frac{388}{9}\*\zeta_4
          - \frac{224}{9}\*\zeta_2\*\zeta_3 
          - 112\*\zeta_5 
          \right)
\\
&& \mbox{}
       + \cf\*\nft \* \left(
          - \frac{16160}{6561}
          - \frac{16}{81} \* \zeta_2
          - \frac{400}{243} \* \zeta_3
          + \frac{128}{27} \* \zeta_4
          \right)
\, ,
\qquad\quad
\eea
with three still unknown coefficients $f_{4,\, {\nf\*\cft}}^{\rm q}$,  $f_{4,\, {\nf\*\cfs\*\ca}}^{\rm q}$ and $f_{4,\, \dfFA}^{\rm q}$.

\vspace*{1mm}
We do have, however, a low number of fixed Mellin moments for the DIS structure functions at four loops at our disposal~\cite{Ruijl:2016pkm}.
As will be explained in the Sec.~\ref{sec:inclusiveSV} below, this information
can be used to constrain the DIS Wilson coefficients at large $x$,
specifically the term proportional to the plus-distribution $[1/(1-x)]_+$, 
so that we can extract numerical values for the unknown colour coefficients of $f_{4}^{\rm q}$ in eq.~(\ref{eq:f4q}).
We obtain
\bea
\label{eq:f4qunknowns}
\nonumber 
f_{4,\, {\nf\*\cft}}^{\rm q} &=& (- 0.4 \pm \,6. )\cdot 10^3
\nonumber\\
f_{4,\, {\nf\*\cfs\*\ca}}^{\rm q} &=& ( \phantom{-} 2. \phantom{0} \pm \,6. )\cdot 10^3
\nonumber\\
f_{4,\, \dfFA}^{\rm q} &=& (- 1. \phantom{0} \pm \,1. )\cdot 10^2
\, ,
\eea
where the errors are correlated due to the known exact results in the large-$\nc$ limit. 
These are the best estimations for these coefficients given our current knowledge.

\vspace*{1mm}
The four-loop contributions in eq.~(\ref{eq:f4qunknowns}) display still a significant uncertainty, 
but in the bare form factor ${\cal F}_4$ at four loops in eq.~(\ref{eq:ff4loop}) they are numerically not dominant.
This leads to the expression for the full colour dependence of the single
pole in $\varepsilon$ at four loops
\bea
\nonumber
  {\cal F}_4^{}\biggr|_{1/\varepsilon} & = & 
         \cff \* \left(  - 2212.8 \pm 0.3 \right)
       + \cft \* \ca \* \left(  - 1601.9 \pm 0.5 \right)
       + \cfs \* \cas \*  \left( 19661.7 \pm 0.5 \right)
\\[-3mm]
&& \nonumber \mbox{\hspace*{-3mm}}
       + \cf \* \cat \* \left( - 13274.1 \pm 1.0\right)
       + \DfRAnc \* \left(  262.3 \pm 12.5 \right)
       + \cft\*\nf \* \left( 2140. \pm 750. \right)
\\[-1mm]
&& \nonumber \mbox{\hspace*{-3mm}}
       + \cfs\*\ca\*\nf \* \left(  - 12800. \pm 750. \right)
       + \cf\*\cas\*\nf \* \left( 10320. \mp 560. \right)
       + \nf \* \DfRRnc \* \left( 53.12744 \right)
\\[1mm]
&& \mbox{\hspace*{-3mm}}
       + \cfs\*\nfs \* \left( 1604.851 \right)
       + \cf\*\ca\*\nfs \* \left( - 2304.682 \right)
       + \cf\*\nft \* \left( 158.0655 \right)
\label{eq:ff4loopsinglepole}
\, ,
\eea
where all exact values have been rounded to seven digits and the errors 
are inherited from the numerical results in Tab.~\ref{tab:Bq} and eq.~(\ref{eq:f4qunknowns}) 
and therefore are correlated.

\vspace*{1mm}
Note that the form factor ${\cal F}$ in eq.~(\ref{eq:FFphoton}) receives
additional corrections due to a new flavour structure starting from three 
loops, where the photon couples to a closed quark loop~\cite{Vermaseren:2005qc}.
This requires the summation over the charges of all quark flavours in the loop, 
leading to additional terms with a relative factor 
$(\sum_{\rm q^\prime} e_{\rm q^\prime})/e_{\rm q} = \nf \langle e \rangle /e_{\rm q}$ 
and proportional to the group invariant $(d^{\,abc} d_{\,abc})/\nc$ in eq.~(\ref{eq:dabc}).
For the analogous expression of a charged-current, i.e., the form factor with
the coupling of a $W^\pm$-boson, such contributions are absent.
For the photon form factor the respective terms proportional to $(d^{\,abc} d_{\,abc})/\nc$ 
are finite at three loops, thus appearing in $f_{03}^{\rm q}$ in $G_3$ in eq.~(\ref{eq:Gff}).
At four loops, those terms enter in $G_4$ through $\beta_0\*f_{03}^{\rm q}$ 
in eq.~(\ref{eq:Gff}) and generate a single pole in $\varepsilon$ in the bare form factor, 
as confirmed by the exact result for the coefficient of $\nf (d^{\,abc} d_{\,abc})/\nc$ in~\cite{vonManteuffel:2019wbj}.
This colour factor has been omitted in eq.~(\ref{eq:ff4loopsinglepole}).

\vspace*{1mm}
Eqs.~(\ref{eq:f4qLnc}), (\ref{eq:f4q}) and (\ref{eq:ff4loopsinglepole})
are new results of the present paper, with the large-$\nc$ results in 
eq.~(\ref{eq:f4qLnc}) being exact. 
Eq.~(\ref{eq:f4q}) for $f_4^{\rm q}$ and, as a consequence, 
eq.~(\ref{eq:ff4loopsinglepole}) for the single pole in $\varepsilon$ 
of ${\cal F}_4$ are based on the (by now well-supported) conjecture that $f_4^{\rm q}$ 
has the same generalized maximal non-Abelian colour structure as the cusp anomalous
dimension.

\subsection{DIS Wilson coefficients at large $x$ }
\label{sec:inclusiveSV} 

The direct link between the QCD form factor and the 
DIS Wilson coefficients near threshold for $x \to 1$ 
through factorization in the soft and collinear limit 
allows to relate the knowledge on either quantity at a given order in
perturbation theory.
To that end, we write the partonic coefficient function $W^{\rm b}$ as a series in
bare coupling $a_s^{\rm b}$, see~\cite{Moch:2005id,Moch:2005ky}. 
With the convention in eq.~(\ref{eq:finiteorderff}) 
and choosing the scale $\mu=Q$, the expansion coefficients $W_n^{\rm b}$ have 
the following structure up to four loops,
\bea
\label{eq:Softexp}
 W^{\rm b}_0
   &=& \delta(1-x)
\, , \nn \\
 W^{\rm b}_1
   &=& {\cal S}_1 + 2\,{\cal F}_1\,\delta(1-x) 
\, ,\nn \\
 W^{\rm b}_2
   &=& {\cal S}_2 + 2\, {\cal F}_1 {\cal S}_1 
   + (2\,{\cal F}_2 + \left({\cal F}_1\right)^2)\, \delta(1-x)      
\, ,\nn \\
 W^{\rm b}_3
   &=& {\cal S}_3 + 2\,{\cal F}_1 {\cal S}_2 + (2\,{\cal F}_2 + \left({\cal F}_1\right)^2) {\cal S}_1 
   + (2\,{\cal F}_3 + 2\, {\cal F}_1 {\cal F}_2 )\, \delta(1-x)  
\, ,\nn \\
 W^{\rm b}_4
   &=& 
{\cal S}_4 + 2\,{\cal F}_1 {\cal S}_3 + (2\,{\cal F}_2 + \left({\cal F}_1\right)^2) {\cal S}_2 
   + (2\,{\cal F}_3 + 2\, {\cal F}_1 {\cal F}_2 )\, {\cal S}_1 
\nn \\ \mbox{}
& &
   + (2\,{\cal F}_4 + \left({\cal F}_2\right)^2 + 2\, {\cal F}_1 {\cal F}_3 )\, \delta(1-x)
\: .
\eea
Here ${\cal F}_n$ are the space-like form factors discussed in the previous
Sec.~\ref{sec:formfactor}, whereas ${\cal S}_n$ denote the real emission contributions. 
In the limit Bjorken $x \ra 1$, the singular terms are proportional to
$\delta(1-x)$ and to the usual plus-distributions, 
\beq
\label{eq:plusdef}
  \DD_{k} 
  \: = \: \left[ \frac{\ln^{\,k} (1-x)}{1-x} \right]_+ \:\: , 
  \quad\quad 
  k \: = \: 1,\,\ldots\, 2n-1
  \, .
\eeq

The dependence of the pure real-emission contributions ${\cal S}_n$ on the
scaling variable $x$ is given by the $d$-dimensional plus-distributions up to 
$f_{n,\varepsilon}$ defined by 
\beq
\label{eq:Dplus}
f_{k,\varepsilon}(x) 
  \; = \; \varepsilon [\,(1-x)^{-1-k\varepsilon}\,]_+
  \; = \; - {1 \over k}\, \delta(1-x) + \sum_{i=0}\, {(-k \varepsilon)^i \over i\, !}\,\varepsilon\,\DD_{\,i} 
  \, .
\eeq

The bare strong coupling in the coefficients $W_n^{\rm b}$ has to be renormalized (in the \MSb\
scheme) according to
\begin{align}
\label{eq:asrenorm}
a_s^b = a_s 
& \bigg\{ 
   1 
   - \frac{\beta_0}{\varepsilon}a_s   
   + \Big( \frac{\beta_0^2}{\varepsilon^2} - \frac{\beta_1}{2\varepsilon}\Big) a_s^2 
   - \Big(  \frac{\beta_0^3}{\varepsilon^3} - \frac{7\beta_1\beta_0}{6\varepsilon^2} + \frac{\beta_2}{3\varepsilon}  \Big) a_s^3 
   \bigg\} \, .
\end{align}

With the ingredients of Secs.~\ref{sec:splittingfn} and \ref{sec:formfactor}, using eq.~(\ref{eq:Softexp}), 
the renormalized coupling in eq.~(\ref{eq:asrenorm}), 
and the known results for the DIS Wilson coefficients up to third order~\cite{Vermaseren:2005qc}
we can then derive the coefficient functions $C_{a,q}$ 
for the DIS structure functions $F_a$ for $a=1,2,3$ 
in the soft and collinear limit up to four-loop order.
In $x$-space, with distributions $f_i$ for the parton $i$, 
the DIS structure functions $F_a$ factorize as 
$F_a(x,Q^2) = \left[C_{a, i}(\alpha_s,\mu^2/Q^2) \otimes f_i(\mu^2)\right](x)$,
cf. eq.~(\ref{eq:cNres}) for the corresponding Mellin space expressions.
With the standard normalization 
\bea
\label{eq:C2DIS}
C_{a,q}
&=& 
\delta(1-x) + \ars c^{(1)}_{i, {\rm q}} + \ars^2 c^{(2)}_{i, {\rm q}} + \ars^3 c^{(3)}_{i, {\rm q}} + \ars^4 c^{(4)}_{2, {\rm q}}
  \, ,
\eea
for $a=1,2,3$, we determine 
the coefficient proportional to $\DD_{0}$ in $c^{(4)}_{a, {\rm q}}$.
For the structure function $F_2$ it takes the following form, 
\bea
\nonumber
\label{eq:c2dis4full}
  c^{(4)}_{2, {\rm q}}\biggr|_{\DD_{0}} &=&
  - f_4^{\rm q} - B_4^{\rm q}
  -\left(B_3^{\rm q}+f_{3}^{\rm q}\right)\*\left(
      \cqonlnN
    - \frac{3}{2}\*\zeta_2\*A_1
    \right)
 +\frac{1}{3}\*\zeta_3\*\left(B_1^{\rm q}+f_{1}^{\rm q}\right)^4
 \nonumber\\[-0.5mm]
&&\mbox{\hspn}
  -\left(B_2^{\rm q}+f_{2}^{\rm q}\right)\*\left(
      \cqsnlnN
    - \frac{3}{2}\*\zeta_2\*A_2
    - \frac{3}{2}\*\zeta_2\*A_1\*\cqonlnN
    + \frac{15}{16}\*\zeta_4\*A_1^2
    \right)
\nonumber\\[0.5mm]
&&\mbox{\hspn}
  +\left(B_1^{\rm q}+f_{1}^{\rm q}\right)^3\*\left(
        \frac{1}{2}\*\zeta_2\*\cqonlnN
      - \frac{5}{8}\*\zeta_4\*A_1
    \right)
  + \frac{3}{2}\*\zeta_2\*\left(B_1^{\rm q}+f_{1}^{\rm q}\right)^2\*\left(
      B_2^{\rm q}+f_{2}^{\rm q}
    \right)
\nonumber\\[0.5mm]
&&\mbox{\hspn}
  + 4\*\zeta_3\*A_1\*\left(B_1^{\rm q}+f_{1}^{\rm q}\right)\*\left(
      B_2^{\rm q}+f_{2}^{\rm q}
    \right)
          + 2\*\zeta_3\*A_1\*A_3
          + \zeta_3\*A_1^2\*\cqsnlnN
\nonumber\\[0.5mm]
&&\mbox{\hspn}
  +\left(B_1^{\rm q}+f_{1}^{\rm q}\right)^2\*\left(
        2\*\zeta_3\*A_2
      + 2\*\zeta_3\*A_1\*\cqonlnN
      - \frac{15}{2}\*\zeta_2\*\zeta_3\*A_1^2
      + 9\*\zeta_5\*A_1^2
    \right)
\nonumber\\[0.5mm]
&&\mbox{\hspn}
  -\left(B_1^{\rm q}+f_{1}^{\rm q}\right)\*\left(
      \cqtnlnN
    - \frac{3}{2}\*\zeta_2\*A_3
    - \frac{3}{2}\*\zeta_2\*A_2\*\cqonlnN
    - \frac{3}{2}\*\zeta_2\*A_1\*\cqsnlnN
\right.
\nonumber\\[0.5mm]
&&\mbox{\hspn}
\left.
    + \frac{15}{16}\*\zeta_4\*A_1^2\*\cqonlnN
    + \frac{15}{8}\*\zeta_4\*A_1\*A_2
    + \frac{35}{6}\*\zeta_3^2\*A_1^3
    - \frac{525}{128}\*\zeta_6\*A_1^3
    \right)
\nonumber\\[0.5mm]
&&\mbox{\hspn}
          + A_2\*\left( 
            \zeta_3\*A_2
          + 9\*\zeta_5\*A_1^2
          - \frac{15}{2}\*\zeta_2\*\zeta_3\*A_1^2
          + 2\*\zeta_3\*A_1\*\cqonlnN
          \right)
\nonumber\\[0.5mm]
&&\mbox{\hspn}
          - A_1^3\*\cqonlnN\*\left(
            \frac{5}{2}\*\zeta_2\*\zeta_3
          - 3\*\zeta_5
         \right)
          + A_1^4\*\left(
            \frac{35}{16}\*\zeta_3\*\zeta_4
          - \frac{21}{2}\*\zeta_2\*\zeta_5
          + 15\*\zeta_7
         \right)
\nonumber\\[0.5mm]
&&\mbox{\hspn}
  - \beta_2\*\left( 
      \fnoq
    + \cqonlnN
    - \frac{1}{2}\*\zeta_2\*A_1
  \right)
  + \beta_0\*\beta_1\*\left( 
      \frac{5}{2}\*\zeta_2\*f_{1}^{\rm q}
    + \frac{5}{2}\*\zeta_2\*B_1^{\rm q}
    + \frac{5}{3}\*\zeta_3\*A_1
  \right)
\nonumber\\[0.5mm]
&&\mbox{\hspn}
  + \beta_0^3\*\left( 
      3\*\zeta_2\*\fnoq
    + 3\*\zeta_2\*\cqonlnN
    + 2\*\zeta_3\*f_{1}^{\rm q}
    + 2\*\zeta_3\*B_1^{\rm q}
    - \frac{3}{8}\*\zeta_4\*A_1
  \right)
\nonumber\\[0.5mm]
&&\mbox{\hspn}
  + \beta_0^2\*\left( 
      \frac{11}{3}\*\zeta_3\*\left(B_1^{\rm q}+f_{1}^{\rm q}\right)^2
    + \left(B_1^{\rm q}+f_{1}^{\rm q}\right)\*\left( 
      \frac{9}{2}\*\zeta_2\*\fnoq
    + \frac{11}{2}\*\zeta_2\*\cqonlnN
    - \frac{5}{2}\*\zeta_4\*A_1
  \right)
\right.
\nonumber\\[0.5mm]
&&\mbox{\hspn}
\left.
    + 3\*\zeta_2\*\left(B_2^{\rm q}+f_{2}^{\rm q}\right)
    + 2\*\zeta_3\*A_2
    + \zeta_3\*A_1\*\left(
      \frac{16}{3}\*\fnoq + 6\*\cqonlnN 
  \right)
    - A_1^2\*\left(
        \frac{20}{3}\*\zeta_2\*\zeta_3
      - 8\*\zeta_5 
  \right)
  \right)
\nonumber\\[0.5mm]
&&\mbox{\hspn}
  + \beta_1\*\bigg( 
        \frac{3}{2}\*\zeta_2\*\left(B_1^{\rm q}+f_{1}^{\rm q}\right)^2
      + \frac{10}{3}\*\zeta_3\*\left(B_1^{\rm q}+f_{1}^{\rm q}\right)\*A_1
          - \fnsq
\nonumber\\[0.5mm]
&&\mbox{\hspn}
          - 2\*\cqsnlnN
          + \zeta_2\*A_2
          - \cqonlnN\*\fnoq
          + A_1\*\left(
            2\*\zeta_2\*\cqonlnN
          + \frac{3}{2}\*\zeta_2\*\fnoq
          - \frac{5}{8}\*\zeta_4\*A_1
          \right)
  \bigg)
\nonumber\\[0.5mm]
&&\mbox{\hspn}
  + \beta_0\*\bigg( 
            \frac{3}{2}\*\zeta_2\*A_3
          - \fntq
          - 3\*\cqtnlnN
          + A_2\*\left(
              \frac{3}{2}\*\zeta_2\*\fnoq
            - \frac{15}{8}\*\zeta_4\*A_1
          \right)
\nonumber\\[0.5mm]
&&\mbox{\hspn}
          + \frac{3}{2}\*\zeta_2\*A_1\*\fnsq
          - \cqsnlnN\*\left( 
              \fnoq 
            - \frac{7}{2}\*\zeta_2\*A_1 
          \right)
          - A_1^3\*\left(
              \frac{35}{6}\*\zeta_3^2
            - \frac{525}{128}\*\zeta_6
          \right)
\nonumber\\[0.5mm]
&&\mbox{\hspn}
          - \frac{15}{16}\*\zeta_4\*A_1^2\*\fnoq
          - \cqonlnN\*\left( 
            \fnsq
            - \frac{5}{2}\*\zeta_2\*A_2 
            - \frac{3}{2}\*\zeta_2\*A_1\*\fnoq 
            + \frac{25}{16}\*\zeta_4\*A_1^2
          \right)
\nonumber\\[0.5mm]
&&\mbox{\hspn}
          + \left(B_2^{\rm q}+f_{2}^{\rm q}\right)\*\left(
            \frac{9}{2}\*\zeta_2\*B_1^{\rm q}+\frac{9}{2}\*\zeta_2\*f_{1}^{\rm q}
            +\frac{16}{3}\*\zeta_3\*A_1
          \right)
          + 2\*\zeta_3\*\left(B_1^{\rm q}+f_{1}^{\rm q}\right)^3
\nonumber\\[0.5mm]
&&\mbox{\hspn}
          + \left(B_1^{\rm q}+f_{1}^{\rm q}\right)^2\*\left(
              \frac{3}{2}\*\zeta_2\*\fnoq
            + 3\*\zeta_2\*\cqonlnN
            - \frac{5}{2}\*\zeta_4\*A_1
          \right)
          + \left(B_1^{\rm q}+f_{1}^{\rm q}\right)\*\left(
              \frac{14}{3}\*\zeta_3\*A_2
\right.
\nonumber\\[0.5mm]
&&\mbox{\hspn}
\left.
            - \frac{35}{2}\*\zeta_2\*\zeta_3\*A_1^2
            + 21\*\zeta_5\*A_1^2
            + 4\*\zeta_3\*A_1\*\fnoq
            + \frac{22}{3}\*\zeta_3\*A_1\*\cqonlnN
          \right)
  \bigg)
\:\: ,
\eea
where $g_{0i}^{\rm DIS}$ denote the coefficients of the perturbative
expansion of the function $g_{0}^{}(Q^2)$ in eq.~(\ref{eq:cNres}), 
which collects the constant terms in $N$. 
Explicit expressions for $g_{0i}^{\rm DIS}$ can be obtained from
eqs.~(4.6)-(4.8) of~\cite{Moch:2005ba} by omitting all terms proportional
to~$\gamma_{E}$ and are collected in eqs.~(\ref{eq:g01})--(\ref{eq:g03}).
The other functions have been defined in eqs.~(\ref{eq:Aqexp}),
(\ref{eq:A4q}), (\ref{eq:Bqexp}), (\ref{eq:B4qLnc}), (\ref{eq:B4q}), (\ref{eq:softfqexp}), (\ref{eq:f4qLnc}) 
and (\ref{eq:f4q}) respectively.

\vspace*{1mm}
The full colour decomposition of $c^{(4)}_{2, {\rm q}}$ is given in 
eq.~(\ref{appB4}) in Appendix~\ref{app:coeff}, where 
also the lower order soft-virtual expressions for $C_{2,q}$ from~\cite{Moch:2005ba}
have been collected for convenience. 
The expressions for the DIS Wilson coefficients in eq.~(\ref{eq:C2DIS}) 
for neutral- or charged-current DIS are identical up to two loops. 
They start to differ from three-loop order onwards, because in the case of neutral-current DIS 
additional corrections need to be considered, where the exchanged virtual
photon couples to a closed quark loop~\cite{Vermaseren:2005qc}.
These contributions gives rise to a new flavour structure, depending on the quark
flavour composition of the nucleon target through their charges $e_{\rm q}$.
In the normalization of~\cite{Vermaseren:2005qc} this leads to a relative factor 
$\floo = 3 \langle e \rangle = (3/\nf) \sum_{\rm q} e_{\rm q}$ in the non-singlet and 
$\floo = (1/\nf) \langle e \rangle^2 / \langle e^2 \rangle = (1/\nf) (\sum_{\rm q^\prime} e_{\rm q^\prime})^2/(\sum_{\rm q} e_{\rm q}^2)$
in the singlet case, respectively.
These $\floo$ terms are implicitly contained in $\cqtnlnN$ 
in eq.~(\ref{eq:c2dis4full}), which is given in eq.~(\ref{eq:g03}),
and spelled out explicitly in eqs.~(\ref{appB3}) and (\ref{appB4}). 

As mentioned already above, the threshold expansion in eq.~(\ref{appB4}) with complete dependence on all
colour factors can then be used as part of an ansatz for the full four-loop
Wilson coefficient $c^{(4)}_{2, {\rm q}}$.
Following a well-established procedure (see, e.g., \cite{Moch:2017uml}) 
the unknown coefficients of a given functional form for $c^{(4)}_{2, {\rm q}}$
can be approximately determined using
the available Mellin moments for $N \leq 9$ of the 
DIS Wilson coefficients at four loops~\cite{Ruijl:2016pkm}.
In this way, fixing the individual colour coefficients in terms proportional to $\DD_{0}$,
the numerical constraints for the remaining unknown coefficients, i.e., 
$f_{4,\, {\nf\*\cft}}^{\rm q}$,  $f_{4,\, {\nf\*\cfs\*\ca}}^{\rm q}$ and $f_{4,\, \dfFA}^{\rm q}$ have been derived.
The terms presented in eq.~(\ref{eq:f4qunknowns}) display still a significant uncertainty, 
whereas the results in large-$\nc$ limit are exact.

\vspace*{1mm}
Therefore, we determine the best estimate for the term proportional to $\DD_{0}$ in $c_{2,\rm q}^{(4)}$ in eq.~(\ref{eq:c2dis4full})
by using the large-$\nc$ limit of eq.~(\ref{eq:f4qLnc}) 
for the terms proportional to $\nfo$ in $f_4^{\rm q}$ 
and keeping the full colour dependence, i.e. putting $\nc = 3$, 
for the $\nfz$, $\nfs$ and $\nft$ terms of $B_4^{\rm q}$ and $f_4^{\rm q}$ given in eqs.~(\ref{eq:B4q}) and (\ref{eq:f4q}) 
and in all lower order terms.
Using this information then leads to the following numerical result  
\bea
\nonumber
\label{eq:c2dis-numerics}
  c^{(4)}_{2, {\rm q}}\biggr|_{\DD_{0}, {\rm best}} &=&
  (3.874 \pm 0.010)\cdot 10^4
  + (- 3.494 \pm 0.032)\cdot 10^4\, \*\nf 
  + 2062.715\, \*\nfs
\\[-2mm]
&& \mbox{}
  - 12.08488\, \*\nft
  + 47.55183\, \*\nf \* \floo 
\:\: ,
\eea
where all exact values have been rounded to seven digits. 
The error on the term proportional to $\nfz$ is dominated by the uncertainty for $f_{4,\, \dfFA}^{\rm q}$
in eq.~(\ref{eq:f4qunknowns}, while for the term proportional to $\nfo$ 
it indicates the effect of varying the large-$\nc$ limit for $f_4^{\rm q}$ 
of eq.~(\ref{eq:f4qLnc}) by $10\%$.

\vspace*{1mm}
Note that the $\delta(1-x)$ coefficient in $c^{(4)}_{2, {\rm q}}$ is still unknown
and requires a complete four-loop computation of DIS structure functions 
including all virtual corrections. 
Numerical estimates based on the currently available number of Mellin
moments are provided in eq.~(\ref{eq:delta4loop}) in Appendix~\ref{app:coeff}.

\subsection{Determination of  $B^{\rm DIS}$ coefficient}
\label{sec:B4}

The resummation coefficient $B_n^{\rm DIS}$ which enters the jet function in eq.~(\ref{eq:Jint}) 
at $n$ loops can finally be derived from the single logarithms $\ln N$ of the Mellin
transforms of the DIS coefficient functions $c^{(n)}_{2, {\rm q}}(x)$. 
Expanding eqs.~(\ref{eq:g1n})--(\ref{eq:g5n}) leads to the relations 
\bea
\label{eq:xB1}
  B_{1}^{\rm DIS} & \! =\!& 
  - f_1^{\rm q} - B_1^{\rm q}
\:\: , \\[1mm]
\label{eq:xB2}
  B_{2}^{\rm DIS}  &\! =\!&
  - f_2^{\rm q} - B_2^{\rm q}
  - \beta_0\*\left( 
      f_{01}^{\rm q}
    + \cqonlnN
    - \frac{1}{2}\*\zeta_2\*A_1
  \right)
\:\: ,
\\[1mm]
\label{eq:xB3}
  B_{3}^{\rm DIS}  &\! =\! & 
  - f_3^{\rm q} - B_3^{\rm q}
  - \beta_1\*\left( 
      f_{01}^{\rm q}
    + \cqonlnN
    - \frac{1}{2}\*\zeta_2\*A_1
  \right)
  + \beta_0^2\*\left( 
      \zeta_2\*f_{1}^{\rm q}
    + \zeta_2\*B_1^{\rm q}
    + \frac{2}{3}\*\zeta_3\*A_1
  \right)
\nonumber\\[0.5mm]
&&\mbox{\hspn}
  - \beta_0\*\left( 
      f_{02}^{\rm q}
    + 2\*\cqsnlnN
    - \left(\cqonlnN\right)^2
    - \zeta_2\*A_2
  \right)
\:\: ,
\\[1mm]
\label{eq:xB4}
  B_{4}^{\rm DIS}  &\! =\! & 
  - f_4^{\rm q} - B_4^{\rm q}
  - \beta_2\*\left( 
      f_{01}^{\rm q}
    + \cqonlnN
    - \frac{1}{2}\*\zeta_2\*A_1
  \right)
\nonumber\\[0.5mm]
&&\mbox{\hspn}
  + \beta_0^3\*\left( 
      3\*\zeta_2\*f_{01}^{\rm q}
    + 3\*\zeta_2\*\cqonlnN
    + 2\*\zeta_3\*f_{1}^{\rm q}
    + 2\*\zeta_3\*B_1^{\rm q}
    - \frac{3}{8}\*\zeta_4\*A_1
  \right)
\nonumber\\[0.5mm]
&&\mbox{\hspn}
  + \beta_0\*\beta_1\*\left( 
      \frac{5}{2}\*\zeta_2\*f_{1}^{\rm q}
    + \frac{5}{2}\*\zeta_2\*B_1^{\rm q}
    + \frac{5}{3}\*\zeta_3\*A_1
  \right)
  + \beta_0^2\*\left( 
      3\*\zeta_2\*f_{2}^{\rm q}
    + 3\*\zeta_2\*B_2^{\rm q}
    + 2\*\zeta_3\*A_2
  \right)
\nonumber\\[0.5mm]
&&\mbox{\hspn}
  - \beta_1\*\bigg( 
      f_{02}^{\rm q}
    + 2\*\cqsnlnN
    - \left(\cqonlnN\right)^2
    - \zeta_2\*A_2
  \bigg)
\nonumber\\[0.5mm]
&&\mbox{\hspn}
  - \beta_0\*\bigg( 
      f_{03}^{\rm q}
    + 3\*\cqtnlnN
    - 3\*\cqsnlnN\*\cqonlnN
    + \left(\cqonlnN\right)^3
    - \frac{3}{2}\*\zeta_2\*A_3
  \bigg)
\:\: ,
\eea
where references for all quantities have been already given below eq.~(\ref{eq:c2dis4full}).
This leads to 
\bea
\label{eq:B1}
  B_{1}^{\rm DIS} & \! =\!& 
  - 3\* \cf 
\:\: ,
\\[1mm]
\label{eq:B2}
  B_{2}^{\rm DIS}  &\! =\!&
  \cfs \* \left( - {3 \over 2} + 12\: \* \zeta_2 - 24\: \* \zeta_3 \right)
  + \cf \* \ca \* \left( - {3155 \over 54} + {44 \over 3}\: \* \zeta_2 
  + 40\: \* \zeta_3 \right) 
  \nn \\ & & \mbox{\hspn}
  + \cf \* \nf \* \left( {247 \over 27} - {8 \over 3}\: \* \zeta_2 \right)
\:\: ,
\\[1mm]
\label{eq:B3}
  B_{3}^{\rm DIS}  &\! =\! & 
       \cft  \*  \left(  - {29 \over 2}\: - 18\: \* \zeta_2 - 68\: \* \zeta_3 
         - 144\: \* \zeta_4 + 32\: \* \zeta_2 \* \zeta_3 
         + 240\: \* \zeta_5 \right)
\nonumber\\[0.5mm]
&&\mbox{\hspn}
       + \ca \* \cfs  \*  \left(  - 46 + 287\: \* \zeta_2 
         - {712 \over 3}\: \* \zeta_3 - 136\: \* \zeta_4
         - 16\: \* \zeta_2 \* \zeta_3 - 120\: \* \zeta_5 \right)
\nonumber\\[0.5mm]
&&\mbox{\hspn}
       + \cas \* \cf  \*  \left(  - {599375 \over 729} 
         + {32126 \over 81}\: \* \zeta_2 + {21032 \over 27}\: \* \zeta_3 
         - {326 \over 3}\: \* \zeta_4 - {176 \over 3}\: \* \zeta_2 \* \zeta_3 
         - 232\: \* \zeta_5 \right)
\nonumber\\[0.5mm]
&&\mbox{\hspn}
       + \cfs \* \nf  \*  \left( {5501 \over 54} - 50\: \* \zeta_2 
         + {32 \over 9}\: \* \zeta_3 \right)
       + \cf \* \nfs  \*  \left(  - {8714 \over 729} 
         + {232 \over 27}\: \* \zeta_2 - {32 \over 27}\: \* \zeta_3 \right)
\nonumber\\[0.5mm]
&&\mbox{\hspn}
       + \ca \* \cf \* \nf  \*  \left( {160906 \over 729} 
         - {9920 \over 81}\: \* \zeta_2 - {776 \over 9}\: \* \zeta_3 
         + {104 \over 3}\: \* \zeta_4 \right)
\:\: .
\eea
Eqs.~(\ref{eq:B1})--(\ref{eq:B3}) are, of course, well-known results~\cite{Moch:2005ba}.
In the large-$\nc$ limit we can determine $B_{4}^{\rm DIS}$ with eqs.~(\ref{eq:B4qLnc}) and (\ref{eq:f4qLnc}) as 
\bea
\label{eq:B4disLnc}
\nonumber 
  B_{4}^{\rm DIS}\biggl|_{{\rm L}\nc} &=& 
  \cf\* \nct \* \biggl(
      -\frac{2040092429}{139968}
      +\frac{23011973}{1944}\*\zeta_2
      +\frac{517537}{36}\*\zeta_3
      -\frac{312481}{36}\*\zeta_4
\\
&& \nonumber
      -\frac{39838}{9}\*\zeta_2\*\zeta_3
      -\frac{50680}{9}\*\zeta_5
      -988\*\zeta_3^2
      +\frac{12467}{6}\*\zeta_6
      +496\*\zeta_2\*\zeta_5
      +688\*\zeta_3\*\zeta_4
      +2260\*\zeta_7
  \biggr)
\\
&& \nonumber
   + \cf\* \ncs\*\nf \* \biggl(
       \frac{83655179}{11664}
      -\frac{5160215}{972}\*\zeta_2
      -\frac{639191}{162}\*\zeta_3
      +\frac{24856}{9}\*\zeta_4
      +\frac{8624}{9}\*\zeta_2\*\zeta_3
\\
&& \nonumber
      +200\*\zeta_5
      -32\*\zeta_3^2
      -\frac{1201}{3}\*\zeta_6
   \biggr)
   + \cf\* \nc\*\nfs \* \biggl(
      -\frac{5070943}{5832}
      +\frac{160903}{243}\*\zeta_2
\\
&& \nonumber
      +\frac{14618}{81}\*\zeta_3
      -\frac{2110}{9}\*\zeta_4
      -\frac{400}{9}\*\zeta_2\*\zeta_3
      +\frac{904}{9}\*\zeta_5
   \biggr)
\\
&& 
   + \cf\* \nft \* \left(
       \frac{50558}{2187}
      +\frac{80}{81}\*\zeta_3
      -\frac{1880}{81}\*\zeta_2
      +\frac{40}{9}\*\zeta_4
   \right)
\, .
\qquad\quad
\eea
Using instead eqs.~(\ref{eq:B4q}) and (\ref{eq:f4q}) with the full colour dependence in $B_4^{\rm DIS}$ gives the following expression
\bea
\label{eq:B4}
  B_{4}^{\rm DIS}  &\! =\!&
       - \cff \* ~b_{4,\, {\cff}}^{\rm q}
       + \cft \* \ca \* \left(
         -\frac{10769}{12}
         -\frac{2717}{6}\*\zeta_2
         -4565\*\zeta_3
         -2200\*\zeta_4
         +1496\*\zeta_2\*\zeta_3
         -616\*\zeta_5
\right.
\nonumber
\\
&& \nonumber
\left. \mbox{}
         +\frac{2992}{3}\*\zeta_3^2
         +\frac{48620}{9}\*\zeta_6
         - ~b_{4,\, {\cft\*\ca}}^{\rm q}~ 
     \right)
      + \cfs \* \cas \* \left(
        -\frac{51325}{72}
        +\frac{307033}{54}\*\zeta_2
        +\frac{160994}{27}\*\zeta_3
\right.
\\
&& \nonumber
\left. \mbox{}
        -\frac{481637}{54}\*\zeta_4
        -\frac{8932}{9}\*\zeta_2\*\zeta_3
        -\frac{17776}{9}\*\zeta_5
        +\frac{616}{3}\*\zeta_3^2
        -\frac{19558}{9}\*\zeta_6
        - ~b_{4,\, {\cfs\*\cas}}^{\rm q}~ 
    \right)
\\
&& \nonumber \mbox{}
       + \cf \* \cat \* \left(
          -\frac{1958802125}{139968}
          +\frac{2213182}{243}\*\zeta_2
          +\frac{676939}{54}\*\zeta_3
          -\frac{198203}{54}\*\zeta_4
          -\frac{38738}{9}\*\zeta_2\*\zeta_3
\right.
\\
&& \nonumber
\left. \mbox{}
          -\frac{40406}{9}\*\zeta_5
          -1340\*\zeta_3^2
          +\frac{10883}{6}\*\zeta_6
          +496\*\zeta_2\*\zeta_5
          +688\*\zeta_3\*\zeta_4
          +2260\*\zeta_7
	  +\frac{1}{2}\* ~b_{4,\, {\cfs\*\cas}}^{\rm q}~
\right.
\\
&& \nonumber
\left. \mbox{}
	  +\frac{1}{4}\* ~b_{4,\, {\cft\*\ca}}^{\rm q}~
	  +\frac{1}{8}\* ~b_{4,\, {\cff}}^{\rm q}~
	  +\frac{1}{24}\* ~f_{4,\, \dfFA}^{\rm q}~
	  +\frac{1}{24}\* ~b_{4,\, \dfFA}^{\rm q}~
          \right)
\\
&& \nonumber \mbox{}
       - \DfRAnc \* \left( ~f_{4,\, \dfFA}^{\rm q}~ + ~b_{4,\, \dfFA}^{\rm q}~
          \right)
       + \cft\*\nf \* \left( 
          \frac{482}{3}
         +\frac{19}{3}\*\zeta_2
         +806\*\zeta_3
         +564\*\zeta_4
\right.
\\
&& \nonumber
\left. \mbox{}
         -272\*\zeta_2\*\zeta_3
         +112\*\zeta_5
         -\frac{544}{3}\*\zeta_3^2
         -\frac{8840}{9}\*\zeta_6
         - ~f_{4,\, {\nf\*\cft}}^{\rm q}~
         - ~b_{4,\, {\nf\*\cft}}^{\rm q}~
    \right)
\\
&& \nonumber \mbox{}
       + \cfs\*\ca\*\nf \* \left( 
          \frac{2465317}{972}
         -\frac{70060}{27}\*\zeta_2
         -\frac{164470}{81}\*\zeta_3
         +\frac{75011}{27}\*\zeta_4
         +\frac{1976}{9}\*\zeta_2\*\zeta_3
         +\frac{2704}{9}\*\zeta_5
\right.
\\
&& \nonumber
\left. \mbox{}
         -\frac{112}{3}\*\zeta_3^2
         +\frac{3556}{9}\*\zeta_6
         - ~f_{4,\, {\nf\*\cfs\*\ca}}^{\rm q}~
         - ~b_{4,\, {\nf\*\cfs\*\ca}}^{\rm q}~
    \right)
       + \cf\*\cas\*\nf \* \left(
           \frac{68301461}{11664}
\right.
\\
&& \nonumber
\left. \mbox{}
          -\frac{1935001}{486}\*\zeta_2
          -\frac{254678}{81}\*\zeta_3
          +\frac{66211}{54}\*\zeta_4
          +\frac{8272}{9}\*\zeta_2\*\zeta_3
          -\frac{772}{27}\*\zeta_5
          +\frac{364}{9}\*\zeta_3^2
          -\frac{9439}{27}\*\zeta_6
\right.
\\
&& \nonumber
\left. \mbox{}
	  +\frac{1}{2}\* ~f_{4,\, {\nf\*\cfs\*\ca}}^{\rm q}~
	  +\frac{1}{2}\* ~b_{4,\, {\nf\*\cfs\*\ca}}^{\rm q}~
	  +\frac{1}{4}\* ~f_{4,\, {\nf\*\cft}}^{\rm q}~
	  +\frac{1}{4}\* ~b_{4,\, {\nf\*\cft}}^{\rm q}~
	  -\frac{1}{48}\* ~b_{4,\, \dfFF}^{\rm q}~
          \right)
\\
&& \nonumber \mbox{}
       + \nf \* \DfRRnc \* \left( 
          384
         -\frac{4544}{3}\*\zeta_2
         +\frac{5312}{9}\*\zeta_3
         +\frac{800}{3}\*\zeta_4
         -128\*\zeta_2\*\zeta_3
         +\frac{21760}{9}\*\zeta_5
         -\frac{1216}{3}\*\zeta_3^2
\right.
\\
&& \nonumber
\left. \mbox{}
         -\frac{1184}{9}\*\zeta_6
         + ~b_{4,\, \dfFF}^{\rm q}~
          \right)
       + \cfs\*\nfs \* \left(
         -\frac{218239}{486}
         +\frac{2150}{9}\*\zeta_2
         +\frac{5684}{27}\*\zeta_3
         -\frac{380}{3}\*\zeta_4
\right.
\\
&& \nonumber
\left. \mbox{}
         -64\*\zeta_5
          \right)
       + \cf\*\ca\*\nfs \* \left(
         -\frac{3761509}{5832}
         +\frac{131878}{243}\*\zeta_2
         +\frac{6092}{81}\*\zeta_3
         -\frac{1540}{9}\*\zeta_4
         -\frac{400}{9}\*\zeta_2\*\zeta_3
\right.
\\
&& 
\left. \mbox{}
         +\frac{1192}{9}\*\zeta_5
          \right)
       + \cf\*\nft \* \left(
       \frac{50558}{2187}
      +\frac{80}{81}\*\zeta_3
      -\frac{1880}{81}\*\zeta_2
      +\frac{40}{9}\*\zeta_4
          \right)
\eea
which can be evaluated numerically using the results in Tab.~\ref{tab:Bq} and eq.~(\ref{eq:f4qunknowns}). 

\vspace*{1mm}
As discussed above, the four-loop terms in eq.~(\ref{eq:f4qunknowns})
are not very well constrained and therefore we follow the same approach as in the
derivation of eq.~(\ref{eq:c2dis-numerics}) to determine the present best estimate for
$B_4^{\rm DIS}$ in eq.~(\ref{eq:xB4}).
We use the large-$\nc$ limit of eq.~(\ref{eq:f4qLnc}) 
for terms proportional to $\nfo$ in $f_4^{\rm q}$ 
and keep full colour dependence in all others terms, i.e., all those
proportional to the QCD beta-function coefficients in eq.~(\ref{eq:xB4})
and in those proportional to $\nfz$, $\nfs$ and $\nft$ in $B_4^{\rm q}$ and $f_4^{\rm q}$.
This leads to 
\bea
\label{eq:B4dis-numerics}
  B_{4}^{\rm DIS}\biggr|_{{\rm appr}} &=&
  (10.68 \pm 0.01)\cdot 10^4
  +(-2.025 \pm 0.032)\cdot 10^4\, \*\nf 
  +798.0698\, \*\nfs
  \nn \\ & & 
  \qquad
  -12.08488\, \*\nft
  \, ,
\eea
and amounts to a perturbative expansion through four loops 
\bea
\label{eq:BDISqExpQ}
  B^{\rm DIS}(\nf\!=\!3) &=&
  - 0.31831\,\als \: ( 1 - 1.1004\, \als - 3.623\, \als^2 - (6.67 \pm 0.13)\, \als^3 + \, \ldots )
\; , \nn \\
  B^{\rm DIS}(\nf\!=\!4) &=&
  - 0.31831\,\als \: ( 1 - 1.2267\, \als - 3.405\, \als^2 - (4.77 \pm 0.17)\, \als^3 + \, \ldots )
\; , \nn \\
  B^{\rm DIS}(\nf\!=\!5) &=&
  - 0.31831\,\als \: ( 1 -  1.3530\, \als - 3.190\, \als^2 - (3.03 \pm 0.21)\, \als^3 + \, \ldots )
\, ,
 \nn \\
\quad
\eea
where the numerical uncertainties at four loops follow from 
varying the large-$\nc$ limit for the 
terms proportional to $\nfo$ in $f_4^{\rm q}$
of eq.~(\ref{eq:f4qLnc}) by $10\%$,
as done also in eq.~(\ref{eq:c2dis-numerics}).
With the knowledge of all resummation coefficients 
we are now ready to perform the numerical analysis up to N$^4$LL accuracy.

\section{Numerical Results}\label{sec:numerical}

\subsection{Resummation at N$^4$LL}\label{sec:resummation}

In the following we present numerical results for the resummed series up to N$^4$LL accuracy. 
We use the approximate value for $B_4^{\rm DIS}$ as provided in eq.~(\ref{eq:B4dis-numerics}) and study the convergence
of the resummed series and also the soft-virtual (SV) results at the fourth
order in strong coupling and beyond. We set the renormalization and factorization 
scales to be equal to the momentum transfer squared ($Q^2=-q^2$).
Since we are interested in the region of (very) large $x$ at moderate 
$Q^2$, the charm and bottom quarks cannot be treated as effectively massless.
Hence we will focus on the results for $\nf=3$ light quark flavours, which
implies $\floo=0$ for the new flavour structure 
due to a vanishing combination of the quark charges.
For $n_f\neq 3$ this flavour structure contributes to neutral-current DIS.
We will come back to the resulting numerical differences between 
neutral- and charged-current DIS at the end. 

\begin{figure}[ht]
\centerline{
\includegraphics[width=7.6cm, height=9.5cm]{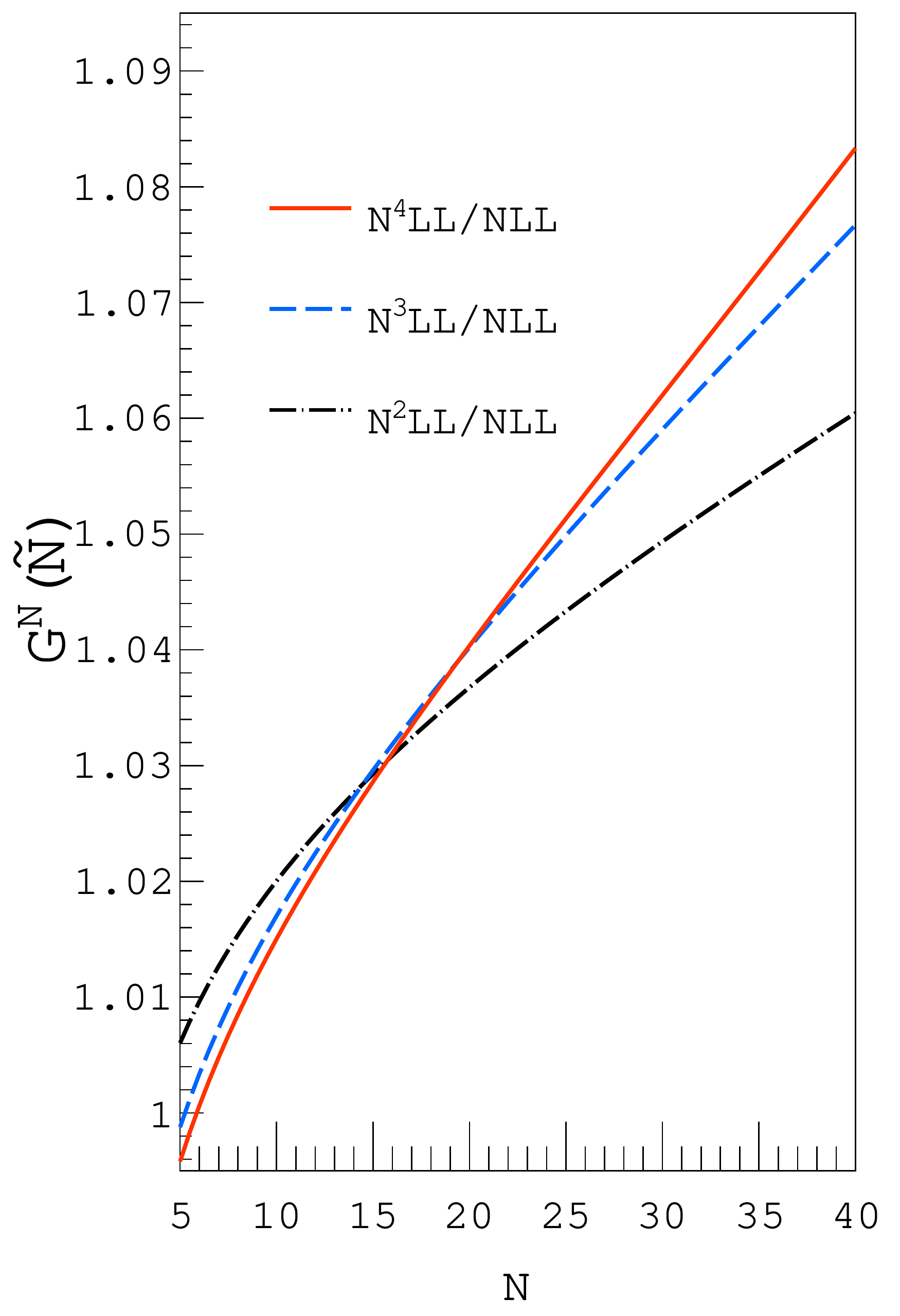}
\includegraphics[width=7.6cm, height=9.5cm]{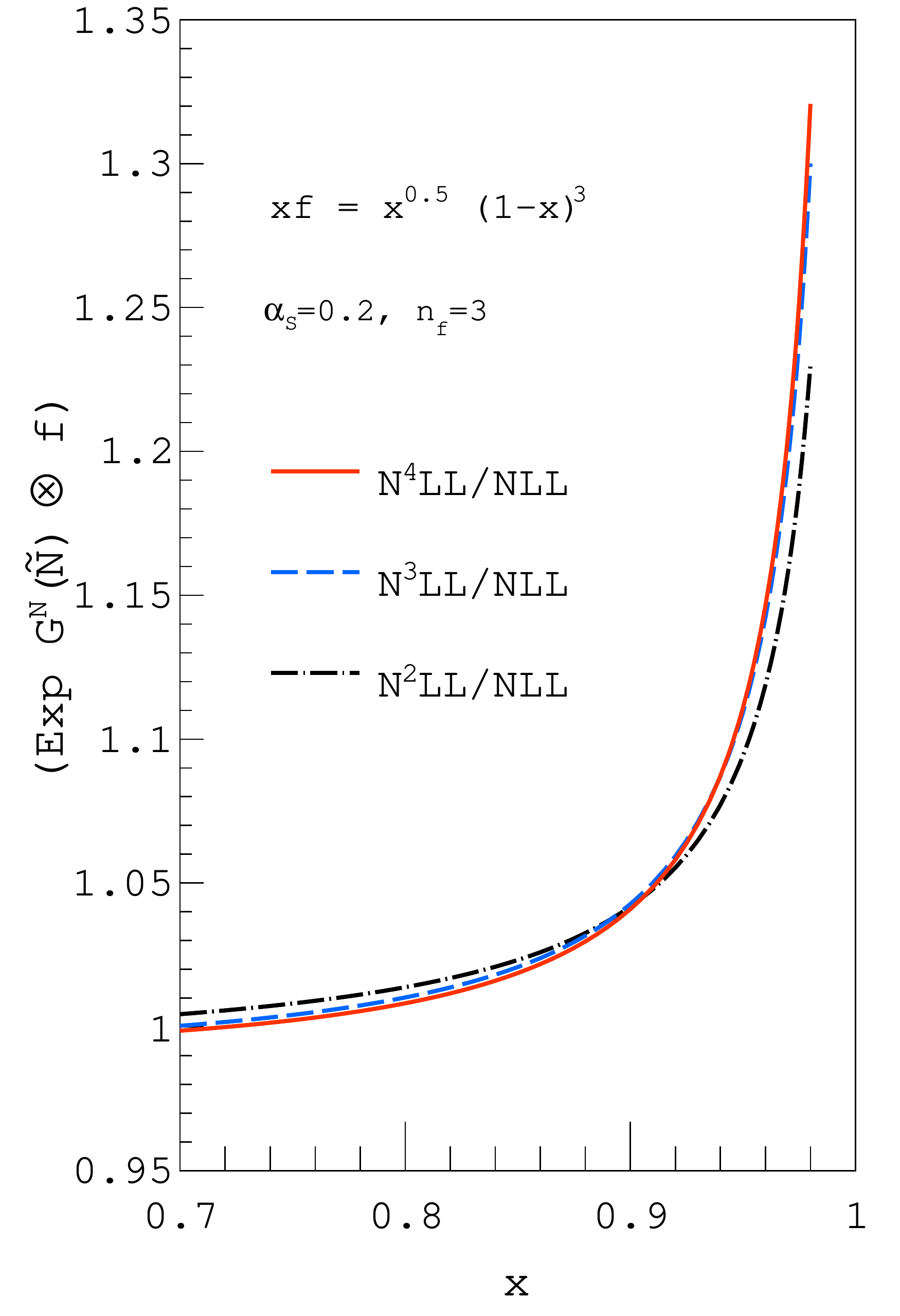}
}
\vspace{-2mm}
\caption{\small{
Left: The resummation exponent $G^N$, normalized to the NLL results, for DIS in 
eq.~(\ref{eq:GN}) plotted successively up to N$^4$LL for $\als = 0.2$ and $\nf = 3$.
Right: The resummed series is convoluted with a 
typical shape for a quark distribution up to N$^4$LL.}}
\label{fig1}
\end{figure}

\vspace{1mm}
In the left panel of Fig.~\ref{fig1}, we present the resummation exponent up to N$^4$LL,
normalized to the NLL results for a better visibility of the small higher-order effects.
Recall that the resummation accuracy is defined by truncating the resummation exponent 
in eq.~(\ref{eq:GN}) to a particular order; i.e., the first term $\ln \widetilde{N} g^{(1)}$ defines the LL
approximation, the first two terms $\ln \widetilde{N} g^{(1)}+g^{(2)}$ define the NLL accuracy
and so on. 
Using the order-independent value  $\als=0.2$ for the strong coupling,
we observe a good perturbative convergence up to N$^4$LL level. 
At the largest $N$-value shown, $N=40$, the exponent increases from LL to NLL by around
$66.7\%$, from NLL to NNLL by around $6.0\%$, from NNLL to N$^3$LL by around
$1.5\%$ and from N$^3$LL to N$^4$LL by around $0.6\%$. 
The resummed series thus shows a good stabilization at this order.  
The uncertainty in $B_4^{\rm DIS}$ quoted in eq.~(\ref{eq:B4dis-numerics})
as a result of changing the respective large-$\nc$ terms by $\pm 10\%$ 
does not affect the cross section much. 
It amounts at most to $0.1\%$, which is small 
compared to the size of the N$^4$LL correction.

\vspace{1mm}
In the right panel of Fig.~\ref{fig1} we plot $\exp(G^N) \otimes f$, 
the resummation exponential convoluted with a schematic but sufficiently typical shape for 
a quark distribution given by $x f = x^{0.5}(1-x)^3$. 
For the Mellin inversion we have used the minimal prescription~\cite{Catani:1996yz} 
with the standard contour choice as in {\sc{QCD-Pegasus}}~\cite{Vogt:2004ns}.
Also here we find a good perturbative convergence up to N$^4$LL. 
At $x=0.9$ the cross section changes from LL to NLL by $67.1\%$, 
from NLL to NNLL by $4.2\%$, 
from NNLL to N$^3$LL by $0.09\%$ and from N$^3$LL to N$^4$LL by $-0.17\%$.  
At $x=0.95$ the corresponding changes  are much larger,
yet the N$^4$LL effect stays well within $0.5\%$.  

\vspace{1mm}
For $\nf=4$ light quark flavours on the other hand, we see an even faster
perturbative convergence for the resummation exponent as well as for the
convolution of exponential with the same input shape. 
For example, at $N=40$ we observe a $61.4\%$ increase from LL to NLL, 
another increase of $1.8\%$ from NLL to NNLL, and changes of $-1.1\%$ 
from NNLL to N$^3$LL and of $-0.9\%$ from N$^3$LL to N$^4$LL accuracy. 
For the case of the convolution of the exponential with the above input 
shape, the corresponding values read $61\%, 0.3\%, -1.7\%$ and $-0.8\%$, 
respectively, at $x=0.9$.

\begin{figure}[ht]
\vspace*{3mm}
\centerline{
\includegraphics[width=7.5cm, height=9cm]{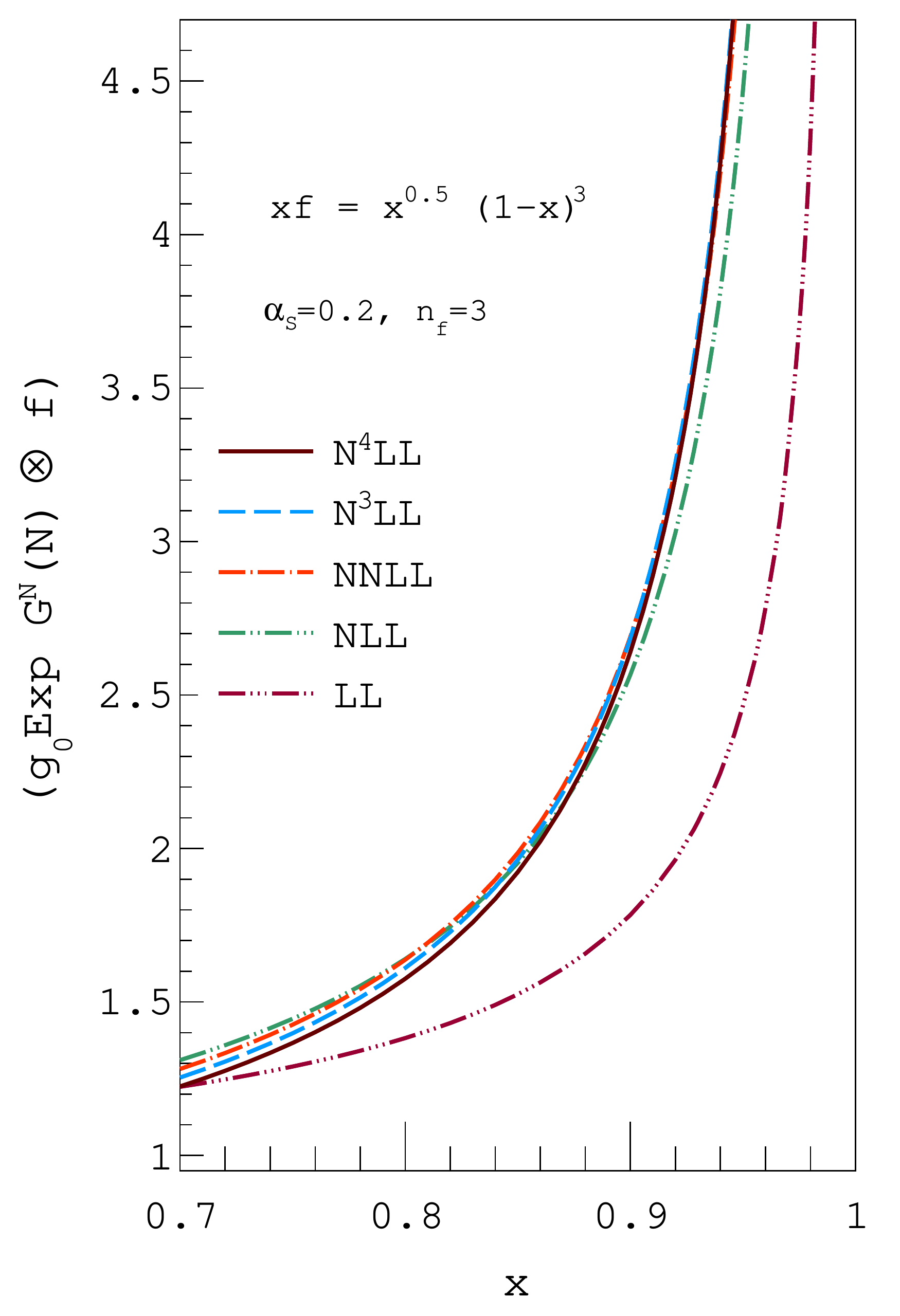}
\includegraphics[width=7.5cm, height=9cm]{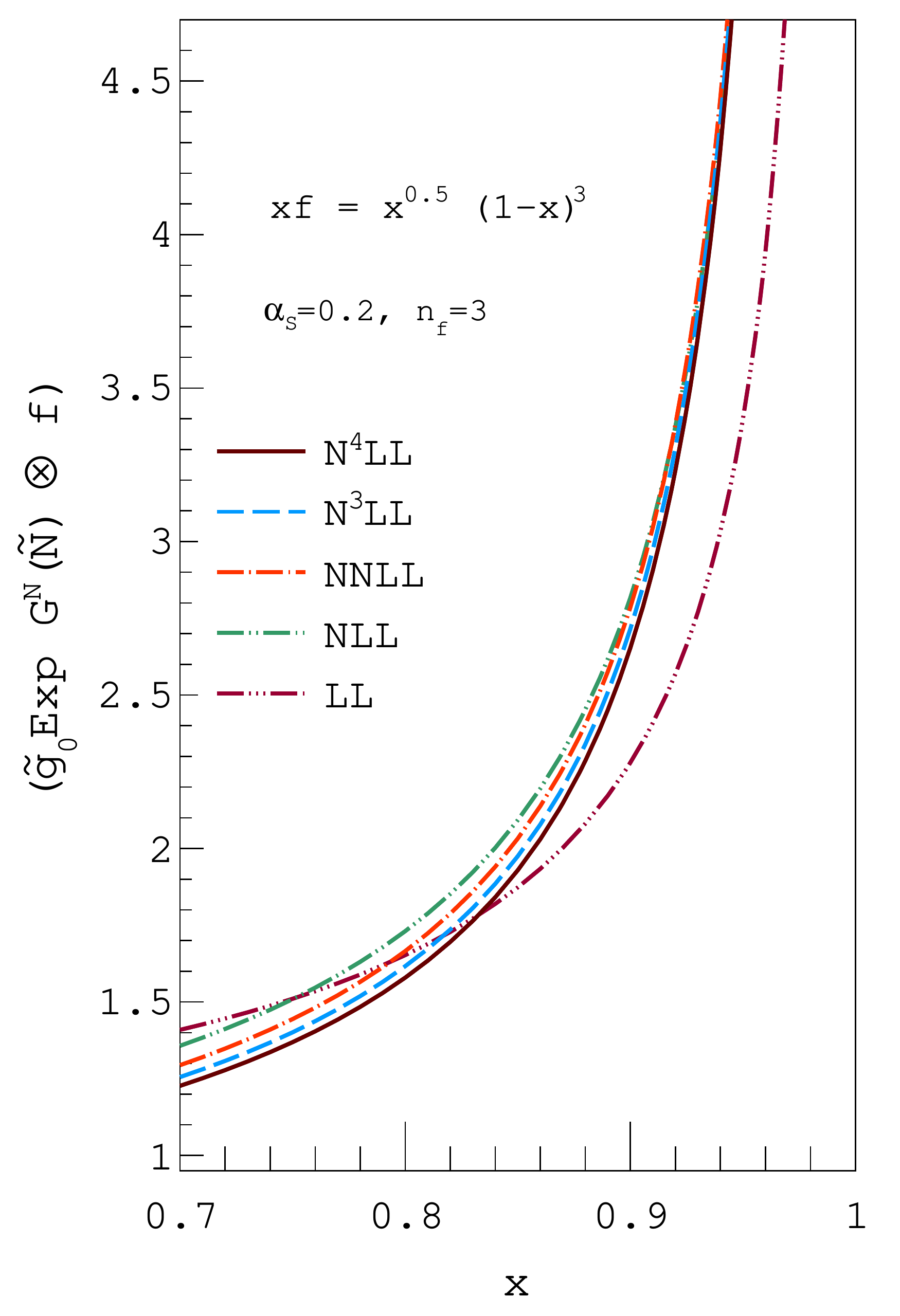}
}
\vspace{-2mm}
\caption{\small{
  Comparison between the $N$-exponentiation (left) and $\nt$-exponentiation (right) as discussed in the text. }}
\label{fig22}
\end{figure}

\vspace{1mm}
The threshold limit in the Mellin-$N$ space is based on the large-$N$ limit,
which we have discussed so far in terms of the variable $\nt = N \exp(\gamma_{E}^{})$, 
including the exponentiation of all $\ln \nt$ terms.
One can, in principle, define the large Mellin variable as $N$ only 
and collect only those terms in the resummed exponent which are enhanced as $N\to \infty$. 
As far as the threshold limit is concerned, both, 
$\nt\to \infty$ and $N\to \infty$ are physically equivalent. 
However, numerically the two options differ since in the former case all
$\gamma_E^{}$ terms associated with $N$ are also exponentiated to all orders. 
Due to this feature the results for both cases are different starting already at LL accuracy. 
The resummed exponent shows a faster convergence for the 
$N$ exponentiation. 
However, note that now also the function $g_0$ in eq.~(\ref{eq:cNres}) differs
between the two cases. Whereas in the $\nt$ exponentiation, the $\gamma_E^{}$
terms are collected by definition in $\nt$ as the large-$N$ variable, 
for the $N$-exponentiation they are appear in the finite function $g_0$ and in
the resummed exponent $G$.
As mentioned already, we collect the explicit results for the perturbative
expansion of $g_0$ in the Appendix~\ref{app:appB}. 

A useful comparison for these two approaches is therefore 
to check the convergence at the cross-section level itself. 
We provide this comparison in Fig.~\ref{fig22}. 
For the $N$-exponentiation satisfactory convergence occurs in the threshold region only beyond NLL accuracy whereas
the $\nt$-exponentiation shows a systematic behaviour for perturbative
convergence for the successive orders of the resummation and good convergence
in the threshold region is achieved already at NLL accuracy. 
At sufficiently high logarithmic order both approaches converge. 
However they start to differ away from the threshold. 
At $x=0.7$, the LL result in $\nt$-exponentiation differs by as large as
  $15\%$ compared to the $N$-exponentiation, whereas at the N$^4$LL level this
  difference decreases to $0.1\%$ showing that at higher logarithmic accuracy
  they tend to converge. We see similar observation in the higher $x$ region
  as well.
In the rest of the article, we apply the standard $\nt$ exponentiation throughout.

\begin{figure}[ht]
\vspace*{3mm}
\centerline{
\includegraphics[width=7.5cm, height=9cm]{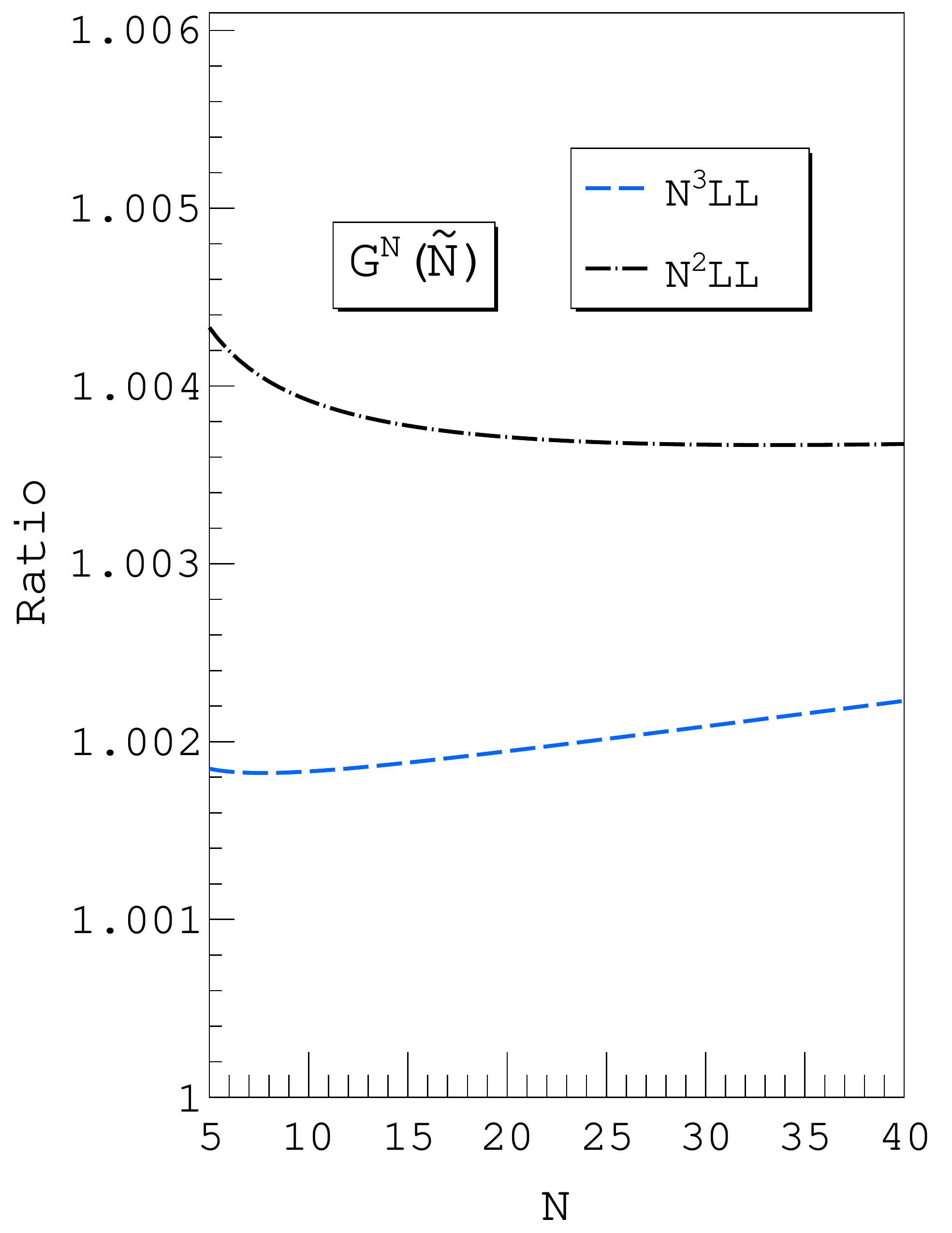}
\includegraphics[width=7.5cm, height=9cm]{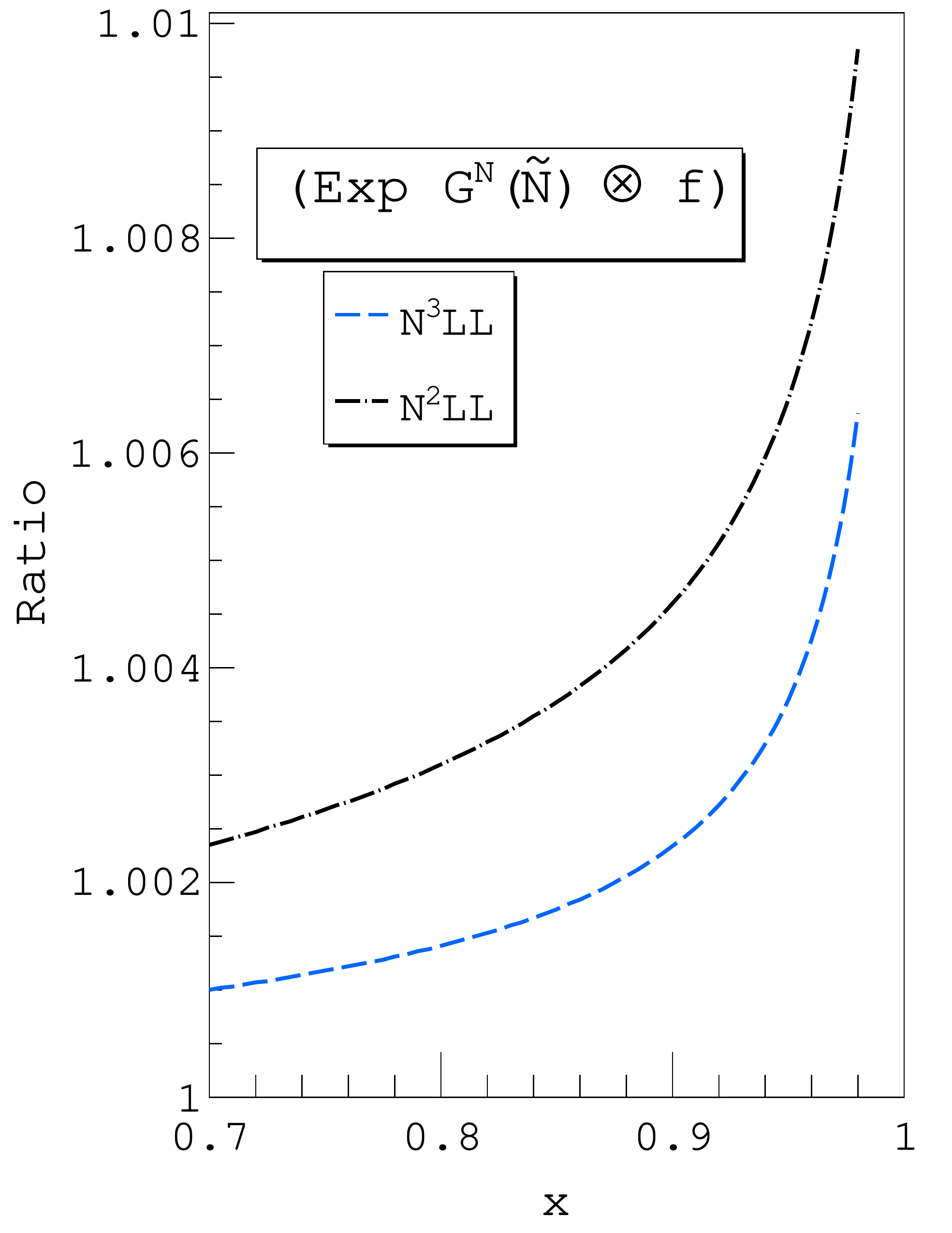}
}
\vspace{-2mm}
\caption{\small{
  The ratio of large-$\nc$ approximation and the exact results at NNLL and N$^3$LL for $\alpha_S=0.2$ and $\nf=3$ light flavours. 
  Left: The ratio for DIS resummed exponent $G^N$ as a function of Mellin-$N$. 
  Right: The ratio for the resummed series convoluted with the
  typical input shape used in Fig.~\ref{fig1} plotted against $x$. }}
\label{fig2}
\end{figure}

\vspace{1mm}
It is instructive to study the quality of the specific large-$\nc$
approximation used to derive the approximate value for $B_4^{\rm DIS}$ in
eq.~(\ref{eq:B4dis-numerics}) at each lower order. 
To that end, we have compared the large-$\nc$ behaviour of results 
at lower loops with the exact full-colour expressions available up to three loops. 
Note that in order to have a meaningful estimate of the resummed coefficient function, 
we take the large-$\nc$ approximation only for the 
process-dependent pieces appearing at that order, whereas all the lower order 
terms are kept exact. 
For example, at the third order, we take the large-$\nc$ approximation only for the
combination $-(f_3^{\rm q} + B_3^{\rm q})$ appearing in eq.~(\ref{eq:xB3})
whereas all the other terms proportional to coefficients of the QCD beta
function are kept exact. 
The LL result is always exact since it only depends on the universal
cusp anomalous dimension $A_1$ while the first process-dependent coefficients 
enter as $-(f_1^{\rm q} + B_1^{\rm q})$ at NLL order.
Note that up to the third order the large-$n_c$ limit
for the terms proportional to $\nfo$ and $\nfz$ in $f_i^{\rm q}$ and $B_i^{\rm q}$ 
coincides with the exact large-$n_c$ expressions of these coefficients 
when restoring the overall factor $\cf$. 
However this is not true anymore at the N$^4$LO level.

\vspace{1mm}
In Fig.~\ref{fig2}, we compare the large-$\nc$
and the exact results for the resummation exponent (left) and the resummed series 
convoluted with the same quark distribution $xf$ used above (right).
In particular we have taken the ratio between the large-$\nc$ and exact results at lower orders.
We have taken $\nf=3$ light quark flavours also for this study. 
Up to NLL the large-$\nc$ result is exact.
The difference in the cross section starts from NNLL and is about 
 $0.4\%$ in the large-$N$ region ($N=40$), whereas at N$^3$LL it is about 
$0.2\%$; the large-$\nc$ approximation always 
 overestimates the exact result in the region of interest. 
 These differences are to be compared with the absolute size of the exact corrections. 
These are much larger with about 
 $6.0\%$ at $N=40$, when increasing the logarithmic accuracy from NLL to NNLL, and 
 $1.5\%$ for  NNLL to N$^3$LL. 
Similar observations also hold at lower $N$ values.

\vspace{1mm}
Using the approximate value for $B_4^{\rm DIS}$ in eq.~(\ref{eq:B4dis-numerics}) 
with the specific large-$\nc$ limit, the step from N$^3$LL to N$^4$LL accuracy
amounts to a correction of about 
$0.6\%$ at $N=40$. 
Based on the lower order studies, we  thus expect that the exact result 
at fourth order for the complete N$^4$LL calculation should not differ 
by more that 
 $\pm 0.2\%$
from our prediction for the resummed exponent at large-$\nc$. 
In summary, the above procedure of taking the specific large-$\nc$ approximation 
provides robust estimates for the cross sections of interest.

\subsection{Soft-virtual cross section at the fourth order}

\begin{figure}[t]
\vspace*{-5mm}
\centerline{
\includegraphics[width=8cm, height=10cm]{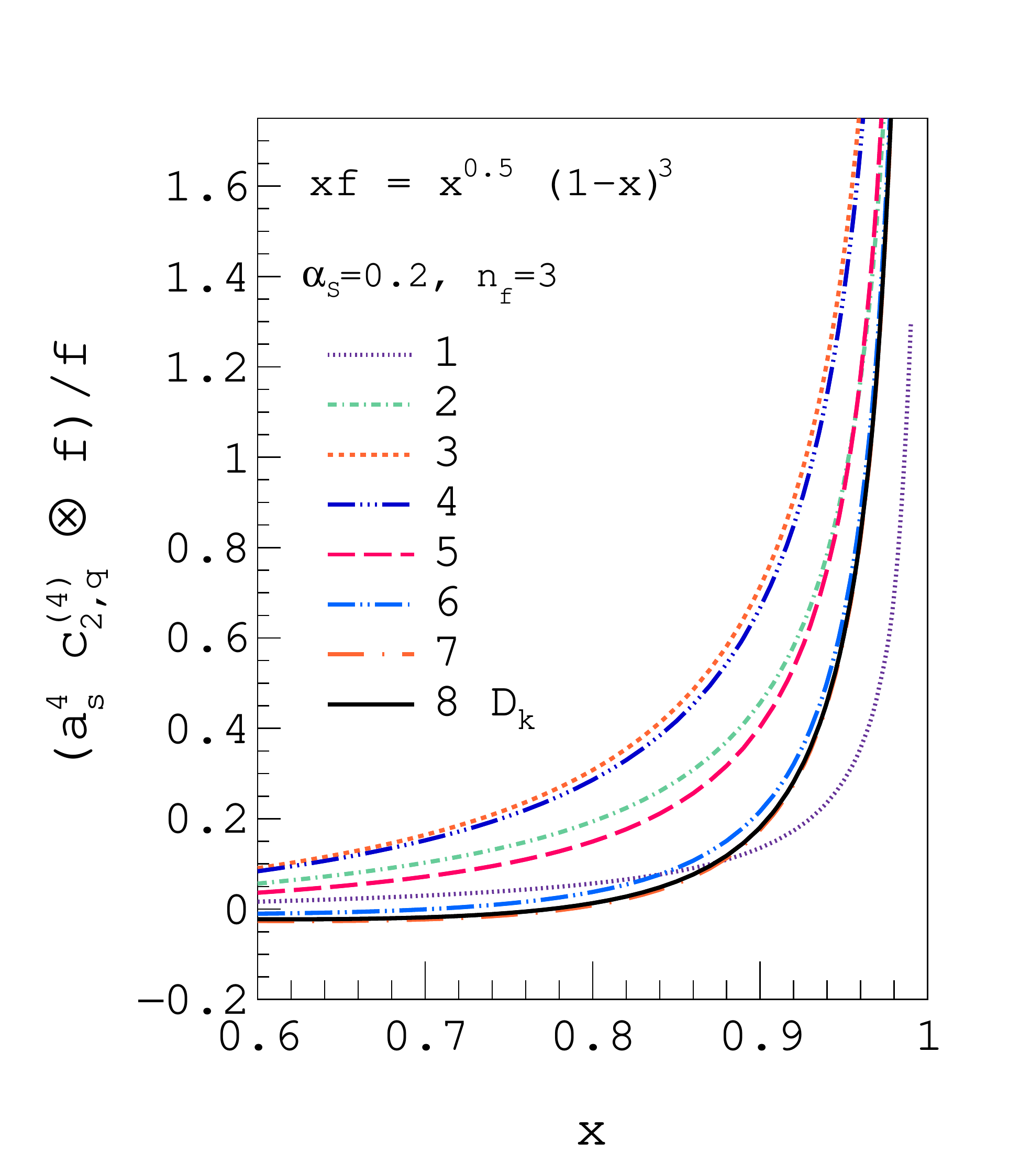}\hspace*{-5mm}
\includegraphics[width=8cm, height=10cm]{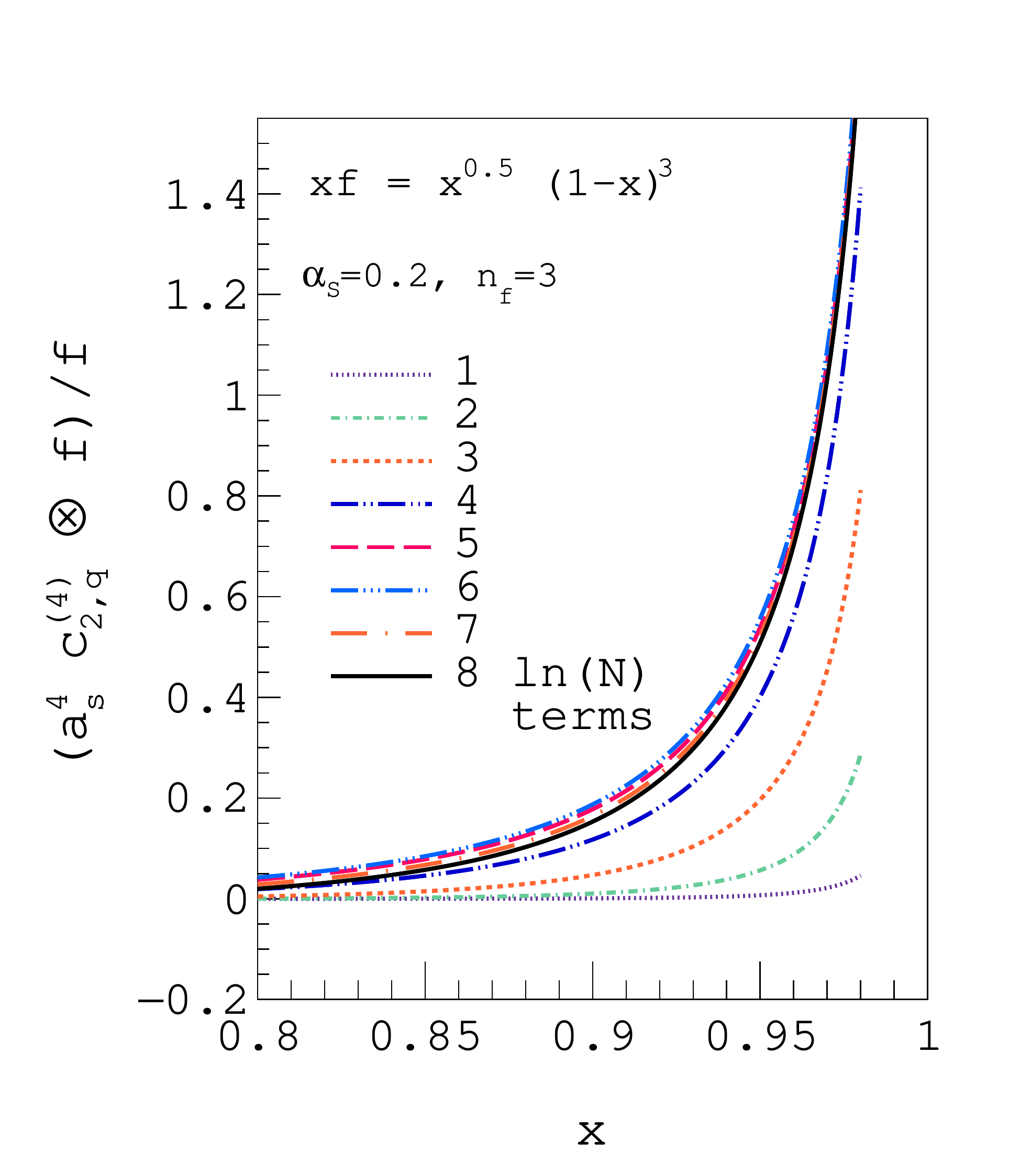}
}
\vspace{-3mm}
\caption{\small{
  Left: The DIS Wilson coefficient $c^{(4)}_{2, {\rm q}}$
  in eq.~(\ref{eq:c2dis4full}) convoluted with the input shape $xf$ of Fig.~\ref{fig1}
  with the successive addition of the plus-distributions ${\cal D}_k$
  starting from the highest term.
  Right: The same with the successive addition of the $N$-space logarithms.}}
\label{fig3}
\end{figure}

The convergence of the threshold expansion in Bjorken-$x$ or Mellin-$N$ space
can be illustrated by means of the successive addition of the plus-distributions
${\cal D}_k$,  $k=2n-1, \dots, 0$ in the SV cross section in $x$-space 
compared to the successive addition of the logarithms $\ln^i N$, $i=2n, \dots, 1$ in $N$-space\footnote{Note that we use powers of $\ln N$, and not $\ln \widetilde{N}$, from now on.}.
The former procedure shows poor convergence, a fact that has been already
observed in~\cite{Catani:1996yz} for two ($n=2$) and in~\cite{Moch:2005ba} for three ($n=3$) loops.
In Fig.~\ref{fig3} we compare the convergence 
of the DIS Wilson coefficient at four loops $c^{(4)}_{2, {\rm q}}$
constructed in this manner either in $x$- or in $N$-space and 
convoluted with the above input shape $xf$. 
As expected we see a very similar behaviour compared to the known lower orders. 
The $x$-space result converges poorly and only stabilizes after the addition
of the six highest plus-distributions ${\cal D}_{k}$ at this order. 
On the other hand in $N$-space, the same convergence is achieved much faster 
after adding only the four highest logarithms $\ln^i N$ in the series. 
This confirms that $N$-space is better suited for studies of the
threshold limit also at the fourth order. 

\vspace{1mm}
The exact expression of ${\cal D}_0$ at fourth order still
contains the poorly constrained coefficients  
$f_{4,\, \dfFA}^{\rm q}$,  $f_{4,\, {\nf\*\cft}}^{\rm q}$ and   
$f_{4,\, {\nf\*\cfs\*\ca}}^{\rm q}$ in eq.~(\ref{eq:f4qunknowns})
which drop out in the large-$\nc$ limit.
We have improved the ${\cal D}_0$ coefficient in eq.~(\ref{eq:c2dis-numerics}) through 
the best estimate for $B_4^{\rm DIS}$ in eq.~(\ref{eq:B4dis-numerics}).
Using instead the central value for the above $f_4$ coefficients (see eq.~(\ref{eq:f4qunknowns})) 
the ${\cal D}_0$ coefficient changes by around $6.5\%$ compared to the value in eq.~(\ref{eq:c2dis-numerics}).
This is a rather small change considering the errors in these coefficients in eq.~(\ref{eq:f4qunknowns}).  
On the other hand setting the $f_4$ coefficients in eq.~(\ref{eq:f4qunknowns}) to zero, which is 
within the uncertainty range quoted, yields a value for the ${\cal D}_0$ coefficient
which differs only by $0.5\%$ from its best prediction in eq.~(\ref{eq:c2dis-numerics}). 
These are small effects given that the best estimate for ${\cal D}_0$ 
in eq.~(\ref{eq:c2dis-numerics}) changes the whole N$^4$LO SV cross-section 
(up to the contribution proportional to $\delta(1-x)$ in eq.~(\ref{eq:delta4loop})) 
by only $0.9\%$ even at $x=0.95$. 
Altogether, this demonstrates a small dependence of the total cross section 
on those, as of yet, still poorly known coefficients. 
At the same time, it corroborates the assumptions made in the derivation of
the best estimate for ${\cal D}_0$ in eq.~(\ref{eq:c2dis-numerics}).

\subsection{Tower expansion vs exponentiation}
\label{sec:tower}

The resummed cross section in Mellin $N$-space can also be reorganized in a
different manner~\cite{Vogt:1999xa,Moch:2005ba} 
by re-expanding the exponential in eq.~(\ref{eq:cNres}) as follows: 
\bea
\label{eq:cNres-tower}
C_{a,q}^{\,N}(Q^2) = 1 + \sum_{k=1}^{\infty} a_s^k \sum_{l=1}^{2k} c_{kl} \ln^{2k-l+1}N 
\, ,
\eea
where the coefficients $c_{kl}$ for DIS are given in Tab.~\ref{table:tower}
for all known terms. 
By successive addition of terms in this expansion, 
one can also predict the resummed cross section up to a certain accuracy. 
The truncation of this series at any particular order will, of course, recover
the fixed-order SV result in $N$-space. 
The coefficients $c_{kl}$ of the logarithms in the above series vanish
factorially at sufficiently higher orders in $\als$ for a fixed logarithmic
structure (i.e., fixed by $l$).  
The coefficients $c_{kl}$ are determined up to $l=1,2,3$ by the complete one loop calculation 
with NLL resummation. 
The complete two-loop calculation along with the NNLL resummation
fixes the next two towers ($l=4,5$), whereas the complete three-loop calculation along
with the N$^3$LL resummation determines again the next two towers, i.e., up to $l=7$. 
The N$^4$LL resummation derived here allows to completely fix the eighth tower, which corresponds to $l=8$. 
Note that the quartic colour factors, i.e., $d^{\,(4)}_{F\!A}$ and $d^{\,(4)}_{F\!F}$
in eq.~(\ref{d4SUn}), appear in this tower for the first time.
To obtain all terms up to the ninth tower ($l=9$), the complete
four-loop calculation is needed, which is currently not available. 
However from the knowledge of higher moments for DIS, we can estimate
approximate values  for the unknown $\delta(1-x)$ coefficient (see Appendix
\ref{app:coeff}) at the four loops which in turn can be used to extract the
coefficient $g_{04}^{\rm DIS}$. 
For the standard $\nt$ exponentiation, then one can have at the fourth order,
\begin{eqnarray}
\label{eq:g04DIS}
g_{04}^{\rm DIS} &=&(-22.51\pm 3.)\cdot 10^4\, ,  \nn\\
              & &(-13.75\pm 3.)\cdot 10^4\, , \nn\\
              & &(-\phantom{0}6.88\pm 3.)\cdot 10^4\, \qquad \text{ for $n_f =3,4,5$, respectively.}
\end{eqnarray}
With the lower orders coefficient the series for $g_0$ looks as follows
\begin{eqnarray}
\label{eq:g04}
 g_0 (n_f = 3) &= 1 - 16.3865 ~a_s  - 147.955 ~a_s^2 - 4528.85 ~a_s^3 - (22.51 \pm 3)\cdot 10^4 ~a_s^4 \nn\\
 g_0 (n_f = 4) &= 1 - 16.3865 ~a_s  - 109.603 ~a_s^2 - 2918.08 ~a_s^3 - (13.75 \pm 3)\cdot 10^4 ~a_s^4 \nn\\
 g_0 (n_f = 5) &= 1 - 16.3865 ~a_s  - 71.2505 ~a_s^2 - 1477.81 ~a_s^3 - (6.88 \pm 3)\cdot 10^4 ~a_s^4  \,.
\end{eqnarray}
The error in the $g_{04}^{\rm DIS}$ coefficient introduces an uncertainty of around $6\%$ in the $\ln^2 N$ 
term of the $a_s^5$ coefficient.  The uncertainty in column $c_{k8}$  is due to only the uncertainty coming from the estimate for $B_4^{\rm DIS}$.  On the other hand the uncertainty due to both $B_4^{\rm DIS}$ and $g_{04}^{\rm DIS}$ are added in quadrature to get the final uncertainty in column $c_{k9}$. Although we note that this uncertainty is mostly dominated by the uncertainty in $g_{04}^{\rm DIS}$ estimate.
\floatsetup[table]{font=tiny}
\begin{table}
\begin{tabular}{|c|r|r|r|r|r|r|r|r|r|}
\hline 
  $ k $ & $ c_{k1}$ & $c_{k2}$ & $c_{k3}$ & $c_{k4}$ & $c_{k5}$ & $c_{k6}$ & $c_{k7}$ & $c_{k8}$\scalebox{0.8}{$(\times 10^{-3})$} & $c_{k9}$ \scalebox{0.8}{$(\times 10^{-4})$}\\ \hline \hline
1   &  2.66667  &  7.0785   &  -  &  -     &  -     &  -       &  -         &  -         &  -         \\  \hline
2   &  3.55556  &  26.8760  &  46.082   &  -52.31   &  -   &  -       &  -       &  -     &  -        \\  \hline
3   &  3.16049  &  46.5013  &  262.409  &  606.02   &  -379.3   &  -1607   &  -     &  -    & -        \\  \hline
4   &  2.10700  &  50.8159  &  526.224  &  2901.29  &  7563.9   &  3839    &  -30240    &  -50.45     $\pm$  1.1  &  -    \\  \hline
5   &  1.12373  &  40.1983  &  643.688  &  5931.23  &  32776.2  &  102186  &  111002    &  -223.83    $\pm$  2.8  &  -135.79    $\pm$  8.5   \\  \hline
6   &  0.49944  &  24.8103  &  562.907  &  7611.78  &  66550.1  &  380864  &  1341323   &  2326.76    $\pm$  3.8  &  -136.09    $\pm$  13.4   \\  \hline
7   &  0.19026  &  12.5251  &  381.022  &  7043.11  &  87178.2  &  749259  &  4455641   &  17525.89   $\pm$  3.4  &  3930.14    $\pm$  15.2   \\  \hline
8   &  0.06342  &  5.3423   &  209.584  &  5057.26  &  83343.1  &  983401  &  8443647   &  52347.85   $\pm$  2.2  &  22576.03   $\pm$  13.2   \\  \hline
9   &  0.01879  &  1.9710   &  96.844   &  2951.41  &  62206.6  &  956176  &  10993901  &  95211.30   $\pm$  1.2  &  61509.98   $\pm$  9.0   \\  \hline
10  &  0.00501  &  0.6404   &  38.511   &  1445.50  &  37852.3  &  731699  &  10762856  &  122209.70  $\pm$  0.5  &  107329.05  $\pm$  5.0  \\  \hline
11  &  0.00121  &  0.1858   &  13.424   &  608.30   &  19356.2  &  458590  &  8364106   &  119639.50  $\pm$  0.2  &  135303.85  $\pm$  2.4   \\  \hline
12  &  0.00027  &  0.0487   &  4.161    &  223.97   &  8508.0   &  242196  &  5352115   &  93780.42   $\pm$  0.1  &  131835.19  $\pm$  1.0   \\  \hline
13  &  0.00006  &  0.0116   &  1.161    &  73.19    &  3271.3   &  110118  &  2895802   &  60869.59   $\pm$  0\phantom{.0}  &  103720.30  $\pm$  0.3   \\  \hline
14  &  0.00001  &  0.0026   &  0.294    &  21.48    &  1115.7   &  43830   &  1351835   &  33532.90   $\pm$  0\phantom{.0}  &  67950.87   $\pm$  0.1  \\  \hline
15  &  0.00000  &  0.0005   &  0.068    &  5.72     &  341.4    &  15479   &  553238    &  15980.28   $\pm$  0\phantom{.0}  &  37933.17   $\pm$  0\phantom{1}      \\  \hline
\end{tabular}
\caption{\small{
  The coefficients of the tower expansion 
  for $C^{\,N}(Q^2)$ in eq.~(\ref{eq:cNres-tower}) for $\nf=3$ light quark 
  flavours, where they are the same for $F_1$, $F_2$ and $F_3$ since $\floo = 0$.
  The first seven columns are exact up to the numerical truncation.
  The eighth  column uses the estimates for $B_4^{\rm DIS}$ and corresponding errors and the ninth
  column in addition requires the estimate for $g_{04}^{\rm DIS}$ in eq.~(\ref{eq:g04DIS}).
}}
\label{table:tower}
\end{table}

\begin{figure}[ht]
\centerline{
\includegraphics[width=7.5cm, height=9cm]{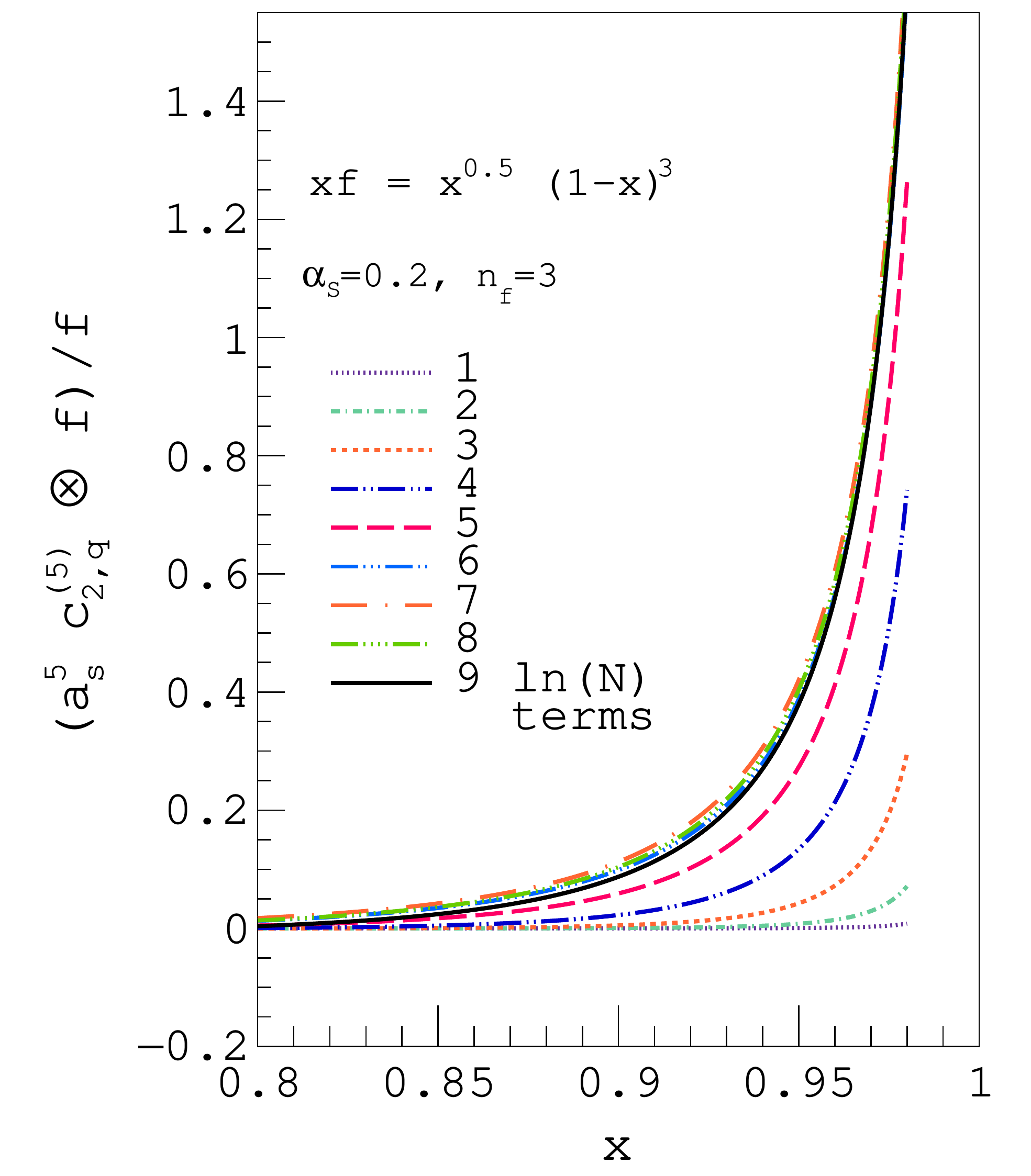}
\includegraphics[width=7.5cm, height=9cm]{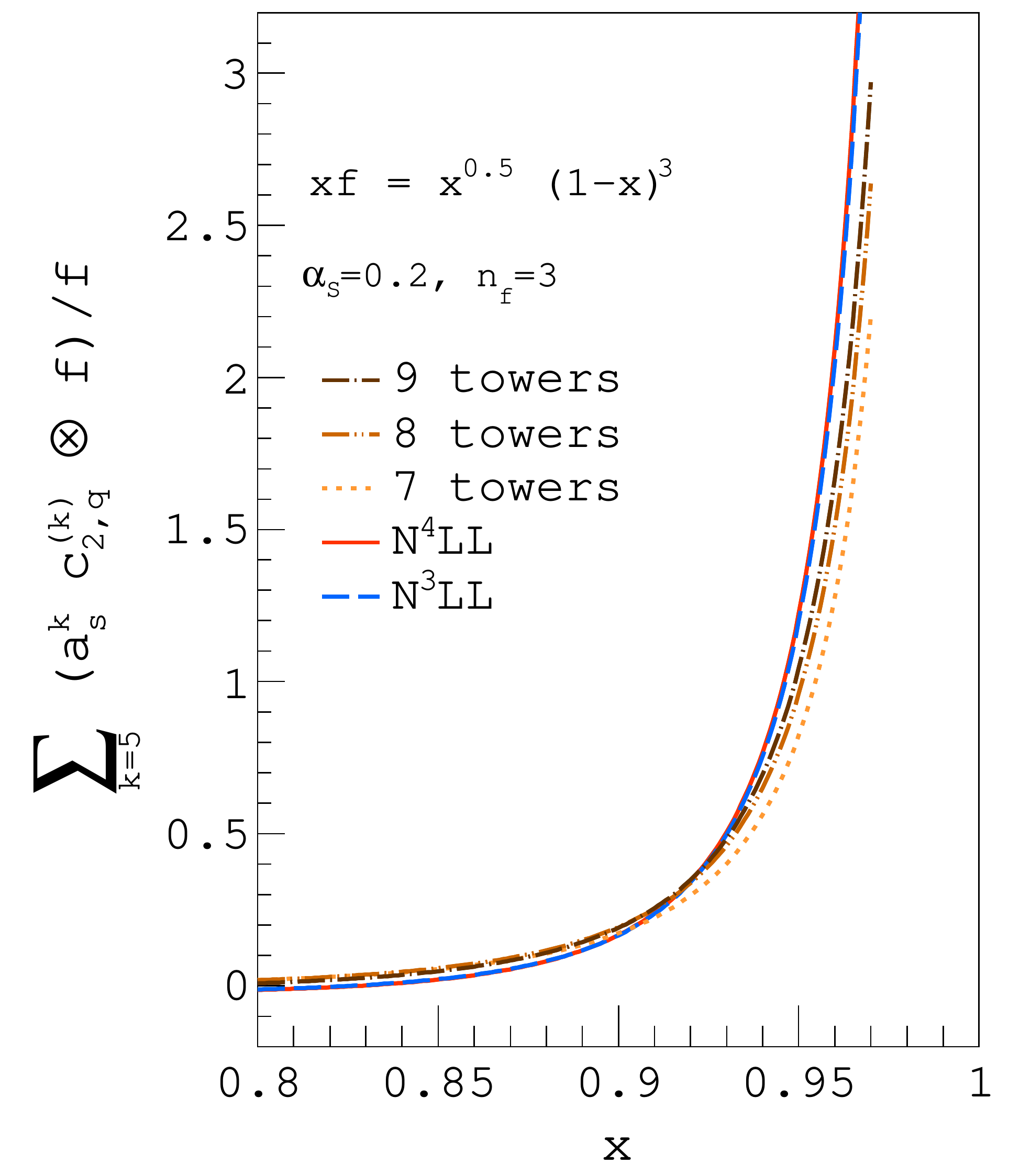}
}
\caption{\small{
  Left: The successive approximations of the five-loop coefficient 
  function $c^{(5)}_{2, {\rm q}}$ by the large-$N$ terms specified in 
  Tab.~\ref{table:tower}, illustrated by the convolution with the input shape 
  $xf$ of Fig.~\ref{fig1}. 
  Right: The corresponding results for the effect of higher terms 
  beyond $\als^4$ as obtained from the
  tower expansion up to nine towers and from the exponentiation up to N$^4$LL
  accuracy on the Wilson coefficient.
  }}
\label{fig4}
\end{figure}
In Fig.~\ref{fig4} on the left we illustrate the expansion in decreasing powers of $\ln\, N$
for the five-loop coefficient function $c^{(5)}_{2, {\rm q}}$ in the same manner 
as for the four-loop coefficient function $c^{(4)}_{2, {\rm q}}$ 
in the right part of Fig.~\ref{fig3} above.
We confirm the general pattern, that the first $l$ logarithms 
provide a good estimate up to the $l$-th order in $\als$, 
i.e., $l=5$ for the case at hand.
In the right panel of Fig.~\ref{fig4}
we compare the resummed result (exponentiated) for the Wilson
coefficient with the tower expansion given in Tab.~\ref{table:tower}. 
Although they both converge in the region of $x\simeq 0.5$, 
they start to diverge at large $x$ 
where exponentiation shows a better stability compared to the tower expansion. 
In fact the expansion with a fixed number of towers severely underestimates 
the effect of the coefficient functions of much higher orders, 
which become more important at the region of very large-$x$ around $x\simeq 0.9$.

So far we have considered $\floo=0$ for the flavour structure, 
which holds for $\nf=3$, due to the vanishing combination of the quark charges. 
Starting from three loops, the neutral-current DIS coefficient functions 
differ due to $\floo \neq 0$, 
which captures the effect of incoming and outgoing photons coupling to quark lines 
of different flavour, see e.g.,~\cite{Vermaseren:2005qc}.
The effect of this contribution is numerically insignificant as already pointed out in \cite{Moch:2005ba}. 
To estimate its numerical impact on the SV cross section as well as in the
tower expansion, we use $\nf=4$ and keep the exact value $\floo=1/10$. 
Note that a non-zero $\floo$ will change the tower expansion 
starting from seventh tower, i.e., the coefficients $c_{k7}$, and from the lowest
order for this towers ($k=3$).
In the seventh tower, the contribution proportional to $\floo$ changes the finite piece 
of third order SV result, whereas at the fourth order in addition it also changes the eight tower.  
Keeping exact value for $\floo$ we get a difference of $0.1\%$ in the
coefficient of the constant term at third order and a change of $0.09\%$ in the
last logarithmic term at fourth order.

The additional new term in the charged-current DIS coefficient function $C_{3,q}$ 
proportional to the group invariant $(d^{\,abc} d_{\,abc})/\nc$ in eq.~(\ref{eq:dabc}) 
which corresponds to incoming and outgoing $W^\pm$-bosons coupling to a closed
internal quark loop~\cite{Moch:2008fj}. 
This term is power suppressed for $x~\to~1$,
hence does not contribute to the threshold approximation considered here. 

\section{Summary}\label{sec:summary}

We have applied recent progress in the calculation of the cusp anomalous dimension, 
the quark (non-singlet) splitting functions and the quark form factor 
to obtain predictions for the DIS cross sections with charged- and neutral-current interactions 
at four loops in perturbation theory. 
We have performed the threshold resummation in the large-$\nc$ limit up to N$^4$LL 
collecting the terms $\als^{\,3} (\als \ln N)^k$ to all orders in $\als$ 
in the threshold resummation exponent $G^N$ in Mellin-$N$ space.
To that end, the relevant quantity $B^{\rm DIS}$ controlling the resummation 
of the quark jet function has been extracted at four-loop order.

We have also derived the full colour dependence of the DIS Wilson coefficients at four loops 
down to the term proportional to the plus-distribution $[1/(1-x)]_+$ using 
a conjecture about the colour structure of the anomalous dimension 
of the eikonal form factor, namely that it exhibits 
the same generalized maximal non-Abelian property regarding quadratic and
quartic Casimir coefficients as the cusp anomalous dimension.
As a by-product, also the single pole of QCD form factor in the parameter of dimensional
regularization $\varepsilon$ has been determined numerically to good accuracy.

Using the analytical expressions for the resummed DIS Wilson coefficient at the
N$^4$LL level, we have shown that the lower order corrections
are sizable and that the threshold logarithms at N$^4$LL resummed order
amount to corrections well within $1\%$ and improve the perturbative convergence.
We have compared the effect of successive plus-distributions in $x$-space
as well as the logarithms in $N$-space for the cross section up to the fourth
order and we confirm known observations that the convergence of the logarithmic series
in $N$-space is much smoother compared to the one in $x$-space.
We have also reorganized the resummed series as a tower of logarithms up to the 
ninth tower. 
While the tower expansion is useful to study the effect of higher logarithms, 
it lacks contributions at very large $x$ compared to the exponentiated result. 
The latter exhibits an even better stability even in the very large-$x$ region. 
Near threshold the difference between DIS Wilson coefficients for 
charged- and neutral-current exchange, which manifests itself in the flavour
structure $\floo$ at three loops and beyond, has only a minor numerical effect
(for $\nf=4$) on the respective cross sections.

The current accuracy can be significantly improved with the knowledge of the
complete QCD form factor at four loops or an exact computation of the corresponding
eikonal anomalous dimension at this order. 
In particular analytic results for all $\nf$ dependent terms in those 
quantities would greatly reduce the present residual numerical uncertainties.
At the accuracy reached with N$^4$LL resummation also power corrections
proportional to $1/N$ in Mellin-$N$ space 
or logarithmic enhancement $\ln^{k}(1-x)$ in $x$-space become numerically relevant, 
see~\cite{Moch:2009hr,Bonocore:2015esa}. 
Finally, the effects from the massive quarks are 
important in the threshold region (see for example \cite{Kawamura:2012cr,Hoang:2015iva}) 
and a proper treatment of heavy quark mass effects becomes essential for
phenomenological applications.

\bigskip
--------------------------------------------------------------------

\subsection*{Acknowledgments}
%
G.D. thanks V. Ravindran for useful discussions.
The algebraic computations have been done with the latest version of the
symbolic manipulation system {\sc Form}~\cite{Vermaseren:2000nd,Ruijl:2017dtg}.
This work has been supported by the {\it Deutsche Forschungsgemeinschaft} (DFG) 
under grant number MO~1801/2-1, and by the COST Action CA16201
PARTICLEFACE supported by {\it European Cooperation in Science and Technology} (COST).
The research of G.D. is supported by the DFG within the Collaborative Research Center TRR 257 (``Particle Physics Phenomenology after the Higgs Discovery'').

\subsection*{Note added in proof}

During review of the present paper, the complete analytic four-loop anomalous dimensions in massless QCD
from form factors have been obtained in \cite{vonManteuffel:2020vjv}.
This provides information on the missing four-loop contributions 
$f_{4,\, {\nf\*\cft}}^{\rm q}$ and $f_{4,\, {\nf\*\cfs\*\ca}}^{\rm q}$ in eq.~(\ref{eq:f4qunknowns}),
which we can determine as 
\bea
\label{eq:f4qknowns}
\nonumber 
f_{4,\, {\nf\*\cft}}^{\rm q} &=& 
            \frac{27949}{108}
          + 322\*\zeta_2
          - \frac{1120}{9}\*\zeta_3
          - 334\*\zeta_4
          - \frac{512}{3}\*\zeta_2\*\zeta_3
          + \frac{3872}{3}\*\zeta_5
          + 368\*\zeta_3^2
\nonumber\\
& & 
          - \frac{14668}{9}\*\zeta_6
          - 2\*b_{4,\, {\nf\*\cft}}^{\rm q}
\nonumber\\[1ex]
&=& 
  - 9.626 \pm 0.010
\, ,
\nonumber\\[1ex]
f_{4,\, {\nf\*\cfs\*\ca}}^{\rm q} &=& 
          - \frac{1092511}{972}
          + \frac{673}{27}\*\zeta_2
          - \frac{23518}{81}\*\zeta_3
          + \frac{52744}{27}\*\zeta_4
          + \frac{3904}{9}\*\zeta_2\*\zeta_3
          - \frac{4472}{3}\*\zeta_5
\nonumber\\
& & 
          - \frac{3400}{3}\*\zeta_3^2
          + 718\*\zeta6
          - 2\*b_{4,\, {\nf\*\cfs\*\ca}}^{\rm q}
\nonumber\\[1ex]
&=& 
  - 2.351 \pm 0.010
\, .
\eea
Their numerical values are in agreement with those quoted in
eq.~(\ref{eq:f4qunknowns}). 
The remaining uncertainties in eq.~(\ref{eq:f4qknowns}) 
are due to the numerical values in Tab.~\ref{tab:Bq}.
We have verfied that the large-$\nc$ results for $\nf = 3$, $4$ or $5$ at
four loops in eq.~(\ref{eq:fqExpQ}) reproduce full QCD within ${\cal O}(1\%)$.

This new information in eq.~(\ref{eq:f4qknowns}) leads to the following numerical result 
for the coefficient proportional to $\DD_{0}$ 
in the DIS Wilson coefficients $c^{(4)}_{a, {\rm q}}$ in eq.~(\ref{eq:c2dis-numerics}),
\bea
\nonumber
\label{eq:new-c2dis-numerics}
  c^{(4)}_{2, {\rm q}}\biggr|_{\DD_{0}} &=&
  (3.874 \pm 0.010)\cdot 10^4
  + (-3.4965 \pm 0.0287)\cdot 10^4\, \*\nf 
  + 2062.715\, \*\nfs
\\[-2mm]
&& \mbox{}
  - 12.08488\, \*\nft
  + 47.55183\, \*\nf \* \floo 
\:\: ,
\eea
where all exact values have been rounded to seven digits. 
The remaining errors come either from the uncertainty for $f_{4,\, \dfFA}^{\rm  q}$ in eq.~(\ref{eq:f4qunknowns}) 
(term proportional to $\nfz$) or 
from the numerical values in Tab.~\ref{tab:Bq} 
(term proportional to $\nfo$).
The new numerical value in eq.~(\ref{eq:new-c2dis-numerics}) agrees very well 
with the best estimate in eq.~(\ref{eq:c2dis-numerics}). 
This also corroborates the arguments based on the large-$\nc$ limit.

With the full colour dependence in eq.~(\ref{eq:new-c2dis-numerics}) at hand, the resummation coefficient 
$B_4^{\rm DIS}$ in eq.~(\ref{eq:B4}) then becomes 
\bea
\label{eq:new-B4dis-numerics}
  B_{4}^{\rm DIS} &=&
  (10.68 \pm 0.01)\cdot 10^4
  +(-2.0275549 \pm 0.0000029)\cdot 10^4\, \*\nf 
  +798.0698\, \*\nfs
  \nn \\ & & 
  \qquad
  -12.08488\, \*\nft
  \, ,
\eea
where the uncertainty in the term proportional to $\nfo$ stems 
from the numerical values in Tab.~\ref{tab:Bq}.
Again, this agrees very well with the previous best estimate in eq.~(\ref{eq:B4dis-numerics}) 
based on the large-$\nc$ limit.
The perturbative expansion of $B^{\rm DIS}$ through four loops (replacing eq.~(\ref{eq:BDISqExpQ}))
reads then
\bea
\label{eq:new-BDISqExpQ}
  B^{\rm DIS}(\nf\!=\!3) &=&
  - 0.31831\,\als \: ( 1 - 1.1004\, \als - 3.623\, \als^2 - (6.659 \pm 0.013)\, \als^3 + \, \ldots )
\; , \nn \\
  B^{\rm DIS}(\nf\!=\!4) &=&
  - 0.31831\,\als \: ( 1 - 1.2267\, \als - 3.405\, \als^2 - (4.753 \pm 0.013)\, \als^3 + \, \ldots )
\; , \nn \\
  B^{\rm DIS}(\nf\!=\!5) &=&
  - 0.31831\,\als \: ( 1 -  1.3530\, \als - 3.190\, \als^2 - (3.010 \pm 0.013)\, \als^3 + \, \ldots )
\, ,
 \nn \\
\quad
\eea
which eliminates any residual uncertainty in the resummation at N$^4$LL accuracy.

\appendix
\renewcommand{\theequation}{\ref{app:coeff}.\arabic{equation}}
\setcounter{equation}{0}
\section{DIS Wilson coefficients}
\label{app:coeff}
For convenience of the reader we have collected all available information 
on the soft-virtual expressions for the DIS coefficient functions $c^{(n)}_{2, {\rm q}}$ 
through four loops $(n\leq 4)$.
The leading order is $c^{(0)}_{2, {\rm q}} = \delta(1-x)$ and  
the one-loop result $c^{(1)}_{2, {\rm q}}$ in eq.~(\ref{appB1}) 
has been obtained in~\cite{Bardeen:1978yd}. 
The two- and three-loop expressions $c^{(2)}_{2, {\rm q}}$ and $c^{(3)}_{2, {\rm q}}$ 
in eqs.~(\ref{appB2}),  (\ref{appB3}) are due to~\cite{vanNeerven:1991nn,Moch:1999eb} and~\cite{Vermaseren:2005qc}, respectively.
At four loops all plus-distributions $\DD_{\,k}$ with $k \geq 2$ in $c^{(4)}_{2, {\rm q}}$ 
have been given in~\cite{Moch:2005ba}. 
The terms $\DD_{\,1}$ with the complete colour dependence of $A_4$ inserted and 
$\DD_{\,0}$ in $c^{(4)}_{2, {\rm q}}$ are new results.
Analytical results for the coefficients $f_{4,\, {\nf\*\cft}}^{\rm q}$,  $f_{4,\, {\nf\*\cfs\*\ca}}^{\rm q}$ and $f_{4,\, \dfFA}^{\rm q}$
can be expected from the computation of the form factor at four loops soon.
The remaining coefficients $b_{4,\, {\cff}}^{\rm q}$, $b_{4,\, {\cft\*\ca}}^{\rm q}$, $b_{4,\, {\cfs\*\cas}}^{\rm q}$,
$b_{4,\, {\nf\*\cft}}^{\rm q}$, $b_{4,\, {\nf\*\cfs\*\ca}}^{\rm q}$, $b_{4,\, \dfFA}^{\rm q}$, 
and $b_{4,\, \dfFF}^{\rm q}$ of the virtual anomalous dimension $B_4^{\rm q}$ 
require an analytical result for the four-loop splitting function, which is not easily available.
The factors $\floo$ in eqs.~(\ref{appB3}), (\ref{appB4}) do not appear in charged-current DIS.
\bea
  c^{(1)}_{2, {\rm q}} &=&
  4\*\cf \* \DD_{1}
  -3\*\cf \* \DD_{0}
  -\cf\*(9+4 \* \zeta_2)\*\delta(1-x)
  \label{appB1}
  \, ,
\\
  \nonumber
  c^{(2)}_{2, {\rm q}} &=&
    8\*\cfs\*\DD_{3}
  -\DD_{2}\*\biggl\{
     18\*\cfs
    +\frac{22}{3}\*\cf\*\ca
    -\frac{4}{3}\*\cf\*\nf
    \biggr\}
  +\DD_{1}\*\biggl\{
    -\cfs\*(27 + 32 \* \zeta_2)
\\
&& \nonumber
    +\cf\*\ca\*\biggl(\frac{367}{9} - 8 \* \zeta_2\biggr)
    -\frac{58}{9}\*\cf\*\nf
    \biggr\}
  +\DD_{0}\*\biggl\{
     \cfs\*\biggl(\frac{51}{2} + 36 \* \zeta_2 - 8 \* \zeta_3 \biggr)
\\
&& \nonumber
    -\cf\*\ca\*\biggl(\frac{3155}{54} - \frac{44}{3} \* \zeta_2 - 40 \* \zeta_3 \biggr)
    +\cf\*\nf\*\biggl(\frac{247}{27}-\frac{8}{3} \* \zeta_2 \biggr)
    \biggr\}
  +\delta(1-x)\*\biggl\{
   \cfs\*\biggl(\frac{331}{8}
\\
&& \nonumber
   + 69 \* \zeta_2-78 \* \zeta_3 +6 \* \zss\biggr)
  -\cf\*\ca\*\biggl(\frac{5465}{72}+ \frac{251}{3} \* \zeta_2  - \frac{140}{3} \* \zeta_3 - \frac{71}{5} \* \zss\biggr)
\\
&&
  +\cf\*\nf\*\biggl(\frac{457}{36} + \frac{38}{3} \* \zeta_2 + \frac{4}{3} \* \zeta_3 \biggr)
  \biggr\}
  \, ,
    \label{appB2}
  \eea
  \bea
  \nonumber
  c^{(3)}_{2, {\rm q}} &=&
  8\*\cft\*\DD_{5}
  +\DD_{4}\*\biggl\{
    -30\*\cft - \frac{220}{9}\*\cfs\*\ca + \frac{40}{9}\*\cfs\*\nf\
    \biggr\}
  +\DD_{3}\*\biggl\{
    -\cft\*(36+96 \* \zeta_2)
\\
&& \nonumber
    +\cfs\*\ca\*\biggl(\frac{1732}{9}-32 \* \zeta_2\biggr)
    +\frac{484}{27}\*\cf\*\cas
    -\frac{280}{9}\*\cfs\*\nf
    -\frac{176}{27}\*\cf\*\ca\*\nf
    +\frac{16}{27}\*\cf\*\nfs
    \biggr\}
\\
&& \nonumber
  +\DD_{2}\*\biggl\{
     \cft\*\biggl(\frac{279}{2}+288 \* \zeta_2+16 \* \zeta_3\biggr)
    -\cfs\*\ca\*\biggl(\frac{8425}{18}-\frac{724}{3} \* \zeta_2-240 \* \zeta_3\biggr)
\\
&& \nonumber
    -\cf\*\cas\*\biggl(\frac{4649}{27}-\frac{88}{3} \* \zeta_2\biggr)
    +\cfs\*\nf\*\biggl(\frac{683}{9}-\frac{112}{3} \* \zeta_2\biggr)
    +\cf\*\ca\*\nf\*\biggl(\frac{1552}{27}-\frac{16}{3} \* \zeta_2\biggr)
\\
&& \nonumber
    -\frac{116}{27}\*\cf\*\nfs
  \biggr\}
  +\DD_{1}\*\biggl\{
     \cft\*\biggl(\frac{187}{2}+240 \* \zeta_2-360 \* \zeta_3+\frac{376}{5}\*\zss\biggr)
    -\cfs\*\ca\*\biggl(\frac{5563}{18}
\\
&& \nonumber
    +972 \* \zeta_2
    +\frac{160}{3} \* \zeta_3-\frac{764}{5} \* \zss\biggr)
    +\cf\*\cas\*\biggl(\frac{50689}{81}-\frac{680}{3} \* \zeta_2-264 \* \zeta_3+\frac{176}{5} \* \zss\biggr)
\\
&& \nonumber
    +\cfs\*\nf\*\biggl(\frac{83}{9}+168 \* \zeta_2+\frac{112}{3} \* \zeta_3\biggr)
    -\cf\*\ca\*\nf\*\biggl(\frac{15062}{81}-\frac{512}{9} \* \zeta_2-16 \* \zeta_3\biggr)
\\
&& \nonumber
    +\cf\*\nfs\*\biggl(\frac{940}{81}-\frac{32}{9} \* \zeta_2\biggr)
    \biggr\}
  +\DD_{0}\*\biggl\{
    -\cft\*\biggl(\frac{1001}{8}+429 \* \zeta_2-274 \* \zeta_3+210 \* \zss-32\* \zeta_2 \* \zeta_3
\\
&& \nonumber
    -432 \* \zeta_5\biggr)
    +\cfs\*\ca\*\biggl(\frac{16981}{24}+\frac{26885}{27} \* \zeta_2-\frac{3304}{9} \* \zeta_3-209 \* \zss-400 \* \zeta_2 \* \zeta_3-120 \* \zeta_5\biggr)
\\
&& \nonumber
    -\cf\*\cas\*\biggl(\frac{599375}{729}-\frac{32126}{81} \* \zeta_2-\frac{21032}{27} \* \zeta_3+\frac{652}{15} \* \zss+\frac{176}{3} \* \zeta_2 \* \zeta_3+232 \* \zeta_5\biggr)
\\
&& \nonumber
    -\cfs\*\nf\*\biggl(\frac{2003}{108}+\frac{4226}{27} \* \zeta_2+60 \* \zeta_3-16 \* \zss\biggr)
    +\cf\*\ca\*\nf\*\biggl(\frac{160906}{729}-\frac{776}{9} \* \zeta_3-\frac{9920}{81} \* \zeta_2
\\
&& \nonumber
    +\frac{208}{15} \* \zss\biggr)
    -\cf\*\nfs\*\biggl(\frac{8714}{729}-\frac{232}{27} \* \zeta_2+\frac{32}{27} \* \zeta_3\biggr)
    \biggr\}
  +\delta(1-x)\*\biggl\{
    -\cft\*\biggl(\frac{7255}{24}+\frac{3379}{6} \* \zeta_2
\\
&& \nonumber
    +318 \* \zeta_3+\frac{2148}{5} \* \zss-808 \* \zeta_2 \* \zeta_3-1240 \* \zeta_5+\frac{304}{3} \* \zts
    -\frac{4184}{315} \* \zst\biggr)
    +\cfs\*\ca\*\biggl(\frac{9161}{12}
\\
&& \nonumber
    +\frac{104117}{54} \* \zeta_2-\frac{6419}{3} \* \zeta_3+\frac{87632}{135}\*\zss-\frac{6644}{9} \* \zeta_2 \* \zeta_3-\frac{4952}{9} \* \zeta_5
    +\frac{1016}{3} \* \zts
    -\frac{33556}{315} \* \zst\biggr)
\\
&& \nonumber
    -\cf\*\cas\*\biggl(\frac{1909753}{1944}+\frac{143255}{81} \* \zeta_2-\frac{105712}{81} \* \zeta_3-\frac{25184}{135} \* \zss
    -540 \* \zeta_2 \* \zeta_3
    +\frac{416}{3} \* \zeta_5
\\
&& \nonumber
    +\frac{248}{3} \* \zts+\frac{3512}{63} \* \zst\biggr)
    -\cfs\*\nf\*\biggl(\frac{341}{36}+\frac{5491}{27} \*\zeta_2-\frac{1348}{3} \* \zeta_3+\frac{16472}{135} \* \zss
    +\frac{352}{9} \* \zeta_2 \* \zeta_3
\\
&& \nonumber
    +\frac{592}{9} \* \zeta_5\biggr)
    +\cf\*\ca\*\nf\*\biggl(\frac{142883}{486}+\frac{40862}{81} \* \zeta_2-\frac{18314}{81} \* \zeta_3-\frac{2488}{135} \* \zss-\frac{56}{3} \* \zeta_2 \* \zeta_3
    +\frac{8}{3} \* \zeta_5\biggr)
\\
&& \nonumber
    -\cf\*\nfs\*\biggl(\frac{9517}{486}+\frac{860}{27} \* \zeta_2+\frac{152}{81} \* \zeta_3+\frac{32}{27} \*\zss\biggr)
+\floo\*\nf\*\dabcnc\*\biggl(
    64
    +160\*\zeta_2
    +\frac{224}{3}\*\zeta_3
\\
&& 
    -\frac{32}{5}\*\zss
    -\frac{1280}{3}\*\zeta_5
    \biggr)
    \biggr\}
  \, ,
      \label{appB3}
  \eea
  \bea
  \nonumber
  c^{(4)}_{2, {\rm q}} &=&
  \frac{16}{3}\*\cff\*\DD_{7}
  +\DD_{6}\*\biggl\{
    -28\*\cff
    -\frac{308}{9}\*\cft\*\ca
    +\frac{56}{9}\*\cft\*\nf
    \biggr\}
  +\DD_{5}\*\biggl\{
    -\cff\*(18+128 \* \zeta_2)
\\
&& \nonumber
    +\cft\*\ca\*\biggl(\frac{998}{3}-48 \* \zeta_2\biggr)
    +\frac{1936}{27}\*\cfs\*\cas
    -\frac{164}{3}\*\cft\*\nf
    -\frac{704}{27}\*\cfs\*\ca\*\nf
    +\frac{64}{27}\*\cfs\*\nfs
    \biggr\}
\\
&& \nonumber
  +\DD_{4}\*\biggl\{
     \cff\*\biggl(210+600 \* \zeta_2+\frac{400}{3} \* \zeta_3\biggr)
    -\cft\*\ca\*\biggl(\frac{27835}{27}-\frac{6800}{9} \* \zeta_2 -400 \* \zeta_3\biggr)
\\
&& \nonumber
    -\cfs\*\cas\*\biggl(\frac{24040}{27}-\frac{440}{3} \* \zeta_2\biggr)
    -\frac{1331}{27}\*\cf\*\cat
    +\cft\*\nf\*\biggl(\frac{4630}{27}-\frac{1040}{9} \* \zeta_2\biggr)
\\
&& \nonumber
    +\cfs\*\ca\*\nf\*\biggl(\frac{8120}{27}-\frac{80}{3} \* \zeta_2\biggr)
    +\frac{242}{9}\*\cf\*\cas\*\nf
    -\frac{640}{27}\*\cfs\*\nfs
    -\frac{44}{9}\*\cf\*\ca\*\nfs
    +\frac{8}{27}\*\cf\*\nft
  \biggr\}
\\
&& \nonumber
  +\DD_{3}\*\biggl\{
     \cff\*\biggl(\frac{113}{2}+264 \* \zeta_2-1072 \* \zeta_3+\frac{1392}{5} \* \zss\biggr)
    -\cft\*\ca\*\biggl(\frac{1534}{3}+\frac{41824}{9} \* \zeta_2
\\
&& \nonumber
    +\frac{8800}{9} \* \zeta_3
    -\frac{3128}{5} \* \zss\biggr)
    +\cfs\*\cas\*\biggl(\frac{2154563}{486}-\frac{52912}{27} \* \zeta_2-\frac{13024}{9} \* \zeta_3+\frac{864}{5} \* \zss\biggr)
\\
&& \nonumber
    +\cf\*\cat\*\biggl(\frac{55627}{81}-\frac{968}{9} \* \zeta_2\biggr)
    -\cfs\*\ca\*\nf\*\biggl(\frac{339134}{243}-\frac{14096}{27} \* \zeta_2-\frac{1216}{9} \* \zeta_3\biggr)
\\
&& \nonumber
    -\cft\*\nf\*\biggl(\frac{280}{3}-\frac{7216}{9} \* \zeta_2-\frac{1888}{9} \* \zeta_3\biggr)
    -\cf\*\cas\*\nf\*\biggl(\frac{9502}{27}-\frac{352}{9} \* \zeta_2\biggr)
\\
&& \nonumber
    +\cfs\*\nfs\*\biggl(\frac{24238}{243}-\frac{928}{27} \* \zeta_2\biggr)
    +\cf\*\ca\*\nfs\*\biggl(\frac{1540}{27}-\frac{32}{9} \* \zeta_2\biggr)
    -\frac{232}{81}\*\cf\*\nft
    \biggr\}
\\
&& \nonumber
  +\DD_{2}\*\biggl\{
    -\cff\*\biggl(\frac{1299}{2}+2808 \* \zeta_2-1392 \* \zeta_3+1836 \* \zss +640 \* \zeta_2 \* \zeta_3-4128 \* \zeta_5\biggr)
\\
&& \nonumber
    +\cft\*\ca\*\biggl(\frac{13990}{3}+\frac{30704}{3} \* \zeta_2+\frac{2716}{3} \* \zeta_3
    -\frac{12906}{5} \* \zss-3648 \*\zeta_2 \* \zeta_3-720 \* \zeta_5\biggr)
\\
&& \nonumber
    -\cfs\*\cas\*\biggl(\frac{2254339}{243}-\frac{86804}{9} \* \zeta_2-\frac{24544}{3} \* \zeta_3+\frac{4034}{3} \* \zss
    +832 \* \zeta_2 \* \zeta_3+1392 \* \zeta_5\biggr)
\\
&& \nonumber
    -\cf\*\cat\*\biggl(\frac{649589}{162}-\frac{4012}{3} \* \zeta_2-1452 \* \zeta_3+\frac{968}{5} \* \zss\biggr)
    -\cft\*\nf\*\biggl(\frac{145}{9}+\frac{5132}{3} \* \zeta_2+936 \* \zeta_3
\\
&& \nonumber
    -\frac{1032}{5} \* \zss\biggr)
    +\cfs\*\ca\*\nf\*\biggl(\frac{713162}{243}-\frac{10600}{9} \* \zeta_3-\frac{82004}{27} \* \zeta_2+\frac{3772}{15} \* \zss\biggr)
\\
&& \nonumber
    +\cf\*\cas\*\nf\*\biggl(\frac{17189}{9}-\frac{5096}{9} \* \zeta_2-352 \* \zeta_3+\frac{176}{5} \* \zss\biggr)
    -\cfs\*\nfs\*\biggl(\frac{52678}{243}-\frac{6104}{27} \* \zeta_2
\\
&& \nonumber
    -\frac{304}{9} \* \zeta_3\biggr)
    -\cf\*\ca\*\nfs\*\biggl(\frac{7403}{27}-\frac{688}{9} \* \zeta_2-16 \* \zeta_3\biggr)
    +\cf\*\nft\*\biggl(\frac{940}{81}-\frac{32}{9} \* \zeta_2\biggr)
    \biggr\}
\\
&& 
\nonumber
  +\DD_{1}\*\biggl\{
    -\cff\*\biggl(\frac{19885}{24}+\frac{3101}{3} \* \zeta_2+1374 \* \zeta_3 +1098 \* \zss -5824 \* \zeta_2 \* \zeta_3
    -64 \* \zeta_5+\frac{256}{3} \* \zts 
\\
&& \nonumber
    + \frac{64192}{315} \* \zst \biggr)
    +\cft\*\ca\*\biggl(\frac{2829}{2}+\frac{202705}{27} \* \zeta_2-\frac{333526}{27} \* \zeta_3+\frac{1043747}{135} \* \zss
    +\frac{23104}{9} \* \zeta_2 \* \zeta_3
\\
&& \nonumber
    -\frac{77216}{9} \* \zeta_5+\frac{5984}{3} \* \zts-\frac{55472}{63} \* \zst\biggr)
    -\cfs\*\cas\*\biggl(\frac{21987073}{5832}+\frac{713440}{27} \* \zeta_2+\frac{238400}{81} \* \zeta_3
\\
&& \nonumber
    -\frac{812819}{135} \* \zss -\frac{24896}{3} \* \zeta_2 \* \zeta_3-\frac{6472}{3} \* \zeta_5-\frac{3808}{3}\*\zts
    +\frac{51848}{63} \* \zst\biggr)
    +\cf\*\cat\*\biggl(\frac{16865531}{1458}
\\
&& \nonumber
    -\frac{156238}{27} \* \zeta_2
    -\frac{78296}{9} \* \zeta_3+\frac{17996}{15} \* \zss+528 \* \zeta_2 \*\zeta_3
    +\frac{19360}{9} \* \zeta_5-16 \* \zts
    -\frac{20032}{105} \* \zst 
    \biggr)
\\
&& \nonumber
    + \dfFAnc\*\biggl(
         -128\*\zeta_2
         +\frac{128}{3}\*\zeta_3
         +\frac{3520}{3}\*\zeta_5
         -384\*\zts
         -\frac{7936}{35}\*\zst
    \biggr)
\\
&& \nonumber
+\cft\*\nf\*\biggl(
    \frac{1775}{18}
    -\frac{7912}{27}\*\zeta_2
    +\frac{78328}{27}\*\zeta_3
    -\frac{195716}{135}\*\zss
    +\frac{6368}{9}\*\zeta_5
    -\frac{6592}{9}\*\zeta_2\*\zeta_3
    \biggr)
\\
&& \nonumber
    -\cf\*\cas\*\nf\*\biggl(
          \frac{2452247}{486}
         -\frac{68836}{27}\*\zeta_2
         -\frac{5752}{3}\*\zeta_3
         +\frac{3944}{15}\*\zss
         -32\*\zeta_2\*\zeta_3
         +\frac{2080}{9}\* \zeta_5
    \biggr)
\\
&& \nonumber
    -\cfs\*\ca\*\nf\*\biggl(
          \frac{326570}{729}
         -\frac{694654}{81}\*\zeta_2
         -\frac{158740}{81}\*\zeta_3
         +\frac{178642}{135}\*\zss
         +\frac{2848}{3}\*\zeta_2\*\zeta_3
         +\frac{208}{3}\*\zeta_5
    \biggr)
\\
&& \nonumber
    + \nf\*\dfFFnc\* \biggl( 256\*\zeta_2 - \frac{256}{3}\*\zeta_3 
    - \frac{1280}{3}\*\zeta_5 \biggr)
    +\cfs\*\nfs\*\biggl(\frac{239633}{1458}-\frac{50140}{81} \* \zeta_2-\frac{19304}{81} \* \zeta_3
\\
&& \nonumber
+\frac{1312}{27} \* \zss\biggr)
    +\cf\*\ca\*\nfs\*\biggl(\frac{315755}{486}-\frac{9848}{27} \* \zeta_2-\frac{688}{9} \* \zeta_3+\frac{64}{5} \* \zss\biggr)
    -\cf\*\nft\*\biggl(\frac{17716}{729}
\\
&& \nonumber
-\frac{464}{27} \* \zeta_2\biggr)
+\floo\*\cf\*\nf\*\dabcnc\*\biggl(
    256
    +640\*\zeta_2
    +\frac{896}{3}\*\zeta_3
    -\frac{128}{5}\*\zss
    -\frac{5120}{3}\*\zeta_5
    \biggr)
    \biggr\}
\\
&& \nonumber
  +\DD_{0}\*\biggl\{
    \cff\*\biggl(\frac{15605}{16}+2803 \* \zeta_2
    +1064 \* \zeta_3
    +\frac{19839}{5} \* \zss-5168 \* \zeta_2 \* \zeta_3-6744 \* \zeta_5+736 \* \zts
\\
&& \nonumber
    +\frac{17624}{21} \* \zst
    -4992 \* \zeta_2 \* \zeta_5-\frac{752}{5} \* \zss \* \zeta_3+3840 \* \zeta_7
    - ~b_{4,\, {\cff}}^{\rm q}~
    \biggr)
    -\cft\*\ca\*\biggl(\frac{137084}{27}
\\
&& \nonumber
    +16683 \* \zeta_2
    -\frac{27860}{3} \* \zeta_3
    +\frac{1975109}{270} \* \zss-\frac{44332}{9} \* \zeta_2 \* \zeta_3
    -\frac{29240}{3} \* \zeta_5
    +\frac{12008}{3} \* \zts-\frac{906268}{315} \* \zst
\\
&& \nonumber
    +192 \* \zeta_2 \* \zeta_5
    -\frac{7992}{5} \* \zss \* \zeta_3
    + ~b_{4,\, {\cft\*\ca}}^{\rm q}~
    \biggr)
    +\cfs\*\cas\*\biggl(\frac{54699043}{3888}+\frac{16063075}{729} \* \zeta_2
\\
&& \nonumber
    -\frac{6500}{3} \* \zeta_3-\frac{7771513}{810} \* \zss-\frac{456400}{27} \* \zeta_2 \* \zeta_3
    +\frac{20248}{9} \* \zeta_5-\frac{1136}{3} \* \zts+\frac{17228}{63} \* \zst
\\
&& \nonumber
    +1856 \* \zeta_2 \* \zeta_5+\frac{25544}{15} \* \zss \* \zeta_3
    - ~b_{4,\, {\cfs\*\cas}}^{\rm q}~
    \biggr)
    -\cf\*\cat\*\biggl(\frac{1958802125}{139968}-\frac{2213182}{243} \* \zeta_2
\\
&& \nonumber
    -\frac{676939}{54} \* \zeta_3+\frac{198203}{135} \* \zss+\frac{38738}{9} \* \zeta_2 \* \zeta_3+\frac{40406}{9} \* \zeta_5
    +1340 \* \zts -\frac{43532}{105} \* \zst- 496 \* \zeta_2 \* \zeta_5
\\
&& \nonumber
    -\frac{1376}{5} \* \zss \* \zeta_3-2260 \* \zeta_7
    - \frac{1}{2}\* ~b_{4,\, {\cfs\*\cas}}^{\rm q}~
    - \frac{1}{4}\* ~b_{4,\, {\cft\*\ca}}^{\rm q}~
    - \frac{1}{8}\* ~b_{4,\, {\cff}}^{\rm q}~
    -\frac{1}{24}\* ~f_{4,\, \dfFA}^{\rm q}~ 
\\
&& \nonumber
    -\frac{1}{24}\* ~b_{4,\, \dfFA}^{\rm q}~ 
    \biggr)
    - \dfFAnc\* \left( ~f_{4,\, \dfFA}^{\rm q}~ + ~b_{4,\, \dfFA}^{\rm q}~ \right)
    -\cft\*\nf\*\biggl(
    \frac{39775}{108}
\\
&& \nonumber
    -\frac{38278}{27} \* \zeta_2+1518 \* \zeta_3-\frac{37502}{27} \* \zss
    -\frac{9664}{9} \* \zeta_2 \* \zeta_3 +\frac{2816}{3} \* \zeta_5+\frac{64}{3} \* \zts
    +\frac{93232}{315} \* \zst
\\
&& \nonumber
    + ~f_{4,\, {\nf\*\cft}}^{\rm q}~ 
    + ~b_{4,\, {\nf\*\cft}}^{\rm q}~ 
   \biggr)
    -\cfs\*\ca\*\nf\*\biggl(
    \frac{859349}{486}
    +\frac{5245922}{729} \* \zeta_2
    +\frac{191764}{81} \* \zeta_3
\\
&& \nonumber
    -\frac{1064717}{405} \* \zss
    -2576 \* \zeta_2 \* \zeta_3
    +\frac{1000}{3} \* \zeta_5
    -\frac{752}{3} \* \zts
    +\frac{2152}{45} \* \zst
    + ~f_{4,\, {\nf\*\cfs\*\ca}}^{\rm q}~ 
    + ~b_{4,\, {\nf\*\cfs\*\ca}}^{\rm q}~ 
    \biggr)
\\
&& \nonumber
    +\cf\*\cas\*\nf\*\biggl(
    \frac{68301461}{11664}
    -\frac{1935001}{486} \* \zeta_2
    -\frac{254678}{81} \* \zeta_3
    +\frac{66211}{135} \* \zss
    +\frac{8272}{9} \* \zeta_2 \* \zeta_3
    -\frac{772}{27} \* \zeta_5
\\
&& \nonumber
    +\frac{364}{9} \* \zts
    -\frac{75512}{945} \* \zst
    +\frac{1}{2}\* ~f_{4,\, {\nf\*\cfs\*\ca}}^{\rm q}~ 
    +\frac{1}{2}\* ~b_{4,\, {\nf\*\cfs\*\ca}}^{\rm q}~ 
    +\frac{1}{4}\* ~f_{4,\, {\nf\*\cft}}^{\rm q}~ 
    +\frac{1}{4}\* ~b_{4,\, {\nf\*\cft}}^{\rm q}~ 
\\
&& \nonumber
    -\frac{1}{48}\* ~b_{4,\, \dfFF}^{\rm q}~ 
    \biggr)
    + \nf\*\dfFFnc\* \biggl( 
          384
         -\frac{4544}{3}\*\zeta_2
         +\frac{5312}{9}\*\zeta_3
         +\frac{320}{3}\*\zss
         -128\*\zeta_2\*\zeta_3
\\
&& \nonumber
         +\frac{21760}{9}\*\zeta_5
         -\frac{9472}{315}\*\zst
         -\frac{1216}{3}\*\zts
         + ~b_{4,\, \dfFF}^{\rm q}~ 
       \biggr)
    -\cfs\*\nfs\*\biggl(
     \frac{161929}{972}
    -\frac{385300}{729} \* \zeta_2
\\
&& \nonumber
    -\frac{3812}{9} \* \zeta_3
    +\frac{19904}{135} \* \zss
    +\frac{1376}{27} \* \zeta_2 \* \zeta_3
    +\frac{64}{9} \* \zeta_5
    \biggr)
    -\cf\*\ca\*\nfs\*\biggl(
     \frac{3761509}{5832}
    -\frac{131878}{243} \* \zeta_2
\\
&& \nonumber
    -\frac{6092}{81} \* \zeta_3
    +\frac{616}{9} \* \zss
    +\frac{400}{9} \* \zeta_2 \* \zeta_3
    -\frac{1192}{9} \* \zeta_5
    \biggr)
    +\cf\*\nft\*\biggl(\frac{50558}{2187}
    -\frac{1880}{81} \* \zeta_2
    +\frac{80}{81} \* \zeta_3
\\
&& \nonumber
    +\frac{16}{9} \* \zss 
    \biggr)
-\floo\*\cf\*\nf\*\dabcnc\*\biggl(
    192
    +480\*\zeta_2
    +224\*\zeta_3
    -\frac{96}{5}\*\zss
    -1280\*\zeta_5
    \biggr)
    \biggr\}
\\
&& 
      +\delta(1-x)\*c_{4, \delta, {\rm DIS}}
\:\: .
    \label{appB4}
\eea
The coefficient $c_{4, \delta, {\rm DIS}}$ of $\delta(1-x)$ in eq.~(\ref{appB4}) is currently unknown. 
However, we estimate the size of these coefficients for fixed values of $\nf$
from the knowledge of moments of the DIS structure functions~\cite{Ruijl:2016pkm},
\begin{eqnarray}    
\label{eq:delta4loop}
c_{4, \delta, {\rm DIS}} (\nf =3) &=& (-13.5 \pm 3)\cdot 10^4\, , \nn\\
c_{4, \delta, {\rm DIS}} (\nf =4) &=& (-\phantom{1}6.0 \pm 3)\cdot 10^4\, , \nn\\
c_{4, \delta, {\rm DIS}} (\nf =5) &=& (-\phantom{1}0.5 \pm 3)\cdot 10^4\, .
\end{eqnarray}
These predictions have been used to extract the $g_{04}^{\rm DIS}$ coefficients
quoted in eq.~(\ref{eq:g04DIS}). 

\setcounter{equation}{0}
\renewcommand{\theequation}{\ref{app:appB}.\arabic{equation}}
\section{$g_0$ coefficients}
\label{app:appB}

Here we collect the $g_0$  coefficients that appear in eq.~(\ref{eq:cNres})
for both $N$-exponentiation and $\nt$-exponentiation. Note that the $g_0$
coefficients have the following perturbative expansion: 
\begin{align}
 g_0 = 1 + \sum_{n=1}^{\infty} a_s^i g_{0i} \,.
\end{align}
For $\nt$-exponentiation the $g_0$ coefficients (denoted as $\tilde{g}_0$) read as
\begin{align} 
\begin{autobreak} 
\gg01DIS = 
  \Cf    \bigg\{ 
- 2  \z2
- 9 \bigg\} ,   
\end{autobreak} 
\\ 
\begin{autobreak} 
\gg02DIS = 
  \Ca  \Cf    \bigg\{ \frac{51}{5}  \z2^2
- \frac{1139}{18}  \z2
+ \frac{464}{9}  \z3
- \frac{5465}{72} \bigg\}      
+ \Cf^2    \bigg\{ \frac{4}{5}  \z2^2
+ \frac{111}{2}  \z2
- 66  \z3
+ \frac{331}{8} \bigg\}      
+ \Cf  \nf    \bigg\{ \frac{85}{9}  \z2
+ \frac{4}{9}  \z3
+ \frac{457}{36} \bigg\} ,   
\end{autobreak} 
\\ 
\begin{autobreak} 
\gg03DIS = 
  \Ca^2  \Cf    \bigg\{ 
- \frac{12016}{315}  \z2^3
+ \frac{13151}{135}  \z2^2
+ \frac{3496}{9}  \z2  \z3
- \frac{78607}{54}  \z2
- \frac{248}{3}  \z3^2
+ \frac{115010}{81}  \z3
- \frac{416}{3}  \z5
- \frac{1909753}{1944} \bigg\}      
+ \Ca  \Cf^2    \bigg\{ 
- \frac{23098}{315}  \z2^3
+ \frac{11419}{27}  \z2^2
- 828  \z2  \z3
+ \frac{191545}{108}  \z2
+ \frac{536}{3}  \z3^2
- \frac{49346}{27}  \z3
- \frac{3896}{9}  \z5
+ \frac{9161}{12} \bigg\}      
+ \Ca  \Cf  \nf    \bigg\{ \frac{164}{135}  \z2^2
- \frac{64}{9}  \z2  \z3
+ \frac{33331}{81}  \z2
- \frac{21418}{81}  \z3
+ \frac{8}{3}  \z5
+ \frac{142883}{486} \bigg\}      
+ \Cf^3    \bigg\{ \frac{8144}{315}  \z2^3
- \frac{1791}{5}  \z2^2
+ 556  \z2  \z3
- \frac{6197}{12}  \z2
- \frac{176}{3}  \z3^2
- 411  \z3
+ 1384  \z5
- \frac{7255}{24} \bigg\}      
+ \Cf^2  \nf    \bigg\{ 
- \frac{10802}{135}  \z2^2
- \frac{40}{3}  \z2  \z3
- \frac{10733}{54}  \z2
+ \frac{10766}{27}  \z3
- \frac{784}{9}  \z5
- \frac{341}{36} \bigg\}      
+ \Cf  \nf^2    \bigg\{ 
- \frac{292}{135}  \z2^2
- \frac{2110}{81}  \z2
+ \frac{80}{81}  \z3
- \frac{9517}{486} \bigg\}      
+ \nf  \floo  \dabcton    \bigg\{ 
- \frac{32}{5}  \z2^2
+ 160  \z2
+ \frac{224}{3}  \z3
- \frac{1280}{3}  \z5
+ 64 \bigg\} \,.
\end{autobreak} 
\end{align}
For $N$-exponentiation, the $g_0$ coefficients are given as below,
\begin{align} 
\begin{autobreak} 
\label{eq:g01}
\g01DISN = 
  \Cf    \bigg\{ 
- 2  \z2
+ 2  \GE^2
+ 3  \GE
- 9 \bigg\} ,   
\end{autobreak} 
\\ 
\begin{autobreak} 
\label{eq:g02}
\g02DISN = 
  \Ca  \Cf    \bigg\{ \frac{51}{5}  \z2^2
- 4  \z2  \GE^2
- \frac{22}{3}  \z2  \GE
- \frac{1139}{18}  \z2
- 40  \z3  \GE
+ \frac{464}{9}  \z3
+ \frac{22}{9}  \GE^3
+ \frac{367}{18}  \GE^2
+ \frac{3155}{54}  \GE
- \frac{5465}{72} \bigg\}      
+ \Cf^2    \bigg\{ \frac{4}{5}  \z2^2
- 4  \z2  \GE^2
- 18  \z2  \GE
+ \frac{111}{2}  \z2
+ 24  \z3  \GE
- 66  \z3
+ 2  \GE^4
+ 6  \GE^3
- \frac{27}{2}  \GE^2
- \frac{51}{2}  \GE
+ \frac{331}{8} \bigg\}      
+ \Cf  \nf    \bigg\{ \frac{4}{3}  \z2  \GE
+ \frac{85}{9}  \z2
+ \frac{4}{9}  \z3
- \frac{4}{9}  \GE^3
- \frac{29}{9}  \GE^2
- \frac{247}{27}  \GE
+ \frac{457}{36} \bigg\} ,   
\end{autobreak} 
\\ 
\begin{autobreak} 
\label{eq:g03}
\g03DISN = 
  \Ca^2  \Cf    \bigg\{ 
- \frac{12016}{315}  \z2^3
+ \frac{88}{5}  \z2^2  \GE^2
+ \frac{212}{15}  \z2^2  \GE
+ \frac{13151}{135}  \z2^2
+ \frac{176}{3}  \z2  \z3  \GE
+ \frac{3496}{9}  \z2  \z3
- \frac{88}{9}  \z2  \GE^3
- \frac{778}{9}  \z2  \GE^2
- \frac{18179}{81}  \z2  \GE
- \frac{78607}{54}  \z2
- \frac{248}{3}  \z3^2
- 132  \z3  \GE^2
- \frac{6688}{9}  \z3  \GE
+ \frac{115010}{81}  \z3
+ 232  \z5  \GE
- \frac{416}{3}  \z5
+ \frac{121}{27}  \GE^4
+ \frac{4649}{81}  \GE^3
+ \frac{50689}{162}  \GE^2
+ \frac{599375}{729}  \GE
- \frac{1909753}{1944} \bigg\}      
+ \Ca  \Cf^2    \bigg\{ 
- \frac{23098}{315}  \z2^3
+ \frac{142}{5}  \z2^2  \GE^2
+ \frac{299}{3}  \z2^2  \GE
+ \frac{11419}{27}  \z2^2
+ 96  \z2  \z3  \GE
- 828  \z2  \z3
- 8  \z2  \GE^4
- \frac{284}{9}  \z2  \GE^3
- \frac{592}{3}  \z2  \GE^2
- \frac{28495}{54}  \z2  \GE
+ \frac{191545}{108}  \z2
+ \frac{536}{3}  \z3^2
- 80  \z3  \GE^3
+ \frac{640}{9}  \z3  \GE^2
+ 752  \z3  \GE
- \frac{49346}{27}  \z3
+ 120  \z5  \GE
- \frac{3896}{9}  \z5
+ \frac{44}{9}  \GE^5
+ \frac{433}{9}  \GE^4
+ \frac{8425}{54}  \GE^3
- \frac{5563}{36}  \GE^2
- \frac{16981}{24}  \GE
+ \frac{9161}{12} \bigg\}      
+ \Ca  \Cf  \nf    \bigg\{ 
- \frac{128}{15}  \z2^2  \GE
+ \frac{164}{135}  \z2^2
- \frac{64}{9}  \z2  \z3
+ \frac{16}{9}  \z2  \GE^3
+ \frac{56}{3}  \z2  \GE^2
+ \frac{5264}{81}  \z2  \GE
+ \frac{33331}{81}  \z2
+ 8  \z3  \GE^2
+ \frac{1976}{27}  \z3  \GE
- \frac{21418}{81}  \z3
+ \frac{8}{3}  \z5
- \frac{44}{27}  \GE^4
- \frac{1552}{81}  \GE^3
- \frac{7531}{81}  \GE^2
- \frac{160906}{729}  \GE
+ \frac{142883}{486} \bigg\}      
+ \Cf^3    \bigg\{ \frac{8144}{315}  \z2^3
+ \frac{8}{5}  \z2^2  \GE^2
+ 84  \z2^2  \GE
- \frac{1791}{5}  \z2^2
- 80  \z2  \z3  \GE
+ 556  \z2  \z3
- 4  \z2  \GE^4
- 36  \z2  \GE^3
+ 66  \z2  \GE^2
+ \frac{579}{2}  \z2  \GE
- \frac{6197}{12}  \z2
- \frac{176}{3}  \z3^2
+ 48  \z3  \GE^3
- 60  \z3  \GE^2
- 346  \z3  \GE
- 411  \z3
- 240  \z5  \GE
+ 1384  \z5
+ \frac{4}{3}  \GE^6
+ 6  \GE^5
- 9  \GE^4
- \frac{93}{2}  \GE^3
+ \frac{187}{4}  \GE^2
+ \frac{1001}{8}  \GE
- \frac{7255}{24} \bigg\}      
+ \Cf^2  \nf    \bigg\{ 
- \frac{8}{3}  \z2^2  \GE
- \frac{10802}{135}  \z2^2
- \frac{40}{3}  \z2  \z3
+ \frac{32}{9}  \z2  \GE^3
+ \frac{112}{3}  \z2  \GE^2
+ \frac{2177}{27}  \z2  \GE
- \frac{10733}{54}  \z2
+ \frac{8}{9}  \z3  \GE^2
- \frac{20}{9}  \z3  \GE
+ \frac{10766}{27}  \z3
- \frac{784}{9}  \z5
- \frac{8}{9}  \GE^5
- \frac{70}{9}  \GE^4
- \frac{683}{27}  \GE^3
+ \frac{83}{18}  \GE^2
+ \frac{2003}{108}  \GE
- \frac{341}{36} \bigg\}      
+ \Cf  \nf^2    \bigg\{ 
- \frac{292}{135}  \z2^2
- \frac{8}{9}  \z2  \GE^2
- \frac{116}{27}  \z2  \GE
- \frac{2110}{81}  \z2
+ \frac{64}{27}  \z3  \GE
+ \frac{80}{81}  \z3
+ \frac{4}{27}  \GE^4
+ \frac{116}{81}  \GE^3
+ \frac{470}{81}  \GE^2
+ \frac{8714}{729}  \GE
- \frac{9517}{486} \bigg\}      
+ \nf  \floo  \dabcton    \bigg\{ 
- \frac{32}{5}  \z2^2
+ 160  \z2
+ \frac{224}{3}  \z3
- \frac{1280}{3}  \z5
+ 64 \bigg\} \,. 
\end{autobreak} 
\end{align}
The resummed exponent for the $N$-exponentiation can be found for example in \cite{Moch:2005ba}.

\bibliography{disbib}

\providecommand{\href}[2]{#2}\begingroup\raggedright\begin{thebibliography}{10}

\bibitem{Vermaseren:2005qc}
J.~A.~M. Vermaseren, A.~Vogt and S.~Moch, \emph{{The Third-order QCD
  corrections to deep-inelastic scattering by photon exchange}},
  \href{http://dx.doi.org/10.1016/j.nuclphysb.2005.06.020}{\emph{Nucl. Phys.}
  {\bf B724} (2005) 3--182}, [\href{http://arxiv.org/abs/hep-ph/0504242}{{\tt
  hep-ph/0504242}}].

\bibitem{Moch:2008fj}
S.~Moch, J.~A.~M. Vermaseren and A.~Vogt, \emph{{Third-order QCD corrections to
  the charged-current structure function $F_3$}},
  \href{http://dx.doi.org/10.1016/j.nuclphysb.2009.01.001}{\emph{Nucl. Phys.}
  {\bf B813} (2009) 220--258}, [\href{http://arxiv.org/abs/0812.4168}{{\tt
  0812.4168}}].

\bibitem{Sterman:1986aj}
G.~F. Sterman, \emph{{Summation of Large Corrections to Short Distance Hadronic
  Cross-Sections}},
  \href{http://dx.doi.org/10.1016/0550-3213(87)90258-6}{\emph{Nucl. Phys.} {\bf
  B281} (1987) 310--364}.

\bibitem{Catani:1989ne}
S.~Catani and L.~Trentadue, \emph{{Resummation of the QCD Perturbative Series
  for Hard Processes}},
  \href{http://dx.doi.org/10.1016/0550-3213(89)90273-3}{\emph{Nucl. Phys.} {\bf
  B327} (1989) 323--352}.

\bibitem{Catani:1990rp}
S.~Catani and L.~Trentadue, \emph{{Comment on QCD exponentiation at large
  $x$}}, \href{http://dx.doi.org/10.1016/0550-3213(91)90506-S}{\emph{Nucl.
  Phys.} {\bf B353} (1991) 183--186}.

\bibitem{Catani:1996yz}
S.~Catani, M.~L. Mangano, P.~Nason and L.~Trentadue, \emph{{The Resummation of
  soft gluons in hadronic collisions}},
  \href{http://dx.doi.org/10.1016/0550-3213(96)00399-9}{\emph{Nucl. Phys.} {\bf
  B478} (1996) 273--310}, [\href{http://arxiv.org/abs/hep-ph/9604351}{{\tt
  hep-ph/9604351}}].

\bibitem{Contopanagos:1996nh}
H.~Contopanagos, E.~Laenen and G.~F. Sterman, \emph{{Sudakov factorization and
  resummation}},
  \href{http://dx.doi.org/10.1016/S0550-3213(96)00567-6}{\emph{Nucl. Phys.}
  {\bf B484} (1997) 303--330}, [\href{http://arxiv.org/abs/hep-ph/9604313}{{\tt
  hep-ph/9604313}}].

\bibitem{Moch:2005ba}
S.~Moch, J.~A.~M. Vermaseren and A.~Vogt, \emph{{Higher-order corrections in
  threshold resummation}},
  \href{http://dx.doi.org/10.1016/j.nuclphysb.2005.08.005}{\emph{Nucl. Phys.}
  {\bf B726} (2005) 317--335}, [\href{http://arxiv.org/abs/hep-ph/0506288}{{\tt
  hep-ph/0506288}}].

\bibitem{Manohar:2003vb}
A.~V. Manohar, \emph{{Deep inelastic scattering as $x \to 1$ using soft
  collinear effective theory}},
  \href{http://dx.doi.org/10.1103/PhysRevD.68.114019}{\emph{Phys. Rev.} {\bf
  D68} (2003) 114019}, [\href{http://arxiv.org/abs/hep-ph/0309176}{{\tt
  hep-ph/0309176}}].

\bibitem{Chay:2005rz}
J.~Chay and C.~Kim, \emph{{Deep inelastic scattering near the endpoint in
  soft-collinear effective theory}},
  \href{http://dx.doi.org/10.1103/PhysRevD.75.016003}{\emph{Phys. Rev.} {\bf
  D75} (2007) 016003}, [\href{http://arxiv.org/abs/hep-ph/0511066}{{\tt
  hep-ph/0511066}}].

\bibitem{Idilbi:2006dg}
A.~Idilbi, X.-d. Ji and F.~Yuan, \emph{{Resummation of threshold logarithms in
  effective field theory for DIS, Drell-Yan and Higgs production}},
  \href{http://dx.doi.org/10.1016/j.nuclphysb.2006.07.002}{\emph{Nucl. Phys.}
  {\bf B753} (2006) 42--68}, [\href{http://arxiv.org/abs/hep-ph/0605068}{{\tt
  hep-ph/0605068}}].

\bibitem{Becher:2006mr}
T.~Becher, M.~Neubert and B.~D. Pecjak, \emph{{Factorization and Momentum-Space
  Resummation in Deep-Inelastic Scattering}},
  \href{http://dx.doi.org/10.1088/1126-6708/2007/01/076}{\emph{JHEP} {\bf 01}
  (2007) 076}, [\href{http://arxiv.org/abs/hep-ph/0607228}{{\tt
  hep-ph/0607228}}].

\bibitem{Accardi:2016ndt}
A.~Accardi et~al., \emph{{A Critical Appraisal and Evaluation of Modern PDFs}},
  \href{http://dx.doi.org/10.1140/epjc/s10052-016-4285-4}{\emph{Eur. Phys. J.}
  {\bf C76} (2016) 471}, [\href{http://arxiv.org/abs/1603.08906}{{\tt
  1603.08906}}].

\bibitem{Collins:1980ih}
J.~C. Collins, \emph{{Algorithm to Compute Corrections to the Sudakov
  Form-factor}}, \href{http://dx.doi.org/10.1103/PhysRevD.22.1478}{\emph{Phys.
  Rev.} {\bf D22} (1980) 1478}.

\bibitem{Sen:1981sd}
A.~Sen, \emph{{Asymptotic Behavior of the Sudakov Form-Factor in QCD}},
  \href{http://dx.doi.org/10.1103/PhysRevD.24.3281}{\emph{Phys. Rev.} {\bf D24}
  (1981) 3281}.

\bibitem{Korchemsky:1988hd}
G.~P. Korchemsky, \emph{{Sudakov Form-factor in {QCD}}},
  \href{http://dx.doi.org/10.1016/0370-2693(89)90799-5}{\emph{Phys. Lett.} {\bf
  B220} (1989) 629--634}.

\bibitem{Magnea:1990zb}
L.~Magnea and G.~F. Sterman, \emph{{Analytic continuation of the Sudakov
  form-factor in QCD}},
  \href{http://dx.doi.org/10.1103/PhysRevD.42.4222}{\emph{Phys. Rev.} {\bf D42}
  (1990) 4222--4227}.

\bibitem{Magnea:2000ss}
L.~Magnea, \emph{{Analytic resummation for the quark form-factor in QCD}},
  \href{http://dx.doi.org/10.1016/S0550-3213(00)00623-4}{\emph{Nucl. Phys.}
  {\bf B593} (2001) 269--288}, [\href{http://arxiv.org/abs/hep-ph/0006255}{{\tt
  hep-ph/0006255}}].

\bibitem{Moch:2005id}
S.~Moch, J.~A.~M. Vermaseren and A.~Vogt, \emph{{The Quark form-factor at
  higher orders}},
  \href{http://dx.doi.org/10.1088/1126-6708/2005/08/049}{\emph{JHEP} {\bf 08}
  (2005) 049}, [\href{http://arxiv.org/abs/hep-ph/0507039}{{\tt
  hep-ph/0507039}}].

\bibitem{Moch:2004pa}
S.~Moch, J.~A.~M. Vermaseren and A.~Vogt, \emph{{The Three loop splitting
  functions in QCD: The Nonsinglet case}},
  \href{http://dx.doi.org/10.1016/j.nuclphysb.2004.03.030}{\emph{Nucl. Phys.}
  {\bf B688} (2004) 101--134}, [\href{http://arxiv.org/abs/hep-ph/0403192}{{\tt
  hep-ph/0403192}}].

\bibitem{Vogt:2004mw}
A.~Vogt, S.~Moch and J.~A.~M. Vermaseren, \emph{{The Three-loop splitting
  functions in QCD: The Singlet case}},
  \href{http://dx.doi.org/10.1016/j.nuclphysb.2004.04.024}{\emph{Nucl. Phys.}
  {\bf B691} (2004) 129--181}, [\href{http://arxiv.org/abs/hep-ph/0404111}{{\tt
  hep-ph/0404111}}].

\bibitem{Kinoshita:1962ur}
T.~Kinoshita, \emph{{Mass singularities of Feynman amplitudes}},
  \href{http://dx.doi.org/10.1063/1.1724268}{\emph{J. Math. Phys.} {\bf 3}
  (1962) 650--677}.

\bibitem{Lee:1964is}
T.~D. Lee and M.~Nauenberg, \emph{{Degenerate Systems and Mass Singularities}},
  \href{http://dx.doi.org/10.1103/PhysRev.133.B1549}{\emph{Phys. Rev.} {\bf
  133} (1964) B1549--B1562}.

\bibitem{Moch:2005ky}
S.~Moch and A.~Vogt, \emph{{Higher-order soft corrections to lepton pair and
  Higgs boson production}},
  \href{http://dx.doi.org/10.1016/j.physletb.2005.09.061}{\emph{Phys. Lett.}
  {\bf B631} (2005) 48--57}, [\href{http://arxiv.org/abs/hep-ph/0508265}{{\tt
  hep-ph/0508265}}].

\bibitem{Laenen:2005uz}
E.~Laenen and L.~Magnea, \emph{{Threshold resummation for electroweak
  annihilation from DIS data}},
  \href{http://dx.doi.org/10.1016/j.physletb.2005.10.038}{\emph{Phys. Lett.}
  {\bf B632} (2006) 270--276}, [\href{http://arxiv.org/abs/hep-ph/0508284}{{\tt
  hep-ph/0508284}}].

\bibitem{Ravindran:2005vv}
V.~Ravindran, \emph{{On Sudakov and soft resummations in QCD}},
  \href{http://dx.doi.org/10.1016/j.nuclphysb.2006.04.008}{\emph{Nucl. Phys.}
  {\bf B746} (2006) 58--76}, [\href{http://arxiv.org/abs/hep-ph/0512249}{{\tt
  hep-ph/0512249}}].

\bibitem{Ravindran:2006cg}
V.~Ravindran, \emph{{Higher-order threshold effects to inclusive processes in
  QCD}}, \href{http://dx.doi.org/10.1016/j.nuclphysb.2006.06.025}{\emph{Nucl.
  Phys.} {\bf B752} (2006) 173--196},
  [\href{http://arxiv.org/abs/hep-ph/0603041}{{\tt hep-ph/0603041}}].

\bibitem{Lee:2016ixa}
J.~Henn, A.~V. Smirnov, V.~A. Smirnov, M.~Steinhauser and R.~N. Lee,
  \emph{{Four-loop photon quark form factor and cusp anomalous dimension in the
  large-$N_c$ limit of QCD}},
  \href{http://dx.doi.org/10.1007/JHEP03(2017)139}{\emph{JHEP} {\bf 03} (2017)
  139}, [\href{http://arxiv.org/abs/1612.04389}{{\tt 1612.04389}}].

\bibitem{Moch:2017uml}
S.~Moch, B.~Ruijl, T.~Ueda, J.~A.~M. Vermaseren and A.~Vogt, \emph{{Four-Loop
  Non-Singlet Splitting Functions in the Planar Limit and Beyond}},
  \href{http://dx.doi.org/10.1007/JHEP10(2017)041}{\emph{JHEP} {\bf 10} (2017)
  041}, [\href{http://arxiv.org/abs/1707.08315}{{\tt 1707.08315}}].

\bibitem{Ruijl:2016pkm}
B.~Ruijl, T.~Ueda, J.~A.~M. Vermaseren, J.~Davies and A.~Vogt, \emph{{First
  Forcer results on deep-inelastic scattering and related quantities}},
  \href{http://dx.doi.org/10.22323/1.260.0071}{\emph{PoS} {\bf LL2016} (2016)
  071}, [\href{http://arxiv.org/abs/1605.08408}{{\tt 1605.08408}}].

\bibitem{Lee:2017mip}
R.~N. Lee, A.~V. Smirnov, V.~A. Smirnov and M.~Steinhauser, \emph{{The $n_f^2$
  contributions to fermionic four-loop form factors}},
  \href{http://dx.doi.org/10.1103/PhysRevD.96.014008}{\emph{Phys. Rev.} {\bf
  D96} (2017) 014008}, [\href{http://arxiv.org/abs/1705.06862}{{\tt
  1705.06862}}].

\bibitem{Grozin:2018vdn}
A.~Grozin, \emph{{Four-loop cusp anomalous dimension in QED}},
  \href{http://dx.doi.org/10.1007/JHEP06(2018)073,
  10.1007/JHEP01(2019)134}{\emph{JHEP} {\bf 06} (2018) 073},
  [\href{http://arxiv.org/abs/1805.05050}{{\tt 1805.05050}}].

\bibitem{Henn:2019rmi}
J.~M. Henn, T.~Peraro, M.~Stahlhofen and P.~Wasser, \emph{{Matter dependence of
  the four-loop cusp anomalous dimension}},
  \href{http://dx.doi.org/10.1103/PhysRevLett.122.201602}{\emph{Phys. Rev.
  Lett.} {\bf 122} (2019) 201602}, [\href{http://arxiv.org/abs/1901.03693}{{\tt
  1901.03693}}].

\bibitem{Bruser:2019auj}
R.~Br{\"u}ser, A.~Grozin, J.~M. Henn and M.~Stahlhofen, \emph{{Matter
  dependence of the four-loop QCD cusp anomalous dimension: from small angles
  to all angles}}, \href{http://dx.doi.org/10.1007/JHEP05(2019)186}{\emph{JHEP}
  {\bf 05} (2019) 186}, [\href{http://arxiv.org/abs/1902.05076}{{\tt
  1902.05076}}].

\bibitem{Moch:2018wjh}
S.~Moch, B.~Ruijl, T.~Ueda, J.~A.~M. Vermaseren and A.~Vogt, \emph{{On quartic
  colour factors in splitting functions and the gluon cusp anomalous
  dimension}},
  \href{http://dx.doi.org/10.1016/j.physletb.2018.06.017}{\emph{Phys. Lett.}
  {\bf B782} (2018) 627--632}, [\href{http://arxiv.org/abs/1805.09638}{{\tt
  1805.09638}}].

\bibitem{Lee:2019zop}
R.~N. Lee, A.~V. Smirnov, V.~A. Smirnov and M.~Steinhauser, \emph{{Four-loop
  quark form factor with quartic fundamental colour factor}},
  \href{http://dx.doi.org/10.1007/JHEP02(2019)172}{\emph{JHEP} {\bf 02} (2019)
  172}, [\href{http://arxiv.org/abs/1901.02898}{{\tt 1901.02898}}].

\bibitem{vonManteuffel:2019wbj}
A.~von Manteuffel and R.~M. Schabinger, \emph{{Quark and gluon form factors in
  four loop QCD: The $N_f^2$ and $N_{q\gamma} N_f$ contributions}},
  \href{http://dx.doi.org/10.1103/PhysRevD.99.094014}{\emph{Phys. Rev.} {\bf
  D99} (2019) 094014}, [\href{http://arxiv.org/abs/1902.08208}{{\tt
  1902.08208}}].

\bibitem{Henn:2019swt}
J.~M. Henn, G.~P. Korchemsky and B.~Mistlberger, \emph{{The full four-loop cusp
  anomalous dimension in $\mathcal{N}=4$ super Yang-Mills and QCD}},
  \href{http://arxiv.org/abs/1911.10174}{{\tt 1911.10174}}.

\bibitem{Huber:2019fxe}
T.~Huber, A.~von Manteuffel, E.~Panzer, R.~M. Schabinger and G.~Yang,
  \emph{{The Four-Loop Cusp Anomalous Dimension from the $\mathcal{N} = 4$
  Sudakov Form Factor}},  \href{http://arxiv.org/abs/1912.13459}{{\tt
  1912.13459}}.

\bibitem{Das:2019uvh}
G.~Das, S.~Moch and A.~Vogt, \emph{{Soft corrections to inclusive DIS at four
  loops and beyond}},  in \emph{{27th International Workshop on Deep Inelastic
  Scattering and Related Subjects (DIS 2019) Torino, Italy, April 8-12, 2019}},
  2019.
\newblock \href{http://arxiv.org/abs/1908.03071}{{\tt 1908.03071}}.

\bibitem{Catani:1998tm}
S.~Catani, M.~L. Mangano and P.~Nason, \emph{{Sudakov resummation for prompt
  photon production in hadron collisions}},
  \href{http://dx.doi.org/10.1088/1126-6708/1998/07/024}{\emph{JHEP} {\bf 07}
  (1998) 024}, [\href{http://arxiv.org/abs/hep-ph/9806484}{{\tt
  hep-ph/9806484}}].

\bibitem{Forte:2002ni}
S.~Forte and G.~Ridolfi, \emph{{Renormalization group approach to soft gluon
  resummation}},
  \href{http://dx.doi.org/10.1016/S0550-3213(02)01034-9}{\emph{Nucl. Phys.}
  {\bf B650} (2003) 229--270}, [\href{http://arxiv.org/abs/hep-ph/0209154}{{\tt
  hep-ph/0209154}}].

\bibitem{Gardi:2002xm}
E.~Gardi and R.~G. Roberts, \emph{{The Interplay between Sudakov resummation,
  renormalons and higher twist in deep inelastic scattering}},
  \href{http://dx.doi.org/10.1016/S0550-3213(03)00035-X}{\emph{Nucl. Phys.}
  {\bf B653} (2003) 227--255}, [\href{http://arxiv.org/abs/hep-ph/0210429}{{\tt
  hep-ph/0210429}}].

\bibitem{Catani:2003zt}
S.~Catani, D.~de~Florian, M.~Grazzini and P.~Nason, \emph{{Soft gluon
  resummation for Higgs boson production at hadron colliders}},
  \href{http://dx.doi.org/10.1088/1126-6708/2003/07/028}{\emph{JHEP} {\bf 07}
  (2003) 028}, [\href{http://arxiv.org/abs/hep-ph/0306211}{{\tt
  hep-ph/0306211}}].

\bibitem{Vogt:2000ci}
A.~Vogt, \emph{{Next-to-next-to-leading logarithmic threshold resummation for
  deep inelastic scattering and the Drell-Yan process}},
  \href{http://dx.doi.org/10.1016/S0370-2693(00)01344-7}{\emph{Phys. Lett.}
  {\bf B497} (2001) 228--234}, [\href{http://arxiv.org/abs/hep-ph/0010146}{{\tt
  hep-ph/0010146}}].

\bibitem{Baikov:2016tgj}
P.~A. Baikov, K.~G. Chetyrkin and J.~H. K{\"u}hn, \emph{{Five-Loop Running of
  the QCD coupling constant}},
  \href{http://dx.doi.org/10.1103/PhysRevLett.118.082002}{\emph{Phys. Rev.
  Lett.} {\bf 118} (2017) 082002}, [\href{http://arxiv.org/abs/1606.08659}{{\tt
  1606.08659}}].

\bibitem{Herzog:2017ohr}
F.~Herzog, B.~Ruijl, T.~Ueda, J.~A.~M. Vermaseren and A.~Vogt, \emph{{The
  five-loop beta function of Yang-Mills theory with fermions}},
  \href{http://dx.doi.org/10.1007/JHEP02(2017)090}{\emph{JHEP} {\bf 02} (2017)
  090}, [\href{http://arxiv.org/abs/1701.01404}{{\tt 1701.01404}}].

\bibitem{Luthe:2017ttg}
T.~Luthe, A.~Maier, P.~Marquard and Y.~Schr{\"o}der, \emph{{The five-loop Beta
  function for a general gauge group and anomalous dimensions beyond Feynman
  gauge}}, \href{http://dx.doi.org/10.1007/JHEP10(2017)166}{\emph{JHEP} {\bf
  10} (2017) 166}, [\href{http://arxiv.org/abs/1709.07718}{{\tt 1709.07718}}].

\bibitem{Chetyrkin:2017bjc}
K.~G. Chetyrkin, G.~Falcioni, F.~Herzog and J.~A.~M. Vermaseren,
  \emph{{Five-loop renormalisation of QCD in covariant gauges}},
  \href{http://dx.doi.org/10.1007/JHEP12(2017)006, 10.3204/PUBDB-2018-02123,
  10.1007/JHEP10(2017)179}{\emph{JHEP} {\bf 10} (2017) 179},
  [\href{http://arxiv.org/abs/1709.08541}{{\tt 1709.08541}}].

\bibitem{Herzog:2018kwj}
F.~Herzog, S.~Moch, B.~Ruijl, T.~Ueda, J.~A.~M. Vermaseren and A.~Vogt,
  \emph{{Five-loop contributions to low-N non-singlet anomalous dimensions in
  QCD}}, \href{http://dx.doi.org/10.1016/j.physletb.2019.01.060}{\emph{Phys.
  Lett.} {\bf B790} (2019) 436--443},
  [\href{http://arxiv.org/abs/1812.11818}{{\tt 1812.11818}}].

\bibitem{Moch:2015usa}
S.~Moch, J.~A.~M. Vermaseren and A.~Vogt, \emph{{On $\gamma_5$ in higher-order
  QCD calculations and the NNLO evolution of the polarized valence
  distribution}},
  \href{http://dx.doi.org/10.1016/j.physletb.2015.07.027}{\emph{Phys. Lett.}
  {\bf B748} (2015) 432--438}, [\href{http://arxiv.org/abs/1506.04517}{{\tt
  1506.04517}}].

\bibitem{Korchemsky:1988si}
G.~P. Korchemsky, \emph{{Asymptotics of the Altarelli-Parisi-Lipatov Evolution
  Kernels of Parton Distributions}},
  \href{http://dx.doi.org/10.1142/S0217732389001453}{\emph{Mod. Phys. Lett.}
  {\bf A4} (1989) 1257--1276}.

\bibitem{Albino:2000cp}
S.~Albino and R.~D. Ball, \emph{{Soft resummation of quark anomalous dimensions
  and coefficient functions in MS-bar factorization}},
  \href{http://dx.doi.org/10.1016/S0370-2693(01)00742-0}{\emph{Phys. Lett.}
  {\bf B513} (2001) 93--102}, [\href{http://arxiv.org/abs/hep-ph/0011133}{{\tt
  hep-ph/0011133}}].

\bibitem{Davies:2016jie}
J.~Davies, A.~Vogt, B.~Ruijl, T.~Ueda and J.~A.~M. Vermaseren,
  \emph{{Large-$n_f$ contributions to the four-loop splitting functions in
  QCD}}, \href{http://dx.doi.org/10.1016/j.nuclphysb.2016.12.012}{\emph{Nucl.
  Phys.} {\bf B915} (2017) 335--362},
  [\href{http://arxiv.org/abs/1610.07477}{{\tt 1610.07477}}].

\bibitem{Gracey:1994nn}
J.~A. Gracey, \emph{{Anomalous dimension of nonsinglet Wilson operators at
  ${\cal O}(1/n_f)$ in deep inelastic scattering}},
  \href{http://dx.doi.org/10.1016/0370-2693(94)90502-9}{\emph{Phys. Lett.} {\bf
  B322} (1994) 141--146}, [\href{http://arxiv.org/abs/hep-ph/9401214}{{\tt
  hep-ph/9401214}}].

\bibitem{Beneke:1995pq}
M.~Beneke and V.~M. Braun, \emph{{Power corrections and renormalons in
  Drell-Yan production}},
  \href{http://dx.doi.org/10.1016/0550-3213(95)00439-Y}{\emph{Nucl. Phys.} {\bf
  B454} (1995) 253--290}, [\href{http://arxiv.org/abs/hep-ph/9506452}{{\tt
  hep-ph/9506452}}].

\bibitem{Baikov:2009bg}
P.~A. Baikov, K.~G. Chetyrkin, A.~V. Smirnov, V.~A. Smirnov and M.~Steinhauser,
  \emph{{Quark and gluon form factors to three loops}},
  \href{http://dx.doi.org/10.1103/PhysRevLett.102.212002}{\emph{Phys. Rev.
  Lett.} {\bf 102} (2009) 212002}, [\href{http://arxiv.org/abs/0902.3519}{{\tt
  0902.3519}}].

\bibitem{Gehrmann:2010ue}
T.~Gehrmann, E.~W.~N. Glover, T.~Huber, N.~Ikizlerli and C.~Studerus,
  \emph{{Calculation of the quark and gluon form factors to three loops in
  QCD}}, \href{http://dx.doi.org/10.1007/JHEP06(2010)094}{\emph{JHEP} {\bf 06}
  (2010) 094}, [\href{http://arxiv.org/abs/1004.3653}{{\tt 1004.3653}}].

\bibitem{Gehrmann:2010tu}
T.~Gehrmann, E.~W.~N. Glover, T.~Huber, N.~Ikizlerli and C.~Studerus,
  \emph{{The quark and gluon form factors to three loops in QCD through to
  $O(\epsilon^2)$}},
  \href{http://dx.doi.org/10.1007/JHEP11(2010)102}{\emph{JHEP} {\bf 11} (2010)
  102}, [\href{http://arxiv.org/abs/1010.4478}{{\tt 1010.4478}}].

\bibitem{Ravindran:2004mb}
V.~Ravindran, J.~Smith and W.~L. van Neerven, \emph{{Two-loop corrections to
  Higgs boson production}},
  \href{http://dx.doi.org/10.1016/j.nuclphysb.2004.10.039}{\emph{Nucl. Phys.}
  {\bf B704} (2005) 332--348}, [\href{http://arxiv.org/abs/hep-ph/0408315}{{\tt
  hep-ph/0408315}}].

\bibitem{Moch:2005tm}
S.~Moch, J.~A.~M. Vermaseren and A.~Vogt, \emph{{Three-loop results for quark
  and gluon form-factors}},
  \href{http://dx.doi.org/10.1016/j.physletb.2005.08.067}{\emph{Phys. Lett.}
  {\bf B625} (2005) 245--252}, [\href{http://arxiv.org/abs/hep-ph/0508055}{{\tt
  hep-ph/0508055}}].

\bibitem{Dixon:2008gr}
L.~J. Dixon, L.~Magnea and G.~F. Sterman, \emph{{Universal structure of
  subleading infrared poles in gauge theory amplitudes}},
  \href{http://dx.doi.org/10.1088/1126-6708/2008/08/022}{\emph{JHEP} {\bf 08}
  (2008) 022}, [\href{http://arxiv.org/abs/0805.3515}{{\tt 0805.3515}}].

\bibitem{Falcioni:2019nxk}
G.~Falcioni, E.~Gardi and C.~Milloy, \emph{{Relating amplitude and PDF
  factorisation through Wilson-line geometries}},
  \href{http://arxiv.org/abs/1909.00697}{{\tt 1909.00697}}.

\bibitem{Vogt:2004ns}
A.~Vogt, \emph{{Efficient evolution of unpolarized and polarized parton
  distributions with QCD-PEGASUS}},
  \href{http://dx.doi.org/10.1016/j.cpc.2005.03.103}{\emph{Comput. Phys.
  Commun.} {\bf 170} (2005) 65--92},
  [\href{http://arxiv.org/abs/hep-ph/0408244}{{\tt hep-ph/0408244}}].

\bibitem{Vogt:1999xa}
A.~Vogt, \emph{{On soft gluon effects in deep inelastic structure functions}},
  \href{http://dx.doi.org/10.1016/S0370-2693(99)01325-8}{\emph{Phys. Lett.}
  {\bf B471} (1999) 97--102}, [\href{http://arxiv.org/abs/hep-ph/9910545}{{\tt
  hep-ph/9910545}}].

\bibitem{Moch:2009hr}
S.~Moch and A.~Vogt, \emph{{On non-singlet physical evolution kernels and
  large-$x$ coefficient functions in perturbative QCD}},
  \href{http://dx.doi.org/10.1088/1126-6708/2009/11/099}{\emph{JHEP} {\bf 11}
  (2009) 099}, [\href{http://arxiv.org/abs/0909.2124}{{\tt 0909.2124}}].

\bibitem{Bonocore:2015esa}
D.~Bonocore, E.~Laenen, L.~Magnea, S.~Melville, L.~Vernazza and C.~D. White,
  \emph{{A factorization approach to next-to-leading-power threshold
  logarithms}}, \href{http://dx.doi.org/10.1007/JHEP06(2015)008}{\emph{JHEP}
  {\bf 06} (2015) 008}, [\href{http://arxiv.org/abs/1503.05156}{{\tt
  1503.05156}}].

\bibitem{Kawamura:2012cr}
H.~Kawamura, N.~A. Lo~Presti, S.~Moch and A.~Vogt, \emph{{On the
  next-to-next-to-leading order QCD corrections to heavy-quark production in
  deep-inelastic scattering}},
  \href{http://dx.doi.org/10.1016/j.nuclphysb.2012.07.001}{\emph{Nucl. Phys.}
  {\bf B864} (2012) 399--468}, [\href{http://arxiv.org/abs/1205.5727}{{\tt
  1205.5727}}].

\bibitem{Hoang:2015iva}
A.~H. Hoang, P.~Pietrulewicz and D.~Samitz, \emph{{Variable Flavor Number
  Scheme for Final State Jets in DIS}},
  \href{http://dx.doi.org/10.1103/PhysRevD.93.034034}{\emph{Phys. Rev.} {\bf
  D93} (2016) 034034}, [\href{http://arxiv.org/abs/1508.04323}{{\tt
  1508.04323}}].

\bibitem{Vermaseren:2000nd}
J.~A.~M. Vermaseren, \emph{{New features of FORM}},
  \href{http://arxiv.org/abs/math-ph/0010025}{{\tt math-ph/0010025}}.

\bibitem{Ruijl:2017dtg}
B.~Ruijl, T.~Ueda and J.~Vermaseren, \emph{{FORM version 4.2}},
  \href{http://arxiv.org/abs/1707.06453}{{\tt 1707.06453}}.

\bibitem{vonManteuffel:2020vjv}
A.~von Manteuffel, E.~Panzer and R.~M. Schabinger, \emph{{Analytic four-loop
  anomalous dimensions in massless QCD from form factors}},
  \href{http://arxiv.org/abs/2002.04617}{{\tt 2002.04617}}.

\bibitem{Bardeen:1978yd}
W.~A. Bardeen, A.~J. Buras, D.~W. Duke and T.~Muta, \emph{{Deep Inelastic
  Scattering Beyond the Leading Order in Asymptotically Free Gauge Theories}},
  \href{http://dx.doi.org/10.1103/PhysRevD.18.3998}{\emph{Phys. Rev.} {\bf D18}
  (1978) 3998}.

\bibitem{vanNeerven:1991nn}
W.~L. van Neerven and E.~B. Zijlstra, \emph{{Order $\alpha_s^2$ contributions
  to the deep inelastic Wilson coefficient}},
  \href{http://dx.doi.org/10.1016/0370-2693(91)91024-P}{\emph{Phys. Lett.} {\bf
  B272} (1991) 127--133}.

\bibitem{Moch:1999eb}
S.~Moch and J.~A.~M. Vermaseren, \emph{{Deep inelastic structure functions at
  two loops}},
  \href{http://dx.doi.org/10.1016/S0550-3213(00)00045-6}{\emph{Nucl. Phys.}
  {\bf B573} (2000) 853--907}, [\href{http://arxiv.org/abs/hep-ph/9912355}{{\tt
  hep-ph/9912355}}].

\end{thebibliography}\endgroup
\bibliographystyle{JHEP}
\end{document}